\documentclass[a4paper,12pt]{article}
\pdfoutput=1
\usepackage{jheppub}
\usepackage[utf8]{inputenc}
\usepackage{url}
\usepackage{mathrsfs}  
\usepackage{pdflscape}
\usepackage{refcount}
\usepackage{booktabs}
\usepackage{todonotes}
\usepackage{enumitem}
\usepackage{adjustbox}
\usepackage{afterpage}
\usepackage{mathdots}

\usepackage{amsfonts}
\usepackage{bbm}

\usepackage{pdfpages}
\usepackage{caption}
\usepackage{subcaption}

\usepackage{tikz}
\usetikzlibrary{positioning}
\usetikzlibrary{arrows}
\usetikzlibrary{decorations.pathreplacing}
\usetikzlibrary{shapes.geometric}
\usetikzlibrary{calc}
\usetikzlibrary{decorations.pathmorphing}
\usetikzlibrary{shapes.misc}

\definecolor{goodgreen}{RGB}{55,169,49}

\definecolor{darkyellow}{RGB}{230,170,10}

\definecolor{brightyellow}{RGB}{255,240,190}

\tikzset{flavour/.style={draw=none,minimum size=0.3mm,fill=white, regular polygon,regular polygon sides=4,draw}}
\tikzset{gaugeBig/.style={inner sep=1mm,draw=none,fill=white,minimum size=2mm,circle, draw}}
\tikzset{bd/.style={circle, draw=black, inner sep=0pt, fill=black, minimum size=2mm}}
\tikzset{wd/.style={circle, draw=black, inner sep=0pt, fill=white, minimum size=2mm}}
\tikzset{Dynkin/.style={circle, draw=black, inner sep=0pt, fill=white, minimum size=2mm}}
\tikzstyle{ligne}=[draw, very thick] 
\tikzstyle{gridline}=[draw, gray] 
\tikzset{gauge/.style={circle, draw,inner sep=2.5pt}}
\tikzset{gaugeo/.style={circle, draw,inner sep=2.5pt,fill=orange}}
\tikzset{gaugec/.style={circle, draw,inner sep=2.5pt,fill=cyan}}
\tikzset{gauger/.style={circle, draw,inner sep=2.5pt,fill=red}}
\tikzset{gaugeb/.style={circle, draw,inner sep=2.5pt,fill=blue}}
\tikzset{gaugeg/.style={circle, draw,inner sep=2.5pt,fill=green}}
\tikzset{gaugem/.style={circle, draw,inner sep=2.5pt,fill=magenta}}
\tikzset{hasse/.style={circle, fill,inner sep=2pt}}
\tikzset{shrinky/.style={circle, fill,inner sep=1pt}}
\tikzset{sized/.style={circle, draw, inner sep=1.5pt}}
\tikzset{seven/.style={circle, draw,inner sep=3pt}}

\tikzset{dotto/.style={circle, orange, draw,inner sep=1.5pt,fill=orange}}
\tikzset{dottp/.style={circle, purple, draw,inner sep=1.5pt,fill=purple}}
\tikzset{dottc/.style={circle, cyan, draw,inner sep=1.5pt,fill=cyan}}
\tikzset{dottr/.style={circle, red, draw,inner sep=1.5pt,fill=red}}
\tikzset{dottb/.style={circle, blue, draw,inner sep=1.5pt,fill=blue}}
\tikzset{dottg/.style={circle, green, draw,inner sep=1.5pt,fill=green}}
\tikzset{dottm/.style={circle, magenta, draw,inner sep=1.5pt,fill=magenta}}

\def\edge#1#2#3#4#5{
    \node (loc1) at ($(#1)!0.25!(#2)$) {$#3$};
    \node (loc2) at ($(#1)!0.75!(#2)$) {$#4$};
    \node[transform canvas={yshift=6pt}] (loc3) at ($(#1)!0.5!(#2)$) {$#5$};
    \draw[double] (#1)--(loc1)--(loc2)--(#2);
}

\def\edger#1#2#3#4#5{
    \node (loc1) at ($(#1)!0.25!(#2)$) {$#3$};
    \node (loc2) at ($(#1)!0.75!(#2)$) {$#4$};
    \node[transform canvas={xshift=6pt}] (loc3) at ($(#1)!0.5!(#2)$) {$#5$};
    \draw[double] (#1)--(loc1)--(loc2)--(#2);
}
\def\edgel#1#2#3#4#5{
    \node (loc1) at ($(#1)!0.25!(#2)$) {$#3$};
    \node (loc2) at ($(#1)!0.75!(#2)$) {$#4$};
    \node[transform canvas={xshift=-6pt}] (loc3) at ($(#1)!0.5!(#2)$) {$#5$};
    \draw[double] (#1)--(loc1)--(loc2)--(#2);
}
\def\edgeur#1#2#3#4#5{
    \node (loc1) at ($(#1)!0.25!(#2)$) {$#3$};
    \node (loc2) at ($(#1)!0.75!(#2)$) {$#4$};
    \node[transform canvas={xshift=6pt,yshift=6pt}] (loc3) at ($(#1)!0.5!(#2)$) {$#5$};
    \draw[double] (#1)--(loc1)--(loc2)--(#2);
}
\def\edgeul#1#2#3#4#5{
    \node (loc1) at ($(#1)!0.25!(#2)$) {$#3$};
    \node (loc2) at ($(#1)!0.75!(#2)$) {$#4$};
    \node[transform canvas={xshift=-6pt,yshift=6pt}] (loc3) at ($(#1)!0.5!(#2)$) {$#5$};
    \draw[double] (#1)--(loc1)--(loc2)--(#2);
}

\usepackage{xstring}
\input{tixxx}

\DeclareMathOperator{\U}{U}
\DeclareMathOperator{\SU}{SU}
\DeclareMathOperator{\SO}{SO}

\DeclareMathOperator{\USp}{USp}
\DeclareMathOperator{\tr}{Tr}

\usepackage{epsfig}
\usepackage{amsmath}
\usepackage{amssymb}
\usepackage{amsfonts}
\usepackage{amsxtra}
\usepackage{amsthm}
\usepackage{mathrsfs}
\usepackage{makeidx}
\usepackage{graphics}
\usepackage{dsfont}
\usepackage{mathtools}
\usepackage{graphicx}
\usepackage{placeins}
\usepackage{bm}
\usepackage[capitalise]{cleveref}
\usepackage{empheq}
\usepackage{colortbl}
\usepackage{xcolor}
\usepackage{titlesec}
\usepackage{longtable}
\usepackage{hhline}
\usepackage{float}
\usepackage{color}
\usepackage{tikz}
\usepackage{xfrac}
\usepackage{footnote}
\usepackage{rotating}
\usepackage{lscape}
\usepackage{makecell}
\usepackage{environ}
\usepackage{tabularx}
\usepackage{subfiles}
\usepackage{ytableau}
\usepackage{tikz-3dplot}
\usepackage{slashed}
\usepackage{pifont}
\usepackage{multirow}
\usepackage{mdframed}
\usepackage{bbm}
\usepackage[normalem]{ulem}
\usepackage{arydshln}
\usepackage{enumitem} 
\usepackage{fancybox}

\usetikzlibrary{positioning,trees,decorations.pathmorphing,decorations.markings,decorations.pathreplacing,calc,shapes,patterns,arrows,chains,arrows.meta,fit,fadings,decorations.markings,graphs,graphs.standard,quotes,plotmarks,calligraphy,decorations.pathreplacing}

\usetikzlibrary{positioning}
\usetikzlibrary{chains}
\usetikzlibrary{arrows, arrows.meta ,fit,decorations.pathreplacing}
\tikzstyle{every picture}+=[remember picture]
\tikzstyle{na} = [baseline]
\tikzstyle{ligne}=[draw, thick]

\usetikzlibrary{arrows, decorations.markings, calc, fadings, decorations.pathreplacing, patterns, decorations.pathmorphing, positioning}
\tikzset{>={Latex[width=1.5mm,length=1.5mm]}}
\tikzset{bd/.style={circle, draw=black, inner sep=0pt, fill=black, minimum size=1.2mm}}
\tikzset{bld/.style={circle, draw=blue, inner sep=0pt, fill=blue, minimum size=1.2mm}}
\tikzset{wd/.style={circle, draw=black, inner sep=0pt, fill=white, minimum size=1.2mm}}
\tikzset{rd/.style={circle, draw=red, inner sep=0pt, fill=red, minimum size=.9mm}}
\tikzset{wrd/.style={circle, draw=red, inner sep=0pt, fill=white, minimum size=.9mm}}
\usetikzlibrary{graphs,graphs.standard,quotes}

\def\node#1#2{\overset{#1}{\underset{#2}{{\color{gray} \bullet}}}}

\def\node#1#2{\overset{#1}{\underset{#2}{\circ}}}

\tikzstyle{every picture}+=[remember picture]
\tikzstyle{na} = [baseline=-.5ex]

\newcommand{\eg}{e.g. }

\newcommand{\ie}{i.e. }

\numberwithin{equation}{section}
\newcommand{\bes}[1]{\begin{equation} \begin{split} #1\end{split} \end{equation}}

\newcommand{\be}{\begin{equation}} \newcommand{\ee}{\end{equation}}
\newcommand{\bea}{\begin{equation} \begin{aligned}} \newcommand{\eea}{\end{aligned} \end{equation}}

\def\tilde{\widetilde}

\def\hat{\widehat}

\def\bar{\overline}

\def\rt2{\sqrt{2}}

\def\mod{{\rm mod}}

\def\tr{\mathop{\rm tr}}

\def\CI{{\cal I}}

\def\CN{{\cal N}}

\def\CZ{{\cal Z}}

\def\1{{\ds 1}}

\newcommand{\fc}{\mathfrak{c}}

\def\SO{\mathrm{SO}}

\def\SU{\mathrm{SU}}

\def\su{\mathfrak{su}}

\def\so{\mathfrak{so}}

\def\usp{\mathfrak{usp}}

\def\repa{\raise4pt\hbox{$\square$}\mkern-14mu\raise-4pt\hbox{$\square$}}
\def\repab{\overline{\raise4pt\hbox{$\square$}\mkern-14mu\raise-4pt\hbox{$\square$}\mkern-1mu}}

\def\smileface{\ensuremath{\hbox{\large$\bigcirc$}\mkern-15mu\raise-1pt\hbox{\scriptsize$\smallsmile$}%
\mkern-10mu\raise4pt\hbox{..}\mkern4mu}}
\def\frownface{\ensuremath{\hbox{\large$\bigcirc$}\mkern-15mu\raise-1pt\hbox{\scriptsize$\smallfrown$}%
\mkern-10mu\raise4pt\hbox{..}\mkern4mu}}

\newcommand{\ba}{\begin{array}}
\newcommand{\ea}{\end{array}}
\newcommand{\bi}{\begin{itemize}}
\newcommand{\ei}{\end{itemize}}
\def\vec#1{\bm{#1}}
\def\bea#1\eea{\allowdisplaybreaks \begin{align}#1\end{align}}
 \newcommand{\ben}{\begin{enumerate}}
\newcommand{\een}{\end{enumerate}}
\newcommand{\bean}{\begin{eqnarray*}}
\newcommand{\eean}{\end{eqnarray*}}
\newcommand{\eref}[1]{(\ref{#1})}

\newcommand{\PE}{\mathop{\rm PE}}

\newcommand{\tQ}{\widetilde{Q}}

\newcommand{\BC}{\mathbb{C}}

\newcommand{\BZ}{\mathbb{Z}}

\newcommand{\Sym}{\mathrm{Sym}}

\definecolor{light-gray}{gray}{0.5}
\definecolor{new-green}{rgb}{0,0.7,0.3}

\newcommand{\purple}{\color{purple}}

\newcommand{\blue}{\color{blue}}

\newcommand{\red}{\color{red}}
\newcommand{\green}{\color{new-green}}
\newcommand{\violet}{\color{violet}}

\def\aup#1 {\overset{#1}{\uparrow} \, \overset{\tilde{#1}}{\downarrow}}

\tikzset{snake it/.style={decorate, decoration={snake, amplitude=.4mm, segment length=2mm,
                       post length=0mm,pre length=0mm}}}

\def\u{\mathfrak{u}}
\DeclareMathAlphabet{\mymathds}{U}{BOONDOX-ds}{m}{n}
\tikzstyle{double_border} = [draw, double, double distance=1pt]

\tikzset{->-/.style={>={Stealth[scale=2.3]},decoration={
  markings,
  mark=at position #1 with {\arrow{>}}},postaction={decorate}}}    
\tikzset{-<-/.style={>={Stealth[scale=2.3]},decoration={
  markings,
  mark=at position #1 with {\arrow{<}}},postaction={decorate}}}

\preprint{\hspace{1cm}}

\title{Generalised-Edged Quivers and Global Forms}
\author[a]{Julius F. Grimminger,}
\author[b,c]{ William Harding,}
\author[c,d]{ and Noppadol Mekareeya }

\affiliation[a]{Mathematical Institute, University of Oxford,\\Andrew Wiles Building, Woodstock Road, Oxford, OX2 6GG, UK}
\affiliation[b]{Dipartimento di Fisica, Universit\`a di Milano-Bicocca, Piazza della Scienza 3, I-20126 Milano, Italy}
\affiliation[c]{INFN, sezione di Milano-Bicocca,
Piazza della Scienza 3, I-20126 Milano, Italy}
\affiliation[d]{Department of Physics, Faculty of Science, Chulalongkorn University, Phayathai Road, Pathumwan, Bangkok 10330, Thailand}

\emailAdd{julius.grimminger@maths.ox.ac.uk}
\emailAdd{w.harding@campus.unimib.it}
\emailAdd{n.mekareeya@gmail.com}

\abstract{
Non-simply laced quivers, despite the lack of complete Lagrangian descriptions, play an important role in characterising moduli spaces of supersymmetric field theories. Notably, the moduli space of instantons in non-simply laced gauge groups can be understood by means of such quivers. We generalise the notion of non-simply laced unitary quivers to those whose edges carry two labels $(p,q)$, dubbed $(p,q)$-edged quivers. The special case of $(p,1)$ corresponds to a conventional non-simply laced edge studied in the literature. In the case of unframed $(p,q)$-edged quivers, we show how to parametrise the lattice of magnetic fluxes upon ungauging the decoupled $\mathrm{U}(1)$, and how one can pick sublattices thereof corresponding to different global forms of the quiver related by discrete gauging. This form of discrete gauging can be applied to any unframed unitary quivers, not just ones with generalised edges. We utilise both the Hilbert series and the superconformal index to study moduli spaces and 't Hooft anomalies. In particular, we study mixed 't Hooft anomalies between a one-form symmetry and a zero-form continuous topological symmetry in various $(p,q)$-edged quivers. We also provide an alternative realisation of the moduli space of $\mathfrak{so}(2n+1)$ instantons via gauging discrete symmetries in supersymmetric QCD with a symplectic gauge group and a large number of flavours.
}

\begin{document}
\maketitle

\section{Introduction and Summary}
Three dimensional $\mathcal{N}=4$ quiver gauge theories and their moduli spaces play a prominent role in the study of superconformal field theories (SCFTs) with 8 supercharges in various spacetime dimensions. Various properties of such SCFTs may not be accessible through the weakly coupled Lagrangian description. Nevertheless, the Higgs branch at the strongly coupled point of a number of such theories can be expressed via the Coulomb branch of certain 3d $\CN=4$ quiver theories, known as \emph{magnetic quivers}, many of which can be obtained from a brane system; see \eg~\cite{Cabrera:2018jxt, Cabrera:2019izd}.

Conventional $\CN=4$ quivers with simply laced edges and unitary gauge groups naturally appear as worldvolume theories of D3 branes in Hanany-Witten D5-D3-NS5 brane configurations \cite{Hanany:1996ie}. Adding an O3 orientifold plane to such a brane system changes the gauge groups in the quiver from unitary to orthosymplectic; see \eg~\cite{Feng:2000eq}. Adding an ON orientifold plane ($\mathrm{ON}^-$ or $\tilde{\mathrm{ON}}^-$ to be precise) to the brane system produces a unitary non-simply laced quiver \cite{Hanany:2001iy, Cremonesi:2014xha}; see also \cite{Kapustin:1998fa, Sen:1998ii, Hanany:1999sj} for a similar type of brane configurations. The non-simply laced edge in the latter is generally not a representation of the accustomed hypermultiplet. In fact, neither a Lagrangian description nor a Higgs branch construction of such quivers is known in general. Nevertheless, the Coulomb branch moduli space can be constructed. In this paper, we generalise the notion of the non-simply laced edge to a more general one, known as a $(p,q)$-edge, carrying a label $(p,q)$. As will be made more precise in the next section, one can think of a non-simply laced edge as that with a label $(p,1)$. 

For an unframed $(p,q)$-edged quiver, there is an overall $\U(1)$ that acts trivially on the generalised edges. Upon computing the Coulomb branch of such a quiver, one needs to ungauge such an overall $\U(1)$. One of the key points of this paper is to provide a general prescription to do so. A crucial step is to find a parametrisation of the lattice of the gauge magnetic fluxes. We also demonstrate how one can pick sublattices that lead to gauging a $\mathbb{Z}_k$ zero-form symmetry.\footnote{Modifying the magnetic lattice in Coulomb branch computations was considered in the context of non-simply laced quivers in \cite{Hanany:2020jzl} and in the context of orthosymplectic quivers, and unitary quivers with a single SU node, in \cite{Bourget:2020xdz}.} This procedure leaves the Higgs branch invariant, but acts as a quotient on the Coulomb branch, as well as produces a dual $\BZ_k$ one-form symmetry. We study different global forms of the unframed $(p,q)$-edged quiver that are related to each other by such discrete gaugings. A key tool we employ is the Coulomb branch Hilbert series, computed via the monopole formula \cite{Cremonesi:2013lqa}, which was generalised in \cite{Cremonesi:2014xha} to allow for non-simply laced quivers. In this paper, we generalise the latter even further to deal with the $(p,q)$-edged quivers. Whenever the theory in question admits a Lagrangian description, we examine it using the superconformal index \cite{Bhattacharya:2008zy,Bhattacharya:2008bja, Kim:2009wb,Imamura:2011su, Kapustin:2011jm, Dimofte:2011py, Aharony:2013dha, Aharony:2013kma}.

We finally remark that non-simply laced quivers have appeared as magnetic quivers not only for brane systems involving orientifold planes, but also to describe certain phases in brane systems with identical branes of self-intersection zero \cite{Bourget:2022ehw,Bourget:2022tmw}. A natural question is to realise general $(p,q)$-edged quivers from the brane perspective. We leave this problem for future work.

\paragraph{Organisation of the paper.}
In Section \ref{sec:(p,q)CB}, we give a detailed introduction to unitary quivers with generalised $(p,q)$-edges and their Coulomb branches for various choices of global forms. We compare this in some examples with \emph{wreathed quivers}, which were first explored in \cite{Bourget:2020bxh} and were recently studied in \cite{Giacomelli:2024sex, Grimminger:2024mks}. In Section \ref{sec:nonsimplylacedanom}, we present additional examples, including an Abelian theory, in which the quiver has a Lagrangian description and we can study it using the superconformal index, and a non-Abelian theory. We also examine mixed 't Hooft anomalies between a one-form symmetry and a zero-form continuous topological symmetry in various $(p,q)$-edged quivers. In Section \ref{sec:Bn}, we demonstrate how to realise the moduli space of $\mathfrak{so}(2n+1)$ instantons via gauging discrete symmetries in supersymmetric QCD with a symplectic gauge group and a large number of flavours. In Appendix \ref{app:LatticeConstraints}, we illustrate in many examples how to parametrise the magnetic lattice. Finally, in Appendix \ref{app:AbelianQuotients}, we present a way to produce quivers which are quotients of the form $\mathbb{H}^n/\mathbb{Z}_k$, where the $\mathbb{Z}_k$ can act with various charges on the different coordinates.

\section{\texorpdfstring{Coulomb Branch of Unitary Quivers with $(p,q)$-Edges}{Coulomb Branch of of Unitary Quivers with (p,q)-Edges}}
\label{sec:(p,q)CB}

Coulomb branches of 3d $\CN=4$ (simply laced) unitary quivers are well studied in the literature. These quivers consist of a collection of gauge nodes (depicted by circles), flavour nodes (depicted by boxes), and (possibly multiple) edges between such nodes. If there are no flavour nodes in the quiver, it is called \emph{unframed}; if there are flavour nodes in the quiver, it is called \emph{framed}.

In this section we introduce unitary quivers with $(p,q)$-edges, which are generalisations of the non-simply laced quivers first introduced in \cite{Cremonesi:2014xha}. Coulomb branches of non-simply laced unitary quivers have also been extensively studied in the literature, and the generalisation to $(p,q)$-edged quivers is natural. However, in the following, we make some necessary clarifications on how to deal with unframed non-simply laced unitary quivers (and, more generally, $(p,q)$-edged ones). In particular, we address how to decouple the freely acting U$(1)$ gauge group, and how this leads to a notion of \emph{global form} of an unframed unitary quiver.

\subsection{\texorpdfstring{$(p,q)$-Edged Unitary Quivers}{(p,q)-Edged Unitary Quivers}}

\sloppy In a simply laced unitary quiver, an edge between two gauge nodes U$(v_1)$ and U$(v_2)$ represents a hypermultiplet in the bifundamental representation $(\mathbf{v_1},\bar{\mathbf{v_2}})$\footnote{The hypermultiplet is made of two chiral multiplets in the representation $(\mathbf{v_1},\bar{\mathbf{v_2}})\oplus(\bar{\mathbf{v_1}},\mathbf{v_2})$.} of U$(v_1)\times$U$(v_2)$:
\bes{
\label{eq:hyper1}
    \begin{tikzpicture}
        \node[gauge,label=below:{$v_1$}] (1) at (0,0) {};
        \node[gauge,label=below:{$v_2$}] (2) at (2,0) {};
        \draw (1)--(2);
        \node[anchor=west] (h) at (3,-1) {hypermultiplet in $(\mathbf{v_1},\bar{\mathbf{v_2}})$};
        \draw[->] (h) .. controls (1.5,-1) .. (1,-0.15);
    \end{tikzpicture}
}
The contribution of this hypermultiplet to the conformal dimension of a monopole operator of magnetic charge $(m_{1,a_1},m_{2,a_2})$, with $a_1=1,\dots,v_1$ and $a_2=1,\dots,v_2$, is
\begin{equation}
    2\Delta_{\mathrm{hyper}_{\eqref{eq:hyper1}}}(m_{1,a_1},m_{2,a_2})=\sum_{a_1=1}^{v_1}\sum_{a_2=1}^{v_2}|m_{1,a_1}-m_{2,a_2}|~.
\end{equation}

Let us consider the case where $v_1=v_2=1$. The edge represents a hypermultiplet of charge $(1,-1)$. Naturally, one can consider a hypermultiplet of charge $(p,-q)$, which we shall denote with a ``$(p,q)$-edge'':
\bes{
\label{eq:hyper2}
    \begin{tikzpicture}
        \node[gauge,label=below:{$1$}] (1) at (0,0) {};
        \node[gauge,label=below:{$1$}] (2) at (2,0) {};
        \edge{1}{2}{p}{q}{}
        \node[anchor=west] (h) at (3,-1) {hyper of charge $(p,-q)$};
        \draw[->] (h) .. controls (1.5,-1) .. (1,-0.15);
    \end{tikzpicture}
}
The contribution of this hypermultiplet to the conformal dimension of a monopole operator of magnetic charge $(m_{1},m_{2})$ is\footnote{In order to avoid cluttered notation, whenever we have $\U(1)$ gauge nodes labelled by $i$, we denote the magnetic charge of the monopole operators simply by $m_i$, instead of $m_{i,1}$. \label{foot:Magweights}}
\begin{equation}
    2\Delta_{\mathrm{hyper}_{\eqref{eq:hyper2}}}(m_{1},m_{2})=|p\,m_{1}-q\,m_{2}|\;.
\end{equation}

One can ask for a generalisation of the $(p,q)$-edge in the case where $v_1,v_2>1$. A natural generalisation would be to consider the representation $(\mathrm{Sym}^p\mathbf{v_1},\mathrm{Sym}^q\bar{\mathbf{v_2}})$. This is of course a perfectly nice representation for a hypermultiplet to be in. We can take for example $v_1=2$, $v_2=1$, $p=2$, and $q=1$. The contribution of this hypermultiplet to the conformal dimension of a monopole operator of magnetic charge $(m_{1,1},m_{1,2},m_{2})$ is
\bes{
\label{eq:hyper3}
    &2\Delta_{\mathrm{hyper}_{(\mathrm{Sym}^2\mathbf{2},\bar{\mathbf{1}})}}(m_{1,1},m_{1,2},m_{2})\\&=|2\,m_{1,1}-m_{2}|+|2\,m_{1,2}-m_{2}|+|(m_{1,1}+m_{1,2})-m_{2}|\;.
}

However, this is not what we have in mind. A notion of non-simply laced edge, that is a $(p,1)$-edge in our language, was introduced in \cite{Cremonesi:2014xha} (based on Hanany-Witten type brane systems with orientifolds, and 3d mirror symmetry). For a $(2,1)$-edge between a rank $v_1=2$ and a rank $v_2=1$ node, the hypermultiplet contribution to the conformal dimension of a monopole operator of magnetic charge $(m_{1,1},m_{1,2},m_{2})$, is
\begin{equation}
    2\Delta_{(2,1)\mathrm{-edge}}(m_{1,1},m_{1,2},m_{2})=|2\,m_{1,1}-m_{2}|+|2\,m_{1,2}-m_{2}|\;,
\end{equation}
which is different of course from \eqref{eq:hyper3}. The edge does not represent a hypermultiplet in a representation of U$(v_1)\times$U$(v_2)$! The theory denoted by a quiver involving such edges is hence non-Lagrangian, and yet remains to be fully understood. Such quivers do however lead to a well defined notion of Coulomb branch \cite{Cremonesi:2014xha,Nakajima:2019olw}.

For generic $v_1$ and $v_2$, we propose a $(p,q)$-edge
\bes{
\label{eq:hyper4}
    \begin{tikzpicture}
        \node[gauge,label=below:{$v_1$}] (1) at (0,0) {};
        \node[gauge,label=below:{$v_2$}] (2) at (2,0) {};
        \edge{1}{2}{p}{q}{}
        \node[anchor=west] (h) at (3,-1) {$(p,q)$-edge};
        \draw[->] (h) .. controls (1.5,-1) .. (1,-0.15);
    \end{tikzpicture}
}
which contributes to the conformal dimension of a monopole operator of magnetic charge $(m_{1,a_1},m_{2,a_2})$, with $a_1=1,\dots,v_1$ and $a_2=1,\dots,v_2$, as
\begin{equation} \label{confdimpqedge}
    2\Delta_{(p,q)\mathrm{-edge}}(m_{1,a_1},m_{2,a_2})=\sum_{a_1=1}^{v_1}\sum_{a_2=1}^{v_2}|p\,m_{1,a_1}-q\,m_{2,a_2}|\;,
\end{equation}
which is a straightforward generalisation of \cite{Cremonesi:2014xha}.

\subsubsection{Notation} Thus far, generic $(p,q)$-edges have not been studied in the literature,\footnote{With the exception of \cite{Nakajima:2019olw}.} and hence no notation has been introduced. We will denote $N$ $(p,q)$-edges between a rank $v_1$ and a rank $v_2$ gauge node as
\begin{equation}
    \begin{tikzpicture}
        \node[gauge,label=below:{$v_1$}] (1) at (0,0) {};
        \node[gauge,label=below:{$v_2$}] (2) at (2,0) {};
        \edge{1}{2}{p}{q}{N}
    \end{tikzpicture}
\end{equation}
To give an idea on how this new notation works, let us connect with the commonly used notation in the literature.
\bi
\item A standard edge, i.e. a $(1,1)$-edge, is typically drawn as a single line between the two nodes:
\begin{equation}
    \vcenter{\hbox{\begin{tikzpicture}
        \node[gauge,label=below:{$v_1$}] (1) at (0,0) {};
        \node[gauge,label=below:{$v_2$}] (2) at (2,0) {};
        \draw (1)--(2);
    \end{tikzpicture}}}\qquad=\qquad\vcenter{\hbox{\begin{tikzpicture}
        \node[gauge,label=below:{$v_1$}] (1) at (0,0) {};
        \node[gauge,label=below:{$v_2$}] (2) at (2,0) {};
        \edge{1}{2}{1}{1}{1}
    \end{tikzpicture}}}
\end{equation}
\item $N$ $(1,1)$-edges are either depicted by $N$ lines if $N$ is low; for example, if $N=3$, we have
\begin{equation}
    \vcenter{\hbox{\begin{tikzpicture}
        \node[gauge,label=below:{$v_1$}] (1) at (0,0) {};
        \node[gauge,label=below:{$v_2$}] (2) at (2,0) {};
        \draw (1)--(2);
        \draw[transform canvas={yshift=2.5pt}] (1)--(2);
        \draw[transform canvas={yshift=-2.5pt}] (1)--(2);
    \end{tikzpicture}}}\qquad=\qquad\vcenter{\hbox{\begin{tikzpicture}
        \node[gauge,label=below:{$v_1$}] (1) at (0,0) {};
        \node[gauge,label=below:{$v_2$}] (2) at (2,0) {};
        \edge{1}{2}{1}{1}{3}
    \end{tikzpicture}}}
\end{equation}
Otherwise, for high or generic $N$, it is depicted by a double (or fat, or etc.) line with a label $N$:
\begin{equation}
    \vcenter{\hbox{\begin{tikzpicture}
        \node[gauge,label=below:{$v_1$}] (1) at (0,0) {};
        \node[gauge,label=below:{$v_2$}] (2) at (2,0) {};
        \draw[transform canvas={yshift=1.5pt},thick] (1)--(2);
        \draw[transform canvas={yshift=-1.5pt},thick] (1)--(2);
        \node at (1,0.3) {$N$};
    \end{tikzpicture}}}\qquad=\qquad\vcenter{\hbox{\begin{tikzpicture}
        \node[gauge,label=below:{$v_1$}] (1) at (0,0) {};
        \node[gauge,label=below:{$v_2$}] (2) at (2,0) {};
        \edge{1}{2}{1}{1}{N}
    \end{tikzpicture}}}
\end{equation}
\item A non-simply laced edge of `non-simply lacedness $p$' from node $v_1$ to node $v_2$, i.e.\ a $(p,1)$-edge, is typically drawn as $p$ lines with a `` \,{\Large $>$}\,'' symbol if $p$ is low. For example, if $p=3$, we have
\begin{equation}
    \vcenter{\hbox{\begin{tikzpicture}
        \node[gauge,label=below:{$v_1$}] (1) at (0,0) {};
        \node[gauge,label=below:{$v_2$}] (2) at (2,0) {};
        \draw (1)--(2);
        \draw[transform canvas={yshift=2.5pt}] (1)--(2);
        \draw[transform canvas={yshift=-2.5pt}] (1)--(2);
        \node (>) at (1,0) {\Large $>$};
    \end{tikzpicture}}}\qquad=\qquad\vcenter{\hbox{\begin{tikzpicture}
        \node[gauge,label=below:{$v_1$}] (1) at (0,0) {};
        \node[gauge,label=below:{$v_2$}] (2) at (2,0) {};
        \edge{1}{2}{3}{1}{1}
    \end{tikzpicture}}}
\end{equation}
Otherwise, for high or generic $p$, it is depicted by a double (or fat, or etc.) line with a `` \,{\Large $>$}\,'' symbol and a label $p$ :
\begin{equation}
    \vcenter{\hbox{\begin{tikzpicture}
        \node[gauge,label=below:{$v_1$}] (1) at (0,0) {};
        \node[gauge,label=below:{$v_2$}] (2) at (2,0) {};
        \draw[transform canvas={yshift=1.5pt},thick] (1)--(2);
        \draw[transform canvas={yshift=-1.5pt},thick] (1)--(2);
        \node (>) at (1,0) {\Large $>$};
        \node at (1,0.4) {$p$};
    \end{tikzpicture}}}\qquad=\qquad\vcenter{\hbox{\begin{tikzpicture}
        \node[gauge,label=below:{$v_1$}] (1) at (0,0) {};
        \node[gauge,label=below:{$v_2$}] (2) at (2,0) {};
        \edge{1}{2}{p}{1}{1}
    \end{tikzpicture}}}
\end{equation}
\item For $N$ $(p,1)$-edges, there is no consistent notation in the literature, as such quivers have barely been studied. For example, in the case of two $(3,1)$-edges, one can write
\begin{equation}
    \vcenter{\hbox{\begin{tikzpicture}
        \node[gauge,label=below:{$v_1$}] (1) at (0,0) {};
        \node[gauge,label=below:{$v_2$}] (2) at (2,0) {};
        \draw[transform canvas={yshift=5+1.5pt}] (1)--(2);
        \draw[transform canvas={yshift=5pt}] (1)--(2);
        \draw[transform canvas={yshift=5-1.5pt}] (1)--(2);
        \node[transform canvas={yshift=5pt}] (>) at (1,0) {\Large $>$};
        \draw[transform canvas={yshift=-5+1.5pt}] (1)--(2);
        \draw[transform canvas={yshift=-5pt}] (1)--(2);
        \draw[transform canvas={yshift=-5-1.5pt}] (1)--(2);
        \node[transform canvas={yshift=-5pt}] (>) at (1,0) {\Large $>$};
    \end{tikzpicture}}}\qquad=\qquad\vcenter{\hbox{\begin{tikzpicture}
        \node[gauge,label=below:{$v_1$}] (1) at (0,0) {};
        \node[gauge,label=below:{$v_2$}] (2) at (2,0) {};
        \edge{1}{2}{3}{1}{2}
    \end{tikzpicture}}}
\end{equation}
\ei

\subsubsection{Unframed Unitary Quivers and their Adjacency Matrix} In the following, we consider unframed unitary quivers with generalised edges.
Let $i,j$, with $i\geq j$, label two nodes in the quiver. Then, we define $e_{ij}$ to be the number of generalised edges from node $i$ to node $j$, and we denote by $(p_k,q_k)_{k=1,\dots,e_{ij}}$ the charges of these generalised edges. We require that
\bes{ \label{condsadj}
&\text{$\frac{p_1}{q_1}=\dots=\frac{p_{e_{ij}}}{q_{e_{ij}}}$~,~ and} \\
&\text{if $i=j$~,~  $\frac{p_1}{q_1}=\dots=\frac{p_{e_{ii}}}{q_{e_{ii}}}=1$~.}
}
These are not necessary conditions for the underlying quiver to make sense. For example, if the ranks of all gauge nodes are one, then any hypermultiplets can have arbitrary charges under all gauge groups, and this still gives a perfectly well-defined Lagrangian. However, these conditions seem to be obeyed by any quivers (electric or magnetic) which are read from brane systems. It also provides a convenient way to define a notation of the adjacency matrix (see below), which can be interpreted as a generalised Cartan matrix, for a quiver.\footnote{We illustrate the significance of conditions \eref{condsadj} via an example as follows. Let us take a quiver with two $\U(1)$ gauge nodes connected by a $(p_1, q_1) = (2,1)$-edge and a $(p_2, q_2) = (1,3)$-edge so that the conditions \eref{condsadj} are violated. Let $m_1$ and $m_2$ be the gauge fluxes of each gauge node.  Then, the conformal dimension of the monopole operator with flux $m_1$ and $m_2$ is given by $2\Delta(m_1, m_2) = |2m_1-m_2| + |m_1-3m_2|$. The adjacency matrix defined in \eref{defadjacency} for this quiver is $A = \begin{pmatrix} -2 & 3 \\ 4 & -2 \end{pmatrix}$. We see that $2\Delta$ cannot be written in the form of \eref{Deltaofm}, namely $|A_{12} m_1 - A_{21} m_2|$.}

We define the \emph{adjacency matrix} $A$ of the quiver by setting
\bes{ \label{defadjacency}
    \begin{split}
        A_{ji}=&\sum_{k=1}^{e_{ij}}q_k~, \quad
        A_{ij}=\sum_{k=1}^{e_{ij}}p_k~ \quad \text{for $i>j$}~, \\
        A_{ii}=&-2+2\sum_{k=1}^{e_{ii}}p_k~ \quad \text{for $i=j$}~.
    \end{split}
} 
We refer to a collection of the ranks of the gauge nodes in the quiver as \emph{rank vector}:
\begin{equation}
    \left(v_i\right)_{i=1,\dots,n}\;.
\end{equation}
The adjacency matrix, together with the rank vector, defines a Coulomb branch, as we will see shortly.

Some examples for quivers, with their corresponding adjacency matrices and rank vectors, are given in the table below.
\bes{ \label{eq:quivToA1}
 \begin{array}{|c|c|c|}
 \hline
 \text{Quiver} & \text{Adjacency matrix} & \text{Rank vector} \\
 \hline
     \raisebox{0.5\height}{\begin{tikzpicture}[baseline]
         \node[gauge,label=below:{$v_1$}] (1) at (0,0) {};
         \node[gauge,label=below:{$v_2$}] (2) at (2,0) {};
         \edge{1}{2}{1}{1}{N}
     \end{tikzpicture}}\; &\begin{pmatrix}
         -2 & N\\
         N & -2
     \end{pmatrix}\;&\begin{pmatrix}
         v_1\\
         v_2
     \end{pmatrix} \\
\hline    
    \raisebox{-.5\height}{\begin{tikzpicture}
        \node[gauge,label=below:{$v_1$}] (1) at (0,0) {};
        \node[gauge,label=below:{$v_2$}] (2) at (2,0) {};
        \edge{1}{2}{N}{N}{1}
    \end{tikzpicture}}\;&\begin{pmatrix}
        -2 & N\\
        N & -2
    \end{pmatrix}\; &\begin{pmatrix}
        v_1\\
        v_2
    \end{pmatrix}\; \\
\hline
    \raisebox{-.5\height}{\begin{tikzpicture}
        \node[gauge,label=left:{$v_1$}] (1) at (0,0) {};
        \node[gauge,label=right:{$v_2$}] (2) at (2,0) {};
        \node[rotate=-45] (l1) at (0.5,0.8) {\tiny$N-M$};
        \node[rotate=45] (r1) at (1.5,0.8) {\tiny$N-M$};
        \node (l2) at (0.5,-0.8) {$M$};
        \node (r2) at (1.5,-0.8) {$M$};
        \draw[double] (1)--(l1)--(r1)--(2);
        \draw[double] (1)--(l2)--(r2)--(2);
        \node at (1,1.1) {$1$};
        \node at (1,-1.1) {$1$};
    \end{tikzpicture}}\; &\begin{pmatrix}
        -2 & N\\
        N & -2
    \end{pmatrix}\;& \begin{pmatrix}
        v_1\\
        v_2
    \end{pmatrix}\; \\
\hline    
    \raisebox{-.5\height}{\begin{tikzpicture}
        \node[gauge,label=below:{$v_1$}] (1) at (0,0) {};
        \node[gauge,label=below:{$v_2$}] (2) at (2,0) {};
        \edge{1}{2}{2}{1}{1}
    \end{tikzpicture}}\; &\begin{pmatrix}
        -2 & 2\\
        1 & -2
    \end{pmatrix}\; &\begin{pmatrix}
        v_1\\
        v_2
    \end{pmatrix}\; \\
\hline    
    \raisebox{-.5\height}{\begin{tikzpicture}
        \node[gauge,label=below:{$v_1$}] (1) at (0,0) {};
        \node[gauge,label=below:{$v_2$}] (2) at (2,0) {};
        \edge{1}{2}{p}{q}{a}
    \end{tikzpicture}}\; &\begin{pmatrix}
        -2 & ap\\
        aq & -2
    \end{pmatrix}\; &\begin{pmatrix}
        v_1\\
        v_2
    \end{pmatrix} \\
\hline    
\end{array}}
Note that the map from quiver to adjacency matrix is not one-to-one! The first three quivers in \eqref{eq:quivToA1}, for example, are distinct, yet they have the same adjacency matrix. This is because the adjacency matrix only captures the information about the singular Coulomb branch.
Hence, one can usually assign many different quivers to a given adjacency matrix. While these quivers have the same singular Coulomb branch, they have different Higgs branches,\footnote{It is still not known how to compute the Higgs branch of a generic non-simply laced or $(p,q)$-edged quiver. In a few cases it is possible to obtain the Higgs branch of a non-simply laced quiver via 3d mirror symmetry. Furthermore, it is always possible to directly compute the Higgs branch of a $(p,q)$-edged quiver with Abelian nodes. In this paper, we simply assume the existence of a Higgs branch, and we rarely discuss it.} and they allow for different sets of mass parameters deforming the Coulomb branch.

To illustrate this, let us set $v_i=1$ in \eqref{eq:quivToA1}, and consider the first three quivers as follows: 
\begin{equation}
    \mathsf{Q}_1=\raisebox{-.5\height}{\begin{tikzpicture}
        \node[gauge,label=below:{$1$}] (1) at (0,0) {};
        \node[gauge,label=below:{$1$}] (2) at (2,0) {};
        \edge{1}{2}{1}{1}{N}
    \end{tikzpicture}}\,\qquad\mathsf{Q}_2=\raisebox{-.5\height}{\begin{tikzpicture}
        \node[gauge,label=below:{$1$}] (1) at (0,0) {};
        \node[gauge,label=below:{$1$}] (2) at (2,0) {};
        \edge{1}{2}{N}{N}{1}
    \end{tikzpicture}}\,\qquad\mathsf{Q}_3=\raisebox{-.5\height}{\begin{tikzpicture}
        \node[gauge,label=left:{$1$}] (1) at (0,0) {};
        \node[gauge,label=right:{$1$}] (2) at (2,0) {};
        \node[rotate=-45] (l1) at (0.5,0.8) {\tiny$N-M$};
        \node[rotate=45] (r1) at (1.5,0.8) {\tiny$N-M$};
        \node (l2) at (0.5,-0.8) {$M$};
        \node (r2) at (1.5,-0.8) {$M$};
        \draw[double] (1)--(l1)--(r1)--(2);
        \draw[double] (1)--(l2)--(r2)--(2);
        \node at (1,1.1) {$1$};
        \node at (1,-1.1) {$1$};
    \end{tikzpicture}}
\end{equation}
Such quivers possess the same Coulomb branch:
\begin{equation}
    \mathrm{CB}\left(\mathsf{Q}_1\right)=\mathrm{CB}\left(\mathsf{Q}_2\right)=\mathrm{CB}\left(\mathsf{Q}_3\right)=\mathbb{C}^2/\mathbb{Z}_N\;.
\end{equation}
Yet, the Higgs branches are different:
\bes{
    \mathrm{HB}\left(\mathsf{Q}_1\right)&=\overline{\mathrm{min.}A_{N-1}}\;, \\
    \mathrm{HB}\left(\mathsf{Q}_2\right)&=\mathrm{point}\;, \\
    \mathrm{HB}\left(\mathsf{Q}_3\right)&=\{(x,y,z)\in\mathbb{C}^3|xy=z^{\frac{N}{\mathrm{gcd}(N,N-M)}}\}\;.
}
For $\mathsf{Q}_1$, there are $N-1$ mass parameters, which allow to completely resolve the Coulomb branch $\mathbb{C}^2/\mathbb{Z}_N$ singularity. For $\mathsf{Q}_2$, there are no mass parameters, and the Coulomb branch cannot be resolved. For $\mathsf{Q}_3$, there is one mass parameter. Turning it on produces a Coulomb branch with two singular points: one of them is locally $\mathbb{C}^2/\mathbb{Z}_{N-M}$, and the other one is locally $\mathbb{C}^2/\mathbb{Z}_{M}$.

\subsubsection{Symmetrisability and Shortness}
We furthermore require $A$ to be \emph{symmetrisable}, i.e.\ there is a shortness vector
\begin{equation}
    \left(s_i\right)_{i=1,\dots,n}\;,
\end{equation}
such that the following conditions are satisfied:
\bes{
    A_{ij}s_i=A_{ji}s_j\quad&\forall \,\, i,j=1,\dots,n~\\
    \mathrm{gcd}(s_i)&=1\;.
}
This implies that the quiver can be \emph{unfolded}, i.e.\ there is a simply laced quiver $\tilde{\mathsf{Q}}$ such that, folding it, gives the non-simply laced quiver $\mathsf{Q}$. A node of shortness $s_i$ will appear $s_i$ times in the unfolded quiver.
\paragraph{Example.} If we take the following quiver
\bes{ \label{Triangleunfold}
    \begin{tikzpicture}
        \node at (-3,1) {$\mathsf{Q}=$};
        \node[gauge,label=below:{$v_3$}] (1) at (-2,0) {};
        \node[gauge,label=below:{$v_1$}] (2) at (2,0) {};
        \node[gauge,label=above:{$v_2$}] (3) at (0,2) {};
        \edge{1}{2}{3}{1}{1}
        \edgeur{2}{3}{2}{3}{1}
        \edgeul{1}{3}{2}{1}{1}
    \end{tikzpicture}
}
with symmetrisable adjacency matrix and shortness vector given by
\begin{equation}
    A=\begin{pmatrix}
        -2 & 2 & 1\\
        3 & -2 & 1\\
        3 & 2 & -2
    \end{pmatrix}\,,\quad s=\begin{pmatrix}
        3\\
        2\\
        1
    \end{pmatrix}\;,
\end{equation}
then \eref{Triangleunfold} can be unfolded to the following simply laced quiver
\bes{
    \begin{tikzpicture}
        \node at (-3,1) {$\tilde{\mathsf{Q}}=$};
        \node[gauge,label=below:{$v_3$}] (1) at (-2,0) {};
        \node[gauge,label=below right:{$v_1$}] (21) at (1.5,-0.5) {};
        \node[gauge,label=below right:{$v_1$}] (22) at (2,0) {};
        \node[gauge,label=below right:{$v_1$}] (23) at (2.5,0.5) {};
        \node[gauge,label=above:{$v_2$}] (31) at (-0.5,2) {};
        \node[gauge,label=above:{$v_2$}] (32) at (0.5,2) {};
        \draw (1)--(21) (1)--(22) (1)--(23) (1)--(31) (1)--(32) (21)--(31) (21)--(32) (22)--(31)  (22)--(32) (23)--(31) (23)--(32);
    \end{tikzpicture}
}
\paragraph{Example.} On the other hand, if we consider the following quiver
\bes{ \label{Trianglenounfold}
    \begin{tikzpicture}
        \node[gauge,label=below:{$v_3$}] (1) at (-2,0) {};
        \node[gauge,label=below:{$v_1$}] (2) at (2,0) {};
        \node[gauge,label=above:{$v_2$}] (3) at (0,2) {};
        \edge{1}{2}{3}{1}{1}
        \edgeur{2}{3}{1}{1}{1}
        \edgeul{1}{3}{2}{1}{1}
    \end{tikzpicture}
}
the corresponding adjacency matrix is non-symmetrisable
\begin{equation}
    A=\begin{pmatrix}
        -2 & 1 & 1\\
        1 & -2 & 1\\
        3 & 2 & -2
    \end{pmatrix}\;,
\end{equation}
meaning that \eref{Trianglenounfold} cannot be unfolded.

\subsubsection{Naive Gauge Group and Magnetic Lattice}
We refer to the gauge group that is manifest in the quiver diagram, namely
\begin{equation}
    \tilde{G}=\prod_{i=1}^n \U(v_i)~,
\end{equation}
as the \emph{naive gauge group}, with magnetic (i.e.\ coweight) lattice $\Gamma_{\tilde{G}}^{mw}$ parametrised by magnetic weights $m_{i,a_i}$, with $a_i=1,\dots,v_i$ and $i=1,\dots,n$.\footnote{As already anticipated in Footnote \ref{foot:Magweights}, we remark that, when $v_i = 1$, we suppress the subscript $a_i = 1$ and we simply denote the magnetic weights by $m_i$.}
Unless stated explicitly, such as around \eref{diracpq}, we generally take the magnetic weights to be integers. We also define
\begin{equation}
    d_i=\sum_{a_i=1}^{v_i}m_{i,a_i}
\end{equation}
as the charge of the topological symmetry associated with the gauge group $\U(v_i)$.

\subsubsection{Conformal Dimension and Shift Symmetry}
The conformal dimension of a bare monopole operator is given by
\begin{equation}\label{Deltaofm}
    2\Delta(\vec{m})=\frac{1}{2}\sum_{i=1}^n\sum_{j=1}^n\sum_{a_i=1}^{v_i}\sum_{b_j=1}^{v_j}\mathrm{sign}(A_{ij})\left|A_{ij}m_{i,a_i}-A_{ji}m_{j,b_j}\right|\;.
\end{equation}
It has a shift symmetry -- corresponding to the U$(1)\subset\tilde{G}$ acting trivially on the edges (i.e.\ generalised hypermultiplets) -- generated by the \emph{shift vector}
\begin{equation}
    \delta\in\Gamma_{\tilde{G}}^{mw}~, \quad \text{with} \quad \mathrm{gcd}(\delta_{i,a_i})=1\;,
\end{equation}
such that
\begin{equation}
2\Delta(\vec{m}+x\delta)=2\Delta(\vec{m})\;,\quad\forall x\in\mathbb{Z}\;.
\end{equation}
It is simple to check that
\begin{equation}
    \delta_{i,a_i}=s_i\;.
\end{equation}
We refer to the one-dimensional sublattice generated by the shift vector $\delta$ as
\begin{equation}
\Lambda_\delta\subset\Gamma_{\tilde{G}}^{mw}\;.
\end{equation}

\subsubsection{Weyl Chamber}
Given the Weyl group of $\tilde{G}$
\begin{equation}
    W_{\tilde{G}}=\prod_{i=1}^nS_{v_i}~,
\end{equation}
then the Weyl chamber of the magnetic lattice of $\tilde{G}$, namely $\Gamma_{\tilde{G}}^{mw}/W_{\tilde{G}}$, can be parametrised by setting
\begin{equation}\label{eq:WeylChambGt}
    -\infty<m_{i,1}\leq m_{i,2}\leq\dots\leq m_{i,v_i}<\infty\quad\forall\,i\in\{1,\dots,n\}\;.
\end{equation}

\subsubsection{Ungauging the Decoupled U(1)} \label{sec:ungauge}
Upon decoupling an overall $\U(1)$, we refer to the resulting gauge group as the {\it true gauge group}:
\begin{equation}
    G=\tilde{G}/\U(1)\;.
\end{equation}
The magnetic lattice of this group is $\Gamma_G^{mw}=\Gamma_{\tilde{G}}^{mw}/\Lambda_\delta$, where the Weyl chamber of the magnetic lattice of $\Gamma_G^{mw}$, namely $\Gamma_G^{mw}/W_G$, can be parametrised via a constraint on the $m_{i,a_i}$, while still obeying \eqref{eq:WeylChambGt}. There are infinitely many such constraints. In Appendix \ref{app:LatticeConstraints}, many examples of such possible constraints are illustrated very explicitly for several quivers.

For instance, a simple choice for a constraint is
\begin{equation}
    m_{i,a_i}\in\{0,\dots,s_i-1\}\;,\quad\textnormal{for some }i\textnormal{ and }a_i\;.
\end{equation}
If $s_i=1$ we can set $m_{i,1}=0$, which is what is commonly done in the literature, and sometimes referred to as ``ungauging on the long side'' (see e.g. \cite{Hanany:2020jzl}). Yet, there seems to be a much more general rule.

\paragraph{General rule.}  Let us describe a general rule for ungauging an overall $\U(1)$ as follows. Let us first choose the following set of non-negative integers:\footnote{Throughout this paper, whenever we encounter a $\U(1)$ gauge node, \ie $v_i = 1$, we suppress the subscript $a_i = 1$ in \eref{cnneginteg}. Explicitly, for a $\U(1)$ gauge node labelled by $i$, we denote the associated non-negative integer in \eref{cnneginteg} simply by $\mathfrak{c}_i$.}
\begin{equation} \label{cnneginteg}
\mathfrak{c}=\mathfrak{c}_{i,a_i}\in\mathbb{Z}_{\geq 0}~, \quad \text{with} \quad a_i=1,\dots,v_i~, \quad i=1,\dots,n~.
\end{equation}
Then, if we define
\begin{equation} \label{lambdac}
\lambda_{\mathfrak{c}}=\left(\sum_{i=1}^n\sum_{a_i=1}^{v_i}\mathfrak{c}_{i,a_i}s_i\right)-1\;,
\end{equation}
we can pick the following constraint on the magnetic weights:
\begin{equation}
\label{eq:MagLatGenCon}
\sum_{i=1}^n\sum_{a_i=1}^{v_i}\mathfrak{c}_{i,a_i}m_{i,a_i}\in\left\{0,1,\dots,\lambda_{\mathfrak{c}}\right\}\;,
\end{equation}
which can also be generalised to
\begin{equation} \label{eq:MagLatGenConxj}
    \sum_{i=1}^n\sum_{a_i=1}^{v_i}\mathfrak{c}_{i,a_i}m_{i,a_i}\in\left\{x_0,x_1,\dots,x_{\lambda_{\mathfrak{c}}}\right\}\;,\qquad\textnormal{where}\qquad (x_{j}\;\mod\;\lambda_{\mathfrak{c}})=j\;.
\end{equation}

\paragraph{Alternative approach.}
Let us discuss an alternative, but equivalent, approach in ungauging an overall $\U(1)$. In this method, the magnetic fluxes are allowed to take integral and non-integral values, subject to the condition that each term in the contribution of the matter fields to the conformal dimension is integral. In particular, for a $(p,q)$-edge between the $\U(v_i) \times \U(v_j)$ gauge group whose contribution is given by \eref{confdimpqedge}, we require that
\bes{ \label{diracpq}
p m_{i,a_i} - q m_{j,a_j} \in \BZ~
}
and allow $m_{i,a_i}$ and $m_{j,a_j}$ to be also non-integral. In the special case in which the gauge groups are Abelian, namely $v_i=v_j=1$, this theory admits a conventional Lagrangian description in terms of a hypermultiplet of charge $(p,-q)$ under $\U(v_i) \times \U(v_j)$, and the above condition is simply the Dirac quantisation condition.  In the next step, we choose the non-negative integers $\fc_{i,a_i}$ and impose the following constraint on the magnetic fluxes:
\bes{ \label{fixU1}
\sum_{i=1}^n\sum_{a_i=1}^{v_i}\mathfrak{c}_{i,a_i}m_{i,a_i} =0~.
}
In contrast to \eref{eq:MagLatGenCon}, we restrict the freedom on the linear combination of the left hand side to be zero, but this lack of freedom is compensated by allowing $m_{i,a_i}$ to take also non-integral values. The two approaches thus yield the same result. In this method, the conditions \eref{diracpq} and \eref{fixU1} determine the magnetic lattices that need to be summed over upon computing the Coulomb branch Hilbert series or the superconformal index. This will be demonstrated explicitly via a number of examples in Section \ref{sec:nonsimplylacedanom}. 

Let us suppose that the allowed values of the magnetic fluxes are fractional and that, upon writing them in the lowest terms, the largest denominator is $r$. There is a natural $\BZ_r$ zero-form symmetry that acts non-trivially on the monopole operators with such fractional magnetic fluxes. Then, we can turn on a discrete fugacity associated with the $\BZ_r$ symmetry in the Coulomb branch Hilbert series or in the index. This allows for gauging such a symmetry by summing this discrete fugacity over all the $r$-th roots of unity. The gauging leads to a dual $\BZ_r$ one-form symmetry. We finally remark that, upon gauging a discrete one-form symmetry, the magnetic fluxes are also summed over non-integral values. This naturally leads to the dual $\BZ_r$ zero-form as realised above. We will discuss the procedure of gauging one-form symmetries in Section \ref{sec:nonAbelianexample}.

\subsubsection{Good, Ugly, and Bad} Following the nomenclature of \cite{Gaiotto:2008ak}, we call a theory:
\begin{itemize}
    \item good, if $2\Delta(\vec{m})>1$ for all $\vec{m}\neq0$;
    \item ugly, if $2\Delta(\vec{m})>0$ for all $\vec{m}\neq0$, with some $2\Delta(\vec{m})=1$; and
    \item bad, if $2\Delta(\vec{m})\leq0$ for some $\vec{m}\neq0$.
\end{itemize}
Good theories have conical Coulomb branches. Ugly theories have Coulomb branches which are a product of a cone and free space. Bad theories have more intricate Coulomb branches. In this paper, we focus mainly on the good and ugly theories.

\paragraph{Balanced nodes.} A useful quantity that can be associated with a node in a quiver is its so-called \emph{balanced}. For a quiver with adjacency matrix $A$ and rank vector $v$, the balance of a node $i$, with $A_{ii}=-2$ (i.e.\ no adjoint loops), is given as
\begin{equation}
    b_i=\sum_{j}A_{ij}v_{j}\;.
\end{equation}
Defining the analogous condition for a balanced node becomes more complex in the presence of adjoint loops. We do not address this case in the present work.

\subsubsection{Coulomb Branch Hilbert Series} The Hilbert series of the Coulomb branch of a good or ugly quiver $\mathsf{Q}$ can be computed via the monopole formula \cite{Cremonesi:2013lqa}
\begin{equation}
    \mathrm{HS}\left[\text{CB of $\mathsf{Q}$}\right](t)=\sum_{\vec{m}\in\frac{\Gamma_{G}^{mw}}{W_G}}P_{G}(\vec{m},t^2)t^{2\Delta(\vec{m})}\;,
\end{equation}
where
\begin{equation}
    P_{G}(\vec{m},t^2)=(1-t^2)\prod_{i=1}^nP_{\mathrm{U}(v_i)}(m_i,t^2)\;,  
\end{equation}
and
\begin{equation}
    P_{\mathrm{U}(v_i)}\left(m_i,t^2\right)=\prod_{l=1}^{N}\left[\prod_{k=1}^{\lambda_l}\frac{1}{\left(1-t^{2k}\right)}\right] 
\end{equation}
for $\lambda=(\lambda_1,\dots,\lambda_N)$ a partition of $v_i$, which encodes how many $m_{i,a_i}$ are equal.

This Hilbert series can be refined with respect to the fugacities $w_i$ associated with the topological symmetries as
\begin{equation}
\label{CBHSref}
    \mathrm{HS}\left[\text{CB of $\mathsf{Q}$}\right](t,w_i)=\sum_{\vec{m}\in\frac{\Gamma_{G}^{mw}}{W_G}}P_{G}(m,t^2)t^{2\Delta(\vec{m})}\prod_{i=1}^n w_i^{\sum_{a_i=1}^{v_i}m_{i,a_i}}\;,
\end{equation}
with the constraint
\begin{equation}
\label{constraintonzi}
    \prod_{i=1}^n w_i^{s_iv_i}=1\;.
\end{equation}
If the quiver is bad, then the monopole formula diverges. This is due to the non-conical structure of the Coulomb branch of a bad theory.

\subsection{Global Forms}\label{sec:globalforms}

It is by now widely appreciated that, in gauge theory, it is important to specify the precise global form of the gauge group. This choice determines amongst other things the magnetic lattice, and hence affects the Coulomb branch of a $3d$ $\mathcal{N}=4$ gauge theory, while typically leaving the Higgs branch unchanged. For a given $3d$ $\mathcal{N}=4$ gauge theory, the hypermultiplet representations restrict the possible global forms of the gauge group.

Let us take an unframed simply laced unitary quiver with naive gauge group $\tilde{G}=\prod_{i=1}^{n}\mathrm{U}(v_i)$, where ungauging the overall U$(1)$ is understood, and pick a node labelled by $i_0$, with associated gauge node U$(v_{i_0})$ in the quiver. One way to parametrise the magnetic lattice of the true gauge group, that is $G=\tilde{G}/\mathrm{U}(1)$, is by setting $d_{i_0}=\sum_{a_{i_0}=1}^{v_{i_0}}m_{i_0,a_{i_0}}\in\{0,\dots,v_{i_0}-1\}$:
\bes{
    \vcenter{\hbox{\begin{tikzpicture}
        \node[gauge,label=below:{U$(v_{i_0})$}] (1) at (0,0) {};
        \draw (-1,0)--(1)--(1,0) (-0.6,0.6)--(1)--(0.6,0.6);
        \draw (1,-1.25)--(2,-0.25);
        \node at (2,-1) {U$(1)$};
    \end{tikzpicture}}}\qquad d_{i_0}\in\{0,\dots,v_i-1\}
}
A different global form of the gauge group compatible with the hypermultiplet representations is
\begin{equation}
    G'=\prod_{\underset{i\neq i_0}{i=1}}^{n}\mathrm{U}(v_i)\times\mathrm{SU}(v_{i_0})\;,
\end{equation}
\ie we replace U$(v_{i_0})$ with an SU$(v_{i_0})$ node, producing a special unitary quiver. The magnetic lattice of $G'$ can be parametrised for example by setting $d_{i_0}=0$:
\begin{equation} \label{di0eq0}
    \vcenter{\hbox{\begin{tikzpicture}
        \node[gauge,label=below:{SU$(v_{i_0})$}] (1) at (0,0) {};
        \draw (-1,0)--(1)--(1,0) (-0.6,0.6)--(1)--(0.6,0.6);
    \end{tikzpicture}}}\qquad d_{i_0}=0
\end{equation}
The newly obtained (special) unitary quiver has a $\mathbb{Z}_{v_{i_0}}$ one-form symmetry. Gauging this $\mathbb{Z}_{v_{i_0}}$ one-form symmetry (at the level of the magnetic lattice, this implies adding $d_{i_0}\in\{1,\dots,v_{i_0}-1\}$) produces the original quiver
\bes{
\scalebox{0.98}{$
\label{SUvmodZveqUcmodU1}
    \vcenter{\hbox{\begin{tikzpicture}
        \node[gauge,label=below:{SU$(v_{i_0})$}] (1) at (0,0) {};
        \draw (-1,0)--(1)--(1,0) (-0.6,0.6)--(1)--(0.6,0.6);
        \draw (1,-1.25)--(2,-0.25);
        \node at (2,-1) {$\mathbb{Z}_{v_{i_0}}$};
    \end{tikzpicture}}}\qquad=\qquad
    \vcenter{\hbox{\begin{tikzpicture}
        \node[gauge,label=below:{U$(v_{i_0})$}] (1) at (0,0) {};
        \draw (-1,0)--(1)--(1,0) (-0.6,0.6)--(1)--(0.6,0.6);
        \draw (1,-1.25)--(2,-0.25);
        \node at (2,-1) {U$(1)$};
    \end{tikzpicture}}}\qquad d_{i_0}\in\{0,\dots,v_i-1\}
$}
}
Any subgroup $\mathbb{Z}_{(v_{i_0}/k)}\subset\mathbb{Z}_{v_{i_0}}$ of the one-form symmetry may be gauged to reach a different global form of the gauge group, $G''=G'/\mathbb{Z}_{(v_{i_0}/k)}$. The magnetic lattice of $G''$ can be parametrised by setting $d_{i_0}\in\{kj|0\leq j\leq (v_{i_0}/k)-1\}$
\begin{equation}
    \vcenter{\hbox{\begin{tikzpicture}
        \node[gauge,label=below:{SU$(v_{i_0})$}] (1) at (0,0) {};
        \draw (-1,0)--(1)--(1,0) (-0.6,0.6)--(1)--(0.6,0.6);
        \draw (1,-1.25)--(2,-0.25);
        \node at (2.1,-1.1) {~$\mathbb{Z}_{(v_{i_0}/k)}$};
    \end{tikzpicture}}}\qquad d_{i_0}\in\{kj|0\leq j\leq (v_{i_0}/k)-1\}
\end{equation}
Note that the special cases of $k=v_{i_0}$ and $k=1$ correspond to \eref{di0eq0} and \eref{SUvmodZveqUcmodU1}, respectively. Such gauging leads to a dual $\BZ_k$ zero-form symmetry in the resulting theory. We will see via the example below that the presence of the $\BZ_k$ zero-form symmetry is manifest in the periodicity in $d_{i_0}$ of the Coulomb branch Hilbert series.

We can thus associate many global structures to an unframed unitary quiver in the following way. Pick a node $i_0$ and impose $d_{i_0}\in\{kj|0\leq j\leq (v_{i_0}/k)-1\}$ for some $k|v_{i_0}$.\footnote{Note that, once a condition has been imposed on $d_{i_0}$ of some node labelled by $i_0$, one cannot put another condition on $d_{i_1}$ of some node $i_1\neq i_0$, as this would reduce the dimension of the magnetic lattice. One could consider picking two distinct nodes, $i_0$ and $i_1$, and impose a condition on a linear combination of $d_{i_0}$ and $d_{i_1}$.} Such a choice has no effect on the Higgs branch of the quiver, yet it does affect the Coulomb branch. In particular, it leads to a $\mathbb{Z}_k$ quotient of the Coulomb branch of the original theory.  This amounts to gauging the $\BZ_k$ zero-form symmetry, which leads to a dual $\mathbb{Z}_k$ one-form symmetry.

For a general $(p,q)$-edged unframed unitary quiver, with symmetrisable adjacency matrix, the story gets slightly more complicated. We can apply a similar method to that described above by picking a node $i_0$ and modifying the condition on $d_{i_0}$. We observe that, in this case, the order of the discrete zero-form symmetry involves the product $s_{i_0} v_{i_0}$ of shortness $s_{i_0}$ and the rank $v_{i_0}$. Various examples will be provided below and in Section \ref{sec:nonsimplylacedanom}.

\subsubsection*{Example} 
Let us consider the following quiver:
\begin{equation}
\label{eq:(3)-{3,1}-(2)-(1)}
    \vcenter{\hbox{\begin{tikzpicture}
        \node[gauge,label=below:{$3$},label=above:{\color{red}$1$}] (1) at (0,0) {};
        \node[gauge,label=below:{$2$},label=above:{\color{blue}$2$}] (2) at (2,0) {};
        \node[gauge,label=below:{$1$},label=above:{\color{new-green}$3$}] (3) at (4,0) {};
        \edge{1}{2}{3}{1}{1}
        \edge{2}{3}{1}{1}{1}
    \end{tikzpicture}}}
\end{equation}
with adjacency matrix, rank vector and shortness vector
\begin{equation}
    A=\begin{pmatrix}
        -2 & 3 & 0\\
        1 & -2 & 1\\
        0 & 1 & -2
    \end{pmatrix}\,,\;v=\begin{pmatrix}
        3\\
        2\\
        1
    \end{pmatrix}\,,\;s=\begin{pmatrix}
        1\\
        3\\
        3
    \end{pmatrix}\;,
\end{equation}
where the true gauge group is
\bes{
G = \tilde{G}/\U(1)~, \quad \text{with} \quad \tilde{G} = \U(3) \times \U(2) \times \U(1)~.
}
Different global structures of this quiver are summarised in Table \ref{tab:(3)-{3,1}-(2)-(1)_gs_compare}. Let us discuss them more in detail.

Let us denote by $\vec{m}_1 = (m_{1,1}, m_{1,2}, m_{1,3})$, $\vec{m}_2 = (m_{2,1}, m_{2,2})$ and $m_3$ the magnetic fluxes associated with the $\U(3)$, $\U(2)$ and $\U(1)$ gauge nodes respectively, with
\bes{
d_1 = \sum_{1=1}^3 m_{1,i}~, \quad d_2 = \sum_{i=1}^2 m_{2,i}~, \quad d_3 = m_3~.
}
Then, the Coulomb branch Hilbert series of quiver \eref{eq:(3)-{3,1}-(2)-(1)} can be obtained via the monopole formula \eref{CBHSref} as
\bes{ \label{CBHSref321}
&\mathrm{HS}\left[\text{CB of \eref{eq:(3)-{3,1}-(2)-(1)}}\right](t, \vec{w})\\&= \left(1-t^2\right) \sum_{j=0}^{\lambda_{\mathfrak{c}}} \sum_{\vec{m}\in\frac{\Gamma_{G}^{mw}}{W_G}} P_{\mathrm{U}(3)}(\vec{m}_1,t^2) P_{\mathrm{U}(2)}(\vec{m}_2,t^2) P_{\mathrm{U}(1)}(m_3,t^2) \\ & \qquad \qquad \qquad \qquad \quad \times \left[\prod_{i=1}^3 w_i^{d_i}\right] t^{2\Delta(\vec{m})} \delta_{f(\vec{m}),j} ~,
}
where we denote by $w_1$, $w_2$ and $w_3$ the topological fugacities associated with the $\U(3)$, $\U(2)$ and $\U(1)$ gauge nodes respectively. From \eref{Deltaofm}, the conformal dimension entering in \eref{CBHSref321} reads
\bes{
    2\Delta(\vec{m}) = - 2 &\left(\sum_{1 \le i < j \le 3} \left|m_{1,i}-m_{1,j}\right|\right) - 2 \left|m_{2,1}-m_{2,2}\right| \\+& \left(\sum_{i=1,2,3} \, \sum_{j=1,2} \left|3 m_{i,1}-m_{2,j}\right|\right) + \sum_{j=1,2} \left|m_{2,1}-m_{3}\right|~,
}
where we have to take into account the constraint coming from the delta function $\delta_{f(\vec{m}),j}$, which can be explained as follows. The quantity $\lambda_{\mathfrak{c}}$, defined in \eref{lambdac}, in this case reads
\begin{equation} \label{lambdac321}
    \lambda_{\mathfrak{c}} = \left(\sum_{i=1}^3 \mathfrak{c}_{1,i}\right) + 3 \left(\sum_{i=1}^2 \mathfrak{c}_{2,i}\right) + 3 \mathfrak{c}_1 - 1~,
\end{equation}
moreover we define the function
\bes{
f(\vec{m}) = \left(\sum_{i=1}^3 \mathfrak{c}_{1,i} m_{1,i}\right) + \left(\sum_{i=1}^2 \mathfrak{c}_{2,i} m_{2,i}\right) + \mathfrak{c}_{3} m_{3}
}
implementing \eref{eq:MagLatGenCon}, which takes values in $\left\{0,1,\dots,\lambda_{\mathfrak{c}}\right\}$ for given $\mathfrak{c}_{i,a_i}$. Explicitly, some convenient choices to ungauge the decoupled $\U(1)$ are the following.
\bi
\item We can take $\mathfrak{c}_{1,i} = (1,1,1)$, $\mathfrak{c}_{2,i} = (0,0)$ and $\mathfrak{c}_{3} = 0$, which is equivalent to $\lambda_{\mathfrak{c}} = 2$ and $f(\vec{m}) = d_1$. Then, from the first summation in \eref{CBHSref321}, we just have to consider the contribution from $d_1 = \{0,1,2\}$, whereas there is no restriction on $d_2$ and $d_3$. Observe that the resulting Coulomb branch Hilbert series is periodic in $d_1$, with periodicity $\lambda_{\mathfrak{c}} + 1 = 3$. Explicitly, we can detect such periodicity by looking at the contributions coming from $d_1 = j = 0, \ldots, 5$ in the first summation in \eref{CBHSref321}:
\bes{ \label{Z3zeroform}
{\violet j = 0~:} \quad &{\violet 1 + 8 t^2 + 100 t^4 + 645 t^6 + 2968 t^8 + 10976  t^{10}~,} \\
j = 1~: \quad &10 t^2 + 100 t^4 + 640 t^6 + 2975 t^8 + 10976  t^{10}~, \\
j = 2~: \quad &10 t^2 + 100 t^4 + 640 t^6 + 2975 t^8 + 10976  t^{10}~, \\
{\violet j = 3~:} \quad & {\violet 1 + 8 t^2 + 100 t^4 + 645 t^6 + 2968 t^8 + 10976  t^{10}~,} \\
j = 4~: \quad &10 t^2 + 100 t^4 + 640 t^6 + 2975 t^8 + 10976  t^{10}~, \\
j = 5~: \quad &10 t^2 + 100 t^4 + 640 t^6 + 2975 t^8 + 10976  t^{10}~,
}
from which it is clear that every contribution coming from $d_1 = j = i+3 k$, with $i = 0,1,2$ and $k = 0, \ldots, \infty$, is independent of $k$. The periodicity of 3 in $d_1$ indicates that the theory has a $\BZ_3$ zero-form symmetry. We will gauge this symmetry later.

\item Another possible option is to set $\mathfrak{c}_{1,i} = (0,0,0)$, $\mathfrak{c}_{2,i} = (1,1)$ and $\mathfrak{c}_{3} = 0$, \ie $\lambda_{\mathfrak{c}} = 5$ and $f(\vec{m}) = d_2$. It follows that, in this case, $d_2$ takes values in $ \{0,1,2,3,4,5\}$, with no restriction on $d_1$ and $d_3$. It follows that the resulting Coulomb branch Hilbert series is periodic in $d_2$, with periodicity $\lambda_{\mathfrak{c}}+1 = 6$.
\item Finally, another convenient parametrisation is $\mathfrak{c}_{1,i} = (0,0,0)$, $\mathfrak{c}_{2,i} = (0,0)$ and $\mathfrak{c}_{3} = 1$, namely $\lambda_{\mathfrak{c}} = 2$ and $f(\vec{m}) = d_3$. This choice corresponds to summing over $d_3 \in \{0,1,2\}$ in \eref{CBHSref321}, with no further restriction on $d_1$ and $d_2$. The resulting Coulomb branch Hilbert series is periodic in $d_3$, with periodicity $\lambda_{\mathfrak{c}}+1 = 3$.
\ei
In all of such equivalent cases, upon setting the topological fugacities to unity, the Coulomb branch Hilbert series of quiver \eref{eq:(3)-{3,1}-(2)-(1)} admits the following closed form:
\bes{
&\mathrm{HS}\left[\text{CB of \eref{eq:(3)-{3,1}-(2)-(1)}}\right](t, w_i=1) \\& = \frac{1 + 18 t^2 + 65 t^4 + 65 t^6 + 18 t^8 + t^{10}}{\left(1 - t^2\right)^{10}} \\&= 1 + 28 t^2 + 300 t^4 + 1925 t^6 + 8918 t^8 + 32928  t^{10} + \ldots \\ &= \PE\left[28 t^2 - 106 t^4 + 833 t^6 - 8400 t^8 + 91392  t^{10} + \ldots\right]~,
}
where the moment map operators appearing at order $t^2$ in the series expansion reveal that the Coulomb branch symmetry of the theory is $\so(8)$. Indeed, the Coulomb branch in question is isomorphic to the closure of the minimal nilpotent orbit of $\so(8)$, denoted by $\overline{\mathrm{min}.D_4}$ \cite{Benvenuti:2010pq}. A Lagrangian description possessing such Coulomb branch is known to be the following affine $D_4$ quiver:
\bes{ \label{affineD4}
\begin{tikzpicture}[baseline]
            \node[gauge,label=right:{\scriptsize $2$}] (1) at (0,0) {};
            \node[gauge,label=left:{\footnotesize$1$}] (2) at (-1,1) {};
            \node[gauge,label=left:{\footnotesize$1$}] (3) at (-1,-1) {};
            \node[gauge,label=right:{\footnotesize$1$}] (4) at (1,1) {};
            \node[gauge,label=right:{\footnotesize$1$}] (5) at (1,-1) {};
            \draw (2)--(1)--(3) (4)--(1)--(5);
            \node (x) at (3,0) {\footnotesize $/\U(1)$};
        \end{tikzpicture} 
}

We can now realise various discrete quotients of quiver \eref{eq:(3)-{3,1}-(2)-(1)} by picking different values of the parameters $\mathfrak{c}_{i,a_i}$ and restricting the first summation in \eref{CBHSref321} to $j$ taking only {\it some} of the values in a subset of $\{0,\ldots,\lambda_{\mathfrak{c}} \}$.

\paragraph{$\BZ_2$ quotient.} Let us take $\mathfrak{c}_{2,i} = (1,1)$ and the other $\mathfrak{c}_{i,a_i} = 0$, \ie $\lambda_{\mathfrak{c}} = 5$ and $f(\vec{m}) = d_2$. The contributions to the Coulomb branch Hilbert series due to the first summation in \eref{CBHSref321} for the various choices of $j = 0, \ldots, 5$ are as follows:
\bes{ \label{varcontrib}
{\red j = 0~:} \quad &{\red 1 + 6 t^2 + 62 t^4 + 341 t^6 + 1558 t^8 + 5596  t^{10}}~, \\
j = 1~: \quad &4 t^2 + 44 t^4 + 312 t^6 + 1448 t^8 + 5440  t^{10}~, \\
{\violet j = 2~:} \quad &{\violet 3 t^2 + 47 t^4 + 304 t^6 + 1464 t^8 + 5410  t^{10}}~, \\
{\blue j = 3~:} \quad &{\blue 8 t^2 + 56 t^4 + 352 t^6 + 1536 t^8 + 5632  t^{10}}~, \\
{\violet j = 4~:} \quad &{\violet 3 t^2 + 47 t^4 + 304 t^6 + 1464 t^8 + 5410  t^{10}}~, \\
j = 5~: \quad &4 t^2 + 44 t^4 + 312 t^6 + 1448 t^8 + 5440  t^{10}~.
}
We can perform a $\BZ_2$ quotient by considering only the contributions highlighted in {\red red} and in {\violet violet} above, namely we take $j \in \{0,2,4\}$ in the first summation in \eref{CBHSref321}.  This also restricts $d_2 \in \{0,2,4\}$.

This $\BZ_2$ quotient admits the following Lagrangian description:
\bes{ \label{affineD4SU2}
\begin{tikzpicture}[baseline]
            \node[gauge,label=right:{\scriptsize SU$(2)$}] (1) at (0,0) {};
            \node[gauge,label=left:{\footnotesize$1$}] (2) at (-1,1) {};
            \node[gauge,label=left:{\footnotesize$1$}] (3) at (-1,-1) {};
            \node[gauge,label=right:{\footnotesize$1$}] (4) at (1,1) {};
            \node[gauge,label=right:{\footnotesize$1$}] (5) at (1,-1) {};
            \draw (2)--(1)--(3) (4)--(1)--(5);
        \end{tikzpicture}
}
The resulting Coulomb branch Hilbert series is
\bes{
&\mathrm{HS}\left[\text{CB of \eref{eq:(3)-{3,1}-(2)-(1)}}/\BZ_2\right](t, w_i=1) \\&=\frac{1 + 7 t^2 + 101 t^4 + 244 t^6 + 666 t^8 + 650 t^{10} + \text{palindrome} + t^{20}}{\left(1-t^2\right)^{10} \left(1+t^2\right)^5}\\
&=1 + 12 t^2 + 156 t^4 + 949 t^6 + 4486 t^8 + 16416  t^{10}+\ldots\\ &= \PE\left[12 t^2 + 78 t^4 - 351 t^6 - 1832 t^8 + 23424 t^{10} + \ldots\right]~,
}
where the moment map operators appearing at order $t^2$ in the above series expansion signal that the Coulomb branch symmetry is $\su(2)^4$.

\paragraph{$\BZ_3$ quotient.} We can realise a $\BZ_3$ quotient instead by setting $\mathfrak{c}_{3} = 1$ and the other $\mathfrak{c}_{i,a_i} = 0$, \ie $\lambda_{\mathfrak{c}} = 2$ and $f(\vec{m}) = d_3$, where we consider only the contribution coming from $j = 0$ in the first summation in \eref{CBHSref321}. This also restricts $d_3=0$. Equivalently, we can realise the same $\BZ_3$ quotient by taking the same choice of $\mathfrak{c}_{i,a_i}$ as in the $\BZ_2$ case discussed above, namely $\mathfrak{c}_{2,i} = (1,1)$ and the other $\mathfrak{c}_{i,a_i} = 0$, where we now consider only the contributions coming from $j = \{0,3\}$ in the first summation in \eref{CBHSref321}, which are highlighted in {\red red} and in {\blue blue} in \eref{varcontrib}. This amounts to restricting $d_2 = \{0,3\}$. The resulting Coulomb branch Hilbert series reads
\bes{
&\mathrm{HS}\left[\text{CB of \eref{eq:(3)-{3,1}-(2)-(1)}}/\BZ_3\right](t, w_i=1) \\& = \frac{1 + 4 t^2 + 23 t^4 + 23 t^6 + 4 t^8 + t^{10}}{\left(1 - t^2\right)^{10}} \\&= 1 + 14 t^2 + 118 t^4 + 693 t^6 + 3094 t^8 + 11228  t^{10} + \ldots \\ &= \PE\left[14 t^2 + 13 t^4 - 49 t^6 - 56 t^8 + 672  t^{10} + \ldots\right]~,
}
where we observe that the Coulomb branch symmetry is reduced to $\mathfrak{g}_2$. The Coulomb branch in question is isomorphic to the $\BZ_2$ cover of the closure of the subregular nilpotent orbit of $G_2$, denoted by $\mathbb{Z}_2 \, \textnormal{cover}\left(\overline{\mathrm{subreg}.G_2}\right)$ \cite[(6.5)-(6.8)]{Hanany:2020jzl}. 
It turns out that this Coulomb branch Hilbert series agrees with that of the $\BZ_3$ wreathing of the affine $D_4$ quiver \cite{Bourget:2020bxh}: 
\bes{ \label{wreathZ3}
\begin{tikzpicture}
            \node[gauge,label=below:{\footnotesize $2$}] (1) at (0,0) {};
            \node[gauge,label=below:{\footnotesize $1$}] (2) at (-1,0) {};
            \node[gauge,label=below:{\footnotesize $[1]\wr\mathbb{Z}_3$}] (3) at (1,0) {};
            \draw (2)--(1)--(3);
        \end{tikzpicture}\;,
}
where $[1]\wr\mathbb{Z}_3$ indicates the wreath product. The Higgs and Coulomb branch Hilbert series of this was studied in \cite[Figures 8, 9, 11]{Bourget:2020bxh} (see also \cite{Giacomelli:2024sex}), and the superconformal index was computed in \cite[Section 3.4]{Grimminger:2024mks}.  It is interesting to point out that the wreathed quiver has the Higgs branch that is isomorphic to $\BC^2/\hat{E}_6$, but we do not have an access to this moduli space from the description \eref{eq:(3)-{3,1}-(2)-(1)}.

\paragraph{Another $\BZ_3$ quotient.} We can also obtain a different $\BZ_3$ quotient, which we denote by $\BZ_{3'}$, by taking $\mathfrak{c}_{1,i} = (1,1,1)$ and the other $\mathfrak{c}_{i,a_i} = 0$, namely $\lambda_{\mathfrak{c}} = 2$ and $f(\vec{m}) = d_1$, and considering only the contribution coming from $j = 0$ in the first summation in \eref{CBHSref321}. Note that this choice restricts $d_1=0$. The resulting Coulomb branch Hilbert series is indicated in {\purple violet} in \eref{Z3zeroform}, and is given by
\bes{
&\mathrm{HS}\left[\text{CB of \eref{eq:(3)-{3,1}-(2)-(1)}}/\BZ_{3'}\right](t, w_i=1) \\& = \frac{1 + t^2 + 65 t^4 + 75 t^6 + 305 t^8 + 309 t^{10} + 309 t^{12} + \text{palindrome} + t^{22}}{\left(1 - t^2\right)^{7} \left(1 - t^6\right)^{3}} \\&= 1 + 8 t^2 + 100 t^4 + 645 t^6 + 2968 t^8 + 10976  t^{10} + \ldots \\ &= \PE\left[8 t^2 + 64 t^4 + 13 t^6 - 1850 t^8 - 636  t^{10} + \ldots\right]~,
}
where the Coulomb branch symmetry is $\su(3)$. Unlike the previously discussed $\BZ_3$ quotient, this $\BZ_{3'}$ quotient does not arise from wreathing the affine $D_4$ quiver, since this is not contained in \cite[Figure 8]{Bourget:2020bxh}. 

As a check, we can compute the volume of the base of the Coulomb branch viewed as a cone using the method described in \cite[(1.17)]{Martelli:2006yb}; see also \cite{Hanany:2018dvd} for the application in the context of discrete gauging. Given a Coulomb branch Hilbert series $H(t)$, the said volume, denoted by $\text{Vol}$, can be computed as $\lim_{t \rightarrow 1} (1-t)^d H(t)$, where $d$ is the complex dimension of the Coulomb branch.

In this case, we obtain
\bes{
\frac{\mathrm{Vol}\left[\text{CB of \eref{eq:(3)-{3,1}-(2)-(1)}}\right]}{\mathrm{Vol}\left[\text{CB of \eref{eq:(3)-{3,1}-(2)-(1)}/$\BZ_{3'}$}\right]} = 3~,
}
which is expected for a $\BZ_3$ quotient

\paragraph{$\BZ_6$ quotient.} Finally, we can realise a $\BZ_6$ quotient by taking the same choice of $\mathfrak{c}_{i,a_i}$ as in the $\BZ_2$ case, namely $\mathfrak{c}_{2,i} = (1,1)$ and the other $\mathfrak{c}_{i,a_i} = 0$. However, this time, we consider only the contribution of $j  = 0$ in the first summation in \eref{eq:(3)-{3,1}-(2)-(1)}, meaning that we restrict to $d_2 = 0$. This $\BZ_6$ quotient admits the following description as a wreathed quiver \cite{Bourget:2020bxh}, but with the $\SU(2)$ central node (see \cite[Section 3.11]{Grimminger:2024mks}):
\bes{ \label{Z6quotientminD4}
\begin{tikzpicture}
            \node[gauge,label=below:{\footnotesize$\SU(2)$}] (1) at (0,0) {};
            \node[gauge,label=below:{\footnotesize$1$}] (2) at (-2,0) {};
            \node[gauge,label=below:{\footnotesize$[1]\wr\mathbb{Z}_3$}] (3) at (2,0) {};
            \draw (2)--(1)--(3);
        \end{tikzpicture}
}
The resulting Coulomb branch Hilbert series reads
\bes{
&\mathrm{HS}\left[\text{CB of \eref{eq:(3)-{3,1}-(2)-(1)}}/\BZ_{6}\right](t, w_i=1) \\&= \frac{1 + t^2 + 37 t^4 + 76 t^6 + 218 t^8 + 230 t^{10} + 218 t^{12} + \text{palindrome} + t^{20}}{\left(1-t^2\right)^{10} \left(1+t^2\right)^5}\\
&=1 + 6 t^2 + 62 t^4 + 341 t^6 + 1558 t^8 + 5596  t^{10}+\ldots\\ &= \PE\left[6 t^2 + 41 t^4 + 39 t^6 - 524 t^8 - 1392 t^{10} + \ldots\right]~,
}
where we observe that the Coulomb branch symmetry becomes $\su(2)^2$. Upon computing the volume, we find that
\bes{
\frac{\mathrm{Vol}\left[\text{CB of \eref{eq:(3)-{3,1}-(2)-(1)}}\right]}{\mathrm{Vol}\left[\text{CB of \eref{eq:(3)-{3,1}-(2)-(1)}/$\BZ_{6}$}\right]} = 6~,
}
in agreement with a $\BZ_6$ quotient.
\begin{table}[h!]
    \makebox[\textwidth][c]{\begin{tabular}{|c|c|c|c|}
        \hline
        \raisebox{0\height}{\eref{eq:(3)-{3,1}-(2)-(1)}} & CB & CB symmetry & Quiver with same CB \\
        \hline\hline
        ${\color{red}d_1}\in\{0,1,2\}$ & \multirow{5}{*}{$\overline{\mathrm{min}.D_4}$} & \multirow{5}{*}{$\mathfrak{so}(8)$} & \multirow{5}{*}{\raisebox{0\height}{\eref{affineD4}}} \\
        or & & & \\
        ${\color{blue}d_2}\in\{0,1,2,3,4,5\}$ & & & \\
        or & & & \\
        ${\color{new-green}d_3}\in\{0,1,2\}$ & & & \\
        \hline
        ${\color{blue}d_2}\in\{0,2,4\}$ & $\overline{\mathrm{min}.D_4}/\mathbb{Z}_2$ & $\mathfrak{su}(2)^4$ & \raisebox{0\height}{\eref{affineD4SU2}}\\
        \hline
        ${\color{blue}d_2}\in\{0,3\}$ & \multirow{3}{*}{
        \begin{tabular}{c}
             $\overline{\mathrm{min}.D_4}/\mathbb{Z}_3$\\            $=\mathbb{Z}_2\, \textnormal{cover}\left(\overline{\mathrm{subreg}.G_2}\right)$ 
        \end{tabular}} & \multirow{3}{*}{$\mathfrak{g}_2$} & \multirow{3}{*}{\raisebox{0\height}{\eref{wreathZ3}}} \\
        or & & & \\
        ${\color{new-green}d_3}\in\{0\}$ & & & \\
        \hline
        ${\color{red}d_1}\in\{0\}$ & $\overline{\mathrm{min}.D_4}/\mathbb{Z}_{3'}\neq \mathbb{Z}_2\, \textnormal{cover}\left(\overline{\mathrm{subreg}.G_2}\right)$ & $\mathfrak{su}(3)$ & \raisebox{0\height}{$\begin{tikzpicture}
            \node (1) at (0,0) {Not known};
        \end{tikzpicture}$} \\
        \hline
        ${\color{blue}d_2}\in\{0\}$ & \begin{tabular}{c}
             $\overline{\mathrm{min}.D_4}/\mathbb{Z}_6$\\
             $=\left(\mathbb{Z}_2\, \textnormal{cover}\left(\overline{\mathrm{subreg}.G_2}\right)\right)/\mathbb{Z}_2$\\
             $\neq\overline{\mathrm{subreg}.G_2}$
        \end{tabular} & $\mathfrak{su}(2)^2$ & \raisebox{0\height}{\eref{Z6quotientminD4}}\\
        \hline
    \end{tabular}}
    \caption{Different global structures of the quiver \eqref{eq:(3)-{3,1}-(2)-(1)} are given in the form of different possible constraints on a $d_i$ of the quiver. The corresponding Coulomb branch, and its global symmetry, are provided. In the last column a quiver, possibly with a special unitary node, and/or a wreathing \cite{Bourget:2020bxh}, which has the same Coulomb branch, is given.}
    \label{tab:(3)-{3,1}-(2)-(1)_gs_compare}
\end{table}

\section{Discrete Gauging and Mixed 't Hooft Anomalies} \label{sec:nonsimplylacedanom}
In this section, we study zero-form and one-form discrete symmetries in various $(p,q)$-edged quivers and gaugings thereof. We first demonstrate this in an Abelian example, where both the Coulomb branch Hilbert series and the superconformal index are computed. We subsequently examine sequential gauging in a non-Abelian example using the Coulomb branch Hilbert series. We observe that, whenever a non-trivial $(p,q)$-edged quiver possesses a discrete one-form symmetry, possibly arising from gauging the dual zero-form symmetry, the former generally has a mixed 't Hooft anomaly with some combinations of the continuous topological symmetries in the quiver. This manifests itself in the Hilbert series or in the index, as described in \cite{Mekareeya:2022spm}.

\subsection{Abelian Example} \label{sec:Abelianex}
Let us consider the following $(p,q)$-edged Abelian quiver
\bes{ \label{Trianglequivggg}
    \vcenter{\hbox{\begin{tikzpicture}
        \node[gauge,label=below:{$1$},label=left:{\red{\scriptsize $m_3$}}] (1) at (-2,0) {};
        \node[gauge,label=below:{$1$},label=right:{\red{\scriptsize $m_1$}}] (2) at (2,0) {};
        \node[gauge,label=above:{$1$},label=right:{\red{\scriptsize $m_2$}}] (3) at (0,2) {};
        \edge{1}{2}{3}{1}{1}
        \edgeur{2}{3}{2}{3}{1}
        \edgeul{1}{3}{2}{1}{1}
    \end{tikzpicture}}}
\quad \text{\footnotesize$/\U(1)$}
}
where we denote with $m_i \in \BZ$, with $i = 1, 2, 3$, the magnetic fluxes associated with the three $\U(1)$ gauge nodes. Note that, due to the $\U(1)$ gauge groups, this quiver admits a conventional Lagrangian description where the matter fields are the hypermultiplets carrying charges ${(0,1,-2)}$, ${(-1,0,3)}$,  ${(2,-3,0)}$ under the $\U(1)$ gauge groups labelled by $(m_1, m_2, m_3)$ respectively.

 The adjacency matrix, rank vector and shortness vector associated with \eref{Trianglequivggg} are given respectively by
\begin{equation} \label{AdjmatTrianglequivggg}
    A=\begin{pmatrix}
        -2 & 2 & 1 \\
        3 & -2 & 1 \\
        3 & 2 & -2 
    \end{pmatrix}\,,\;v=\begin{pmatrix}
        1\\
        1\\
        1
    \end{pmatrix}\,,\;s=\begin{pmatrix}
        3\\
        2\\
        1
    \end{pmatrix}~,
\end{equation}
from which the quantity $\lambda_{\mathfrak{c}}$, defined in \eref{lambdac}, in the present case reads
\bes{
\lambda_{\mathfrak{c}} = 3 \fc_{1} + 2 \fc_{2} + \fc_{3} - 1~.
}
Any choices of $\fc_{i} \in \BZ_{\geq 0}$ lead to the same Coulomb branch Hilbert series.  However, if we would like to consider a specific discrete gauging, an appropriate choice must be made.  We will shortly state our choices of $\fc_{i}$ in subsequent calculations. 

Let us define the function
\bes{ \label{deffm}
f(\vec{m}) = \sum_{i=1}^3 \fc_{i} m_i
}
as in \eref{eq:MagLatGenCon}, which takes values $0, 1, \ldots, \lambda_\fc$ for given $\fc_{i}$. Then, the Coulomb branch Hilbert series of \eref{Trianglequivggg} can be obtained via the monopole formula \eref{CBHSref} as follows:
\bes{ \label{CBHSrefTrianglequivggg}
&\mathrm{HS}\left[\text{CB of \eref{Trianglequivggg}}\right](t, \vec{w}) \\
&= \left(1-t^2\right) \sum_{j=0}^{\lambda_{\mathfrak{c}}} \sum_{\vec{m}\in\frac{\Gamma_{G}^{mw}}{W_G}} \left[\prod_{i=1}^3 P_{\mathrm{U}(1)}(m_i,t^2) w_i^{m_i}\right] \times t^{2\Delta(\vec{m})} \delta_{f(\vec{m}),j} ~,
}
where $w_i$ are the topological fugacities associated with the three $\U(1)$ gauge nodes, and $G=\tilde{G}/\U(1)$ with $\tilde{G} = \U(1)^3$. Using \eref{AdjmatTrianglequivggg}, the conformal dimension appearing in \eref{CBHSrefTrianglequivggg} reads
\bes{ \label{DeltaofmTrianglequivggg}
    2\Delta(\vec{m})=& \left|2 m_1-3 m_2\right| + \left|m_1-3 m_3\right| + \left|m_2-2 m_3\right|~,
}
where it is important to take into account the constraint coming from the delta function $\delta_{f(\vec{m}),j}$. 

A convenient choice to ungauge the decoupled $\U(1)$ gauge group is to set $\fc_{1} = \fc_{2} = 0$ and $\fc_{3} = 1$, such that $\lambda_{\mathfrak{c}} = 0$ and $f(\vec{m}) = m_3$. Adopting this parametrisation, the conformal dimension is given by \eref{DeltaofmTrianglequivggg}, with the constraint $m_3 = 0$. Upon setting the topological fugacities to unity, the Coulomb branch Hilbert series admits the following closed form:
\bes{  \label{CBHStriangle}
\begin{split}
&\mathrm{HS}\left[\text{CB of \eref{Trianglequivggg}}\right](t, w_i=1) \\ & = \frac{1 + t^2 + 3 t^3 + 4 t^4 + 4 t^5 + 4 t^6 + 3 t^7 + t^8 + t^{10}}{\left(1 - t^2\right) \left(1 - t^3\right) \left(1 - t^4\right) \left(1 - t^5\right)} \\&= 1 + 2 t^2 + 4 t^3 + 7 t^4 + 10 t^5 + 16 t^6 + 22 t^7 + 31  t^8 + 40  t^9 + 54  t^{10} + \ldots \\ &= \PE\left[2 t^2 + 4 t^3 + 4 t^4 + 2 t^5 - 6 t^6 - 10 t^7 - 12 t^8 +2  t^9 + 29  t^{10} + \ldots\right]~.
\end{split}
}
We give an algebraic description of this Coulomb branch by computing the equations for the Higgs branch of the mirror quiver in \ref{sec:mirrorTriangle}.
Note that the Coulomb branch symmetry of the theory, which can be read from the moment map operators appearing at order $t^2$, is $\u(1)^2$, reflecting that there are two independent $\U(1)$ topological symmetries in theory \eref{Trianglequivggg}. This is consistent with the relation \eref{constraintonzi}.

Let us now demonstrate how to realise various $\BZ_r$ quotients, \ie~ gauging $\BZ_r$ zero-form symmetries, which we will refer to as $\BZ^{[0]}_r$.

\paragraph{$\BZ_2$ quotient.} Suppose that we would like to perform a $\BZ_2$ quotient, \ie~ gauge the $\BZ_2$ zero-form symmetry. We can take $(\fc_{1}, \fc_{2}, \fc_{3}) = (0,1,0)$ and consider only the contribution of $j=0$ in the first summation in \eref{CBHSrefTrianglequivggg}. This amounts to setting $m_2 =0$ in \eref{CBHSrefTrianglequivggg}. The resulting Coulomb branch Hilbert series is
\bes{ \label{CBHSZ2}
\begin{split}
&\mathrm{HS}\left[\text{CB of \eref{Trianglequivggg}/$\BZ^{[0]}_2$}\right](t, w_i=1) \\ & = \frac{1 + t^2 + t^3 + t^4 + 2 t^5 + 3 t^6 + 4 t^7 + 3 t^8 + 2 t^9 + t^{10} + t^{11} + t^{12} + t^{14}}{\left(1 - t^2\right) \left(1 - t^3\right) \left(1 - t^5\right) \left(1 - t^8\right)} \\&= 1 + 2 t^2 + 2 t^3 + 3 t^4 + 6 t^5 + 8 t^6 + 12 t^7 + 17  t^8 +20  t^9 + 28  t^{10} + \ldots \\ &= \PE\left[2 t^2 + 2 t^3 + 2 t^5 + t^6 + 2  t^7 - 4  t^9 - 5  t^{10} + \ldots\right]~.
\end{split}
}
In this particular case, we obtain
\bes{
\frac{\mathrm{Vol}\left[\text{CB of \eref{Trianglequivggg}}\right]}{\mathrm{Vol}\left[\text{CB of \eref{Trianglequivggg}/$\BZ^{[0]}_2$}\right]} = 2~,
}
as expected from the $\BZ_2$ quotient.

\paragraph{$\BZ_3$ quotient.} This can be realised using $(\fc_{1}, \fc_{2}, \fc_{3}) = (1,0,0)$ and consider only the contribution of $j=0$ in the first summation in \eref{CBHSrefTrianglequivggg}. The resulting Coulomb branch Hilbert series is
\bes{ \label{CBHSZ3quot}
\begin{split}
&\mathrm{HS}\left[\text{CB of \eref{Trianglequivggg}/$\BZ^{[0]}_3$}\right](t, z_i=1) \\ & = \frac{1 + t^2 + 2 t^4 + t^5 + 2 t^6 + 3 t^7 + 2 t^8 + 3 t^9 + 2 t^{10} + t^{11} + 2 t^{12} + t^{14} + t^{16}}{\left(1 - t^2\right) \left(1 - t^4\right) \left(1 - t^5\right) \left(1 - t^9\right)} \\&= 1 + 2 t^2 + 5 t^4 + 2 t^5 + 8 t^6 + 6 t^7 + 13 t^8 + 14 t^9 + 20  t^{10} + \ldots \\ &= \PE\left[2 t^2 + 2 t^4 + 2 t^5 + 2 t^7 - t^8 - t^{10} + \ldots\right]~.
\end{split}
}
We also obtain
\bes{
\frac{\mathrm{Vol}\left[\text{CB of \eref{Trianglequivggg}}\right]}{\mathrm{Vol}\left[\text{CB of \eref{Trianglequivggg}/$\BZ^{[0]}_3$}\right]} = 3~,
}
as expected from the $\BZ_3$ quotient.

\paragraph{Another $\BZ_{3}$ quotient.} We can also realise a $\BZ_3$ quotient by setting $(\fc_{1}, \fc_{2}, \fc_{3}) = (0,1,1)$ and consider only the contribution of $j=0$ in the first summation in \eref{CBHSrefTrianglequivggg}. We point out that this $\BZ_3$ quotient is not equivalent to the one presented above, as can be seen from the resulting Coulomb branch Hilbert series, which does not coincide with \eref{CBHSZ3quot}:
\bes{ \label{CBHSZ3quotanother}
\begin{split}
&\mathrm{HS}\left[\text{CB of \eref{Trianglequivggg}/$\BZ^{[0]}_{3'}$}\right](t, w_i=1) \\ & = \frac{1}{\left(1 - t\right) \left(1 - t^2\right) \left(1 - t^{12}\right) \left(1 - t^{15}\right)} \left(1 - t + t^2 + t^3 - t^4 + t^5 + t^6 - t^7 \right. \\&\left. \quad \, \, \, \, \, + 3  t^8 + 3  t^9 - t^{10} + 3  t^{11} + 2  t^{12} - 2  t^{13} + 2  t^{14} + \text{palindrome} + t^{26}\right) \\&= 1 + 2 t^2 + 2 t^3 + 3 t^4 + 4 t^5 + 6 t^6 + 6 t^7 + 11 t^8 + 14 t^9 + 
 18  t^{10} + \ldots \\ &= \PE\left[2 t^2 + 2 t^3 - t^6 + 2 t^8 + 4 t^9 + 2  t^{10} + \ldots\right]~,
\end{split}
}
where, in order to avoid confusion with the previous $\BZ_3$ quotient, we denote by $\BZ^{[0]}_{3'}$ the zero-form symmetry which is gauged in the present case. This $\BZ_3$ quotient is consistent with
\bes{
\frac{\mathrm{Vol}\left[\text{CB of \eref{Trianglequivggg}}\right]}{\mathrm{Vol}\left[\text{CB of \eref{Trianglequivggg}/$\BZ^{[0]}_{3'}$}\right]} = 3~.
}

\paragraph{$\BZ_{13}$ quotient.} This can be realised using $(\fc_{1}, \fc_{2}, \fc_{3}) = (3,2,0)$ and consider only the contribution of $j=0$ in the first summation in \eref{CBHSrefTrianglequivggg}. The resulting Coulomb branch Hilbert series is
\bes{ \label{CBHSZ13}
\begin{split}
&\mathrm{HS}\left[\text{CB of \eref{Trianglequivggg}/$\BZ^{[0]}_{13}$}\right](t, w_i=1) \\ & = \frac{1}{\left(1 - t^2\right) \left(1 - t^5\right) \left(1 - t^{39}\right) \left(1 - t^{52}\right)} \left(1 + t^2 + t^4 + t^5 + t^6 + t^7 + t^8 + t^9 \right. \\&\left. \quad \, \, \, \, \,+ t^{10} + t^{11} + t^{12} + \
t^{13} + t^{14} + t^{15} + t^{16} + t^{17} + 3  t^{18} + 5  t^{19} + 3  t^{20} + 5  t^{21} \right. \\&\left. \quad \, \, \, \, \,+ 5  t^{22} + 3  t^{23} + 3  t^{24} + 3  t^{25} + 3  t^{26} + 3  t^{27} + 3  t^{28} + 3  t^{29} + 3  t^{30} + 3  t^{31} \right. \\&\left. \quad \, \, \, \, \,+ 3  t^{32} + 3  t^{33} + 3  t^{34} + 5  t^{35} + 5  t^{36} + 7  t^{37} + 9  t^{38} + 8  t^{39} + 7  t^{40} + 6  t^{41} \right. \\&\left. \quad \, \, \, \, \,+ 5  t^{42} + 4  t^{43} + 4  t^{44} + 4  t^{45} + 4  t^{46} + 4  t^{47} + 4  t^{48} + \text{palindrome} + t^{94}\right) \\&= 1 + 2 t^2 + 3 t^4 + 2 t^5 + 4 t^6 + 4 t^7 + 5 t^8 + 6 t^9 + 8  t^{10} + \ldots \\ &= \PE\left[2 t^2 + 2 t^5 - t^{10} + \ldots\right]~.
\end{split}
}
We also obtain
\bes{
\frac{\mathrm{Vol}\left[\text{CB of \eref{Trianglequivggg}}\right]}{\mathrm{Vol}\left[\text{CB of \eref{Trianglequivggg}/$\BZ^{[0]}_{13}$}\right]} = 13~,
}
which is expected for a $\BZ_{13}$ quotient.

We will discuss such quotients, the resulting dual one-form symmetries and mixed 't Hooft anomalies more systematically using the index.

\subsubsection{Superconformal Index of Theory \eref{Trianglequivggg}}
As described in Section \ref{sec:ungauge}, an overall $\U(1)$ in \eref{Trianglequivggg} can be modded out in many ways. In this subsection, we adopt the alternative method described around \eref{diracpq}, which is more conventional for computing superconformal indices. In particular, we achieve this by restricting the gauge fugacities and the corresponding magnetic fluxes in the following way:
\bes{ \label{constr0}
\prod_{i=1}^3 z_i^{\fc_i} =1~, \quad \sum_{i=1}^3  {\fc_i} m_i  = 0~,
}
where we also allow the magnetic fluxes $m_i$ to be non-integers. As we discussed earlier around \eref{diracpq}, this amounts to reducing the freedom on $\sum_{i=1}^3  {\fc_i} m_i$ in \eref{deffm} by setting it to zero, but increasing the freedom on the magnetic fluxes by allowing them to take non-integral values. We will soon discuss the conditions that determine the allowed values of $m_i$. We will also determine the discrete symmetry that acts on the monopole operators with fractional magnetic fluxes, refine it with an appropriate fugacity in the index, and eventually gauge it. 

The superconformal index of theory \eref{Trianglequivggg} can be computed from the following expression:
\bes{ \label{indextriangle}
\scalebox{0.92}{$
\begin{split}
\CI_{\text{\eref{Trianglequivggg}}} = &\sum_{m_1, m_2, m_3}    w_1^{m_1} w_2^{m_2} w_3^{m_3}
\oint \frac{d z_1}{ 2\pi i z_1} \oint \frac{d z_2}{ 2\pi i z_2} \oint \frac{d z_3}{ 2\pi i z_3}   \,\, \delta\text{\eref{constr0}} \times \CI_{\text{matter}}~, \\
\CI_{\text{matter}} =& \left[\CI_\chi^1\left(a^{-2}; 0; x\right) \right]^2 \times \prod_{s= \pm1} \CI_{\chi}^{\frac{1}{2}} \left( (z_1^{-1} z_3^3)^s a;  s(-m_1+ 3 m_3) ; x\right)\\
& \times \CI_{\chi}^{\frac{1}{2}} \left( (z_1^2 z_2^{-3} a)^s;  s(2m_1- 3 m_2) ; x\right) \times \CI_{\chi}^{\frac{1}{2}} \left( (z_2 z_3^{-2})^s a;  s(m_2- 2 m_3) ; x\right) ~,
\end{split}$}
}
where the contribution of the chiral multiplet of $R$-charge $R$ is
\bes{
\CZ^R_{\chi}(z; l;x) = \left( x^{1-R} z^{-1} \right)^{|l|/2} \prod_{j=0}^\infty \frac{1-(-1)^l z^{-1} x^{|l|+2-R+2j}}{1-(-1)^l z\,  x^{|l|+R+2j}}~.
}
For simplicity, we have turned off the background magnetic fluxes for the axial symmetry $a$ and for the topological symmetries $w_{1,2,3}$. The notation $\delta\text{\eref{constr0}}$ means that we impose the constraint \eref{constr0} upon integrating over the gauge fugacities and summing over the gauge fluxes.  A set of conditions that needs to be satisfied in order to make the index well-defined is the Dirac quantisation conditions which require that the magnetic fluxes in the arguments of each factor $\CI_\chi^{\frac{1}{2}}$ must be integers:
\bes{ \label{dirac0}
m_1- 3 m_3 \in \BZ~, \quad 2m_1- 3 m_2 \in \BZ~, \quad m_2- 2 m_3 \in \BZ~.
}
In \eref{indextriangle}, we sum the gauge magnetic fluxes $m_i$ over all possible values (including the fractional ones) that are allowed by \eref{dirac0}.  

The conditions \eref{constr0} allow us to write, for example, 
\bes{ \label{constr1}
z_2 = z_1^{-\frac{\fc_1}{\fc_2}} z_3^{-\frac{\fc_3}{\fc_2}}~, \quad m_2 = -\frac{\fc_1}{\fc_2} m_1  -\frac{\fc_3}{\fc_2} m_3 ~, \quad \fc_2 \neq 0~.
}
Imposing these in the above index amounts to ungauge the overall $\U(1)$ from the node associated with $m_2$. Using \eref{constr1}, we see that the Dirac quantisation conditions become 
\bes{ \label{dirac2}
m_1 - 3 m_3 &\in \BZ~, \\ 
\frac{3 \fc_1 + 2 \fc_2}{\fc_2} m_1 + \frac{3 \fc_3}{\fc_2} m_3 &\in \BZ~, \\
\frac{\fc_1}{\fc_2} m_1 + \frac{2 \fc_2 + \fc_3}{\fc_2} m_3 &\in \BZ~.
}
Due to the condition \eref{constr1}, we are left with two independent topological fugacities $v_1$ and $v_3$ that appear in the index as $v_1^{m_1} v_3^{m_3}$, where\footnote{The restriction of three topological fugacities to two was also achieved in \eref{constraintonzi} in a different way. In this case, we impose the condition \eref{constr0} on the magnetic fluxes and then seek two independent combinations of topological fugacities which turn out to be $v_1$ and $v_3$.}
\bes{ \label{v1v2topfug}
v_1 = w_1 w_2^{- \frac{\fc_1}{\fc_2}}~, \quad   v_3 = w_3 w_2^{- \frac{\fc_3}{\fc_2}}~.
}

Suppose that the allowed magnetic fluxes in \eref{dirac2} include those with non-integral values and, upon writing them in the lowest terms, the largest denominator is $r$. This indicates the presence of a $\BZ_r$ zero-form magnetic symmetry that acts non-trivially on the monopole operators with fractional magnetic fluxes. In particular, whenever $m_1$ or $m_3$ is {\it not} an integer, if there is an integer linear combination of $m_1$ and $m_3$ such that
\bes{ \label{powerg}
r (x_{1} m_1 + x_3 m_3) \neq 0 ~(\mod~ r)~, \qquad \text{for some $x_1, x_3 \in \BZ$}~,
}
then the index can then be refined with respect to a discrete fugacity $g$ associated with $\BZ_r$ such that $g^r=1$ by inserting the factor 
\bes{ \label{powerg1}
g^{r (x_{1} m_1 + x_3 m_3)}
}
inside the summation in \eref{indextriangle}. For convenience, we will refer to this $\BZ_r$ zero-form symmetry as $(\BZ_r^{[0]})_g$. Explicitly, the index \eref{indextriangle} up to order $x^2$ reads
\bes{ \label{explicitindextriangle}
\scalebox{0.89}{$
\begin{split}
\CI_{\text{\eref{Trianglequivggg}}} &=1+ (a^{2} + 2a^{-2}) x + \left(g^{-2} u_1 +g^{-1} u_2 + g u_2^{-1} + g^2 u_1^{-1} \right) a^{-3} x^{\frac{3}{2}} \\
& \quad \, \, \, \, \, +\left[a^{4} -4 + \left(g^{-3} u_1 u_2 + g^{-1} u_1 u_2^{-1} +3+ g u_1^{-1} u_2 + g^3 u_1^{-1} u_2^{-1} \right) a^{-4}\right] x^2 +\ldots~,
\end{split}
$}
}
where $u_1$ and $u_2$ are some functions of the topological fugacities $v_1$ and $v_3$, and at least one of which, say $u_2$, takes the form $u_2 = (v_1^{y_1} v_3^{y_3})^\frac{1}{r}$, with $y_1$ and $y_3$ integers coprime to $r$.  Setting $g=1$ and $u_i=1$, the Coulomb branch limit of \eref{explicitindextriangle} is in agreement with \eref{CBHStriangle}. We will discuss the Higgs branch below. 

This refinement allows us to gauge $(\BZ_r^{[0]})_g$ by summing $g$ over all of the $r$-th roots of unity and dividing the result by $r$. After gauging, the resulting theory, which we refer to as \eref{Trianglequivggg}$/(\BZ_r^{[0]})_g$, has a dual $\BZ^{[1]}_r$ one-form symmetry, denoted by $(\BZ^{[1]}_r)_{\hat{g}}$.  The fractional powers of $v_1$ and $v_3$ in $u_2 = (v_1^{y_1} v_3^{y_3})^\frac{1}{r}$ in the index of theory \eref{Trianglequivggg}, in fact, imply the following mixed 't Hooft anomaly in theory \eref{Trianglequivggg}$/(\BZ_r^{[0]})_g$:
\bes{ \label{mixedanomZr}
\exp \left( \frac{2\pi i}{r}\int_{M_4} B^2_{\hat{g}} \cup \left[c_1(\U(1)_{v_1^{y_1} v_3^{y_3}}) \, \, \mod \,\, r\right] \right)~,
}
where $B^{(2)}_{\hat{g}}$ is the two-form background field for $(\BZ^{[1]}_2)_{\hat{g}}$ and $c_1(\U(1)_{v_1^{y_1} v_3^{y_3}})$ is the first Chern-class associated with the linear combination $y_1 \U(1)_{v_1} + y_3 \U(1)_{v_3}$ of the $\U(1)_{v_1}$ and $\U(1)_{v_3}$ topological symmetries associated with the fugacities $v_1$ and $v_3$ respectively. Here $M_4$ is the 4d bulk such that $\partial M_4 = M_3$, with $M_3$ being the 3d spacetime in which the theory in question lives. Indeed, \eref{mixedanomZr} manifests itself in the following way: upon gauging $(\BZ^{[1]}_r)_{\hat{g}}$, we go back to theory \eref{Trianglequivggg} whose index contains the terms involving $(v_1^{y_1} v_3^{y_3})^{\frac{1}{r}}$, which contains an improper quantisation of the charge of $\U(1)_{v_1^{y_1} v_3^{y_3}}$. We will demonstrate this in the following examples below:
\bi
\item $(\fc_1, \fc_2, \fc_3) = (0,1,0)$ for $\BZ_2$~,
\item $(\fc_1, \fc_2, \fc_3) = (1,0,0)$ for $\BZ_3$~,
\item $(\fc_1, \fc_2, \fc_3) = (0,1,1)$ for another $\BZ_3 \equiv \BZ_{3'}$~,
\item $(\fc_1, \fc_2, \fc_3) = (3,2,0)$ for $\BZ_{13}$~.
\ei

Let us briefly mention that one can take the Higgs and Coulomb branch limits of the index to obtain the Higgs and Coulomb branch Hilbert series \cite{Razamat:2014pta}.  In particular, we define 
\bes{ \label{CBHBlimits}
&h = x^{1/2} a~, ~ c = x^{1/2} a^{-1}~, \\ 
&\text{or equivalently} \quad  x = h c ~,~ a = (h/c)^{1/2}~,
}
and substitute them in the index.  In the Higgs branch limit we send $c \rightarrow 0$ and keep $h$ fixed, whereas in the Coulomb branch limit we send $h \rightarrow 0$ and keep $c$ fixed. Note that, given a series expansion of the index in terms of $x$, the Higgs ({\it resp.} Coulomb) branch limit can be obtained by reading off the coefficients of the terms $a^{p} x^{\frac{p}{2}}$ ({\it resp.} $a^{-p} x^{\frac{p}{2}}$), with $p \in \BZ_{\geq 0}$, and multiplying each of them to $h^{p}$ ({\it resp.} $c^{p}$). Conventionally, the corresponding Hilbert series can be obtained by simply setting $h$ or $c$ in each of the aforementioned limits to $t$.

The Coulomb branch limits of the indices of \eref{Trianglequivggg}$/(\BZ_r^{[0]})_g$ in these examples, upon setting $u_i=1$, are in agreement with the Coulomb branch Hilbert series \eref{CBHSZ2}, \eref{CBHSZ3quot}, \eref{CBHSZ3quotanother} and \eref{CBHSZ13}, as expected.

On the other hand, the Higgs branch limit of the index \eref{explicitindextriangle} gives $\PE [t^2 + 2t^6 -t^{12}]$.  This indicates that the Higgs branch of \eref{Trianglequivggg} is isomorphic to $\BC^2/\BZ_6$. The generators are as follows:
\bes{ \label{gentriangle}
G_0 &= Q_{(0,1,-2)} \tQ_{(0,1,-2)} = Q_{(-1,0,3)} \tQ_{(-1,0,3)} = Q_{(2,-3,0)} \tQ_{(2,-3,0)}~,\\
G_+ &= Q_{(0,1,-2)}^3 Q_{(-1,0,3)}^2 Q_{(2,-3,0)}~, \\
G_- &= \tQ_{(0,1,-2)}^3 \tQ_{(-1,0,3)}^2 \tQ_{(2,-3,0)}~,
}
where we denote by $Q_{(q_1, q_2,q_3)}$ the chiral field that carries charges $(q_1,q_2,q_3)$ under the gauge groups labelled by $i=1,2,3$ in \eref{Trianglequivggg}, and by $\tQ_{(q_1,q_2,q_3)}$ the chiral field in the same hypermultiplet with the opposite gauge charge. These generators satisfy the defining relation $G_+ G_- = G_0^6$ of $\BC^2/\BZ_6$, as expected.  This moduli space is unaffected upon gauging the $(\BZ_r^{[0]})_g$ zero-form symmetry.  This can also be seen from the Higgs branch limit of the index \eref{explicitindextriangle}: the coefficients of the terms $a^{p} x^{\frac{p}{2}}$, with $p \in \BZ_{\geq 0}$, never contain $g$.

\subsubsection*{Example 1. $(\fc_1, \fc_2, \fc_3) = (0,1,0)$ for $\BZ_2$}
The solutions to \eref{dirac2} form the two-dimensional lattices $\mathbf{\Lambda}_p = \left(\BZ+\frac{p}{2} \right)^2$, with $p=0,1$. In \eref{powerg}, we can choose, for example, $x_1 =1$ and $x_3=0$, so that, when $p=1$, we have a non-trivial power of $g$ in \eref{powerg1}. The index \eref{indextriangle} therefore can be rewritten as
\bes{
\CI_{\text{\eref{Trianglequivggg}}} = \sum_{p=0}^1 \,\, \sum_{(m_1, m_3) \in \Lambda_p} \,\,g^p \,\, v_1^{m_1} v_3^{m_3} \oint \frac{dz_1}{2\pi i z_1} \oint \frac{dz_3}{2\pi i z_3} \left[ \CI_{\text{matter}} \right]_{\text{\eref{constr1}}}~,
}
where we denote by $\left[ \CI_{\text{matter}} \right]_{\text{\eref{constr1}}}$ the matter contribution in \eref{indextriangle} subject to the constraint \eref{constr1}.
It can be easily seen that the fugacity $g$ associated with the $(\BZ^{[0]}_2)_g$ zero-form symmetry always appears with the topological fugacities as $g (v_1 v_3)^\frac{1}{2} v_1^{r_1} v_3^{r_3}$ for some $r_1, r_3 \in \BZ$.  The explicit form of the index up to order $x^2$ is given in \eref{explicitindextriangle} with 
\bes{u_1 =  v_1~, \qquad u_2 = (v_1 v_3)^{\frac{1}{2}}~.}
Gauging $(\BZ^{[0]}_2)_g$ leads to theory \eref{Trianglequivggg}/$(\BZ^{[0]}_2)_{g}$, with the mixed anomaly given by \eref{mixedanomZr} with $r=2$ and $y_1=y_3=1$.  Note that theory \eref{Trianglequivggg}/$(\BZ^{[0]}_2)_g$ corresponds to opening up the quiver \eref{Trianglequivggg} in the following way:
\bes{
    \vcenter{\hbox{\begin{tikzpicture}
        \node[gauge,label=below:{$1$},label=left:{\red{\scriptsize $m_3$}}] (1) at (-2,0) {};
        \node[gauge,label=below:{$1$},label=right:{\red{\scriptsize $m_1$}}] (2) at (2,0) {};
        \node[flavour,label=above:{$1$},label=right:{}] (f1) at (-2,3) {};
        \node[flavour,label=above:{$1$},label=right:{}] (f2) at (2,3) {};
        \edge{1}{2}{3}{1}{1}
        \edger{2}{f2}{2}{3}{1}
        \edgel{1}{f1}{2}{1}{1}
    \end{tikzpicture}}}
}
where the gauge magnetic fluxes are such that $(m_1, m_3) \in \BZ^2$. 

\subsubsection*{Example 2. $(\fc_1, \fc_2, \fc_3) = (1,0,0)$ for $\BZ_3$}
In this case, since $\fc_2 = 0$, in order to solve the conditions \eref{constr0} in a valid way, we have to modify \eref{constr1} by swapping $z_1 \leftrightarrow z_2$, $\fc_1 \leftrightarrow \fc_2$ and $m_1 \leftrightarrow m_2$. It then follows that we have two independent topological fugacities $v_2$ and $v_3$ appearing in the index as $v_2^{m_2} v_3^{m_3}$, which are defined by swapping the subscript $1$ with the subscript $2$, and viceversa, in \eref{v1v2topfug}. Moreover, the Dirac quantisation conditions can be expressed as
\bes{ \label{diracZ3}
m_2 - 2 m_3 &\in \BZ~, \\ 
\frac{3 \fc_1 + 2 \fc_2}{\fc_1} m_2 + \frac{2 \fc_3}{\fc_1} m_3 &\in \BZ~, \\
\frac{\fc_2}{\fc_1} m_2 + \frac{3 \fc_1 + \fc_3}{\fc_1} m_3 &\in \BZ~.
}
In particular, for $(\fc_1, \fc_2, \fc_3) = (1,0,0)$, these equations become
\bes{
m_2-2m_3 \in \BZ~, \qquad 3m_2 \in\BZ~, \qquad 3 m_3 \in \BZ~.
}
The solutions form the two-dimensional lattices $\mathbf{\Lambda}_p =  (\BZ + \frac{p}{3}) \times (\BZ - \frac{p}{3})$, with $p = 0,1,2$. The index can be refined with a fugacity $g$ by inserting the factor \eref{powerg1}, with the substitution $x_1 \rightarrow x_2$ and $m_1 \rightarrow m_2$. In particular, we can choose $x_2 = 1$ and $x_3 = 0$ such that, when $p = 1, 2$, a non-trivial power of $g$ appears in the index \eref{indextriangle}. The explicit form of the index up to order $x^2$ is given in \eref{explicitindextriangle} with 
\bes{
u_1 = (v_2 v_3^{-1})^{\frac{1}{3}}~, \qquad u_2 = (v_2^2 v_3)^{\frac{1}{3}}~.}
Note that the index \eref{explicitindextriangle} can also be rewritten in terms of $u_2$ and {\it integral} powers of $v_2$, since $u_1 = v_2 u_2^{-1}$. Gauging $(\BZ_3^{[0]})_g$ leads to theory \eref{Trianglequivggg}/$(\BZ^{[0]}_3)_g$ with the mixed anomaly \eref{mixedanomZr} with $r=3$, $y_2=2$ and $y_3=1$.

We remark that \eref{Trianglequivggg}/$(\BZ^{[0]}_3)_g$ is also be described by the following quiver, which amounts to opening up the quiver \eref{Trianglequivggg} at node labelled by $m_1$:
\bes{
    \vcenter{\hbox{\begin{tikzpicture}
        \node[gauge,label=below:{$1$},label=left:{\red{\scriptsize $m_3$}}] (1) at (-2,0) {};
        \node[gauge,label=below:{$1$},label=right:{\red{\scriptsize $m_2$}}] (2) at (2,0) {};
        \node[flavour,label=above:{$1$},label=right:{}] (f1) at (-2,3) {};
        \node[flavour,label=above:{$1$},label=right:{}] (f2) at (2,3) {};
        \edge{1}{2}{2}{1}{1}
        \edger{2}{f2}{3}{2}{1}
        \edgel{1}{f1}{3}{1}{1}
    \end{tikzpicture}}}
}
where the gauge magnetic fluxes are such that $(m_2, m_3) \in \BZ^2$. 

\subsubsection*{Example 3. $(\fc_1, \fc_2, \fc_3) = (0,1,1)$ for $\BZ_{3'}$} Let us also examine another $\BZ_3$ quotient, which we previously denoted by $\BZ_{3'}$. In this case, the two-dimensional lattices which arise from solving the Dirac quantisation conditions \eref{dirac2} are given by $\mathbf{\Lambda}_p =\BZ \times \left(\BZ+\frac{p}{3} \right)$, with $p=0,1,2$. Therefore, we can choose $x_1 = 0$ and $x_3 = 1$ in \eref{powerg} such that, when $p \neq 0$, the index is refined with a non-trivial power of $g$, given by \eref{powerg1}. The explicit expression of the index up to order $x^2$ is given by \eref{explicitindextriangle}, with
\bes{
u_1 = g^2 v_1~, \qquad u_2 = v_3^{-\frac{1}{3}}~.
}
It follows that, upon gauging $(\BZ^{[0]}_{3'})_g$, we reach theory \eref{Trianglequivggg}/$(\BZ^{[0]}_{3'})_g$ with the mixed anomaly given by \eref{mixedanomZr} with $r=3$, $y_1 = 0$ and $y_3=1$.

\subsubsection*{Example 4. $(\fc_1, \fc_2, \fc_3) = (3,2,0)$ for $\BZ_{13}$}
Finally, let us consider the $\BZ_{13}$ discrete symmetry. The solutions to the conditions \eref{dirac2} form the two-dimensional lattices $\mathbf{\Lambda}_p =\left(\BZ+\frac{p}{13} \right) \times \left(\BZ-\frac{4 p}{13} \right)$, with $p=0,\ldots,12$. We can choose $x_1 = 1$ and $x_3 = 2$ in \eref{powerg} in such a way that, when $p \neq 0$, a non-trivial power of $g$, given by \eref{powerg1}, appears in the index. Up to order $x^2$, the index is given explicitly by \eref{explicitindextriangle} with
\bes{
u_1 = (v_1^4 v_3^{-3})^{\frac{1}{13}}~, \qquad u_2 = (v_1^2 v_3^5)^{\frac{1}{13}}~.
}
Note that the index \eref{explicitindextriangle} can also be rewritten in terms of $u_2$ and {\it integral} powers of $v_3$, since $u_1  = v_3^{-1} u_2^{2}$. Gauging $(\BZ_{13}^{[0]})_g$ leads to theory \eref{Trianglequivggg}/$(\BZ^{[0]}_{13})_g$ with the mixed anomaly \eref{mixedanomZr} with $r=13$, $y_1=2$ and $y_3=5$.

\subsubsection{Mirror Dual of Theory \eref{Trianglequivggg}}
\label{sec:mirrorTriangle}
The 3d mirror of \eqref{Trianglequivggg} is \cite{deBoer:1996mp}
\bes{ \label{quivmirr1}
    \begin{tikzpicture}
        \node[gauge,label=below:{$1$}] (1) at (0,0) {};
        \node[gauge,label=above:{$1$}] (2) at (0,3) {};
        \node (11) at (-1,1) {$1$};
        \node (12) at (-1,2) {$1$};
        \node (21) at (0,1) {$2$};
        \node (22) at (0,2) {$2$};
        \node (31) at (1,1) {$3$};
        \node (32) at (1,2) {$3$};
        \draw[double] (1)--(11)--(12)--(2);
        \draw[double] (1)--(21)--(22)--(2);
        \draw[double] (1)--(31)--(32)--(2);
        \node at (3,1.5) {$/\U(1)$};
    \end{tikzpicture}
}
where we can ungauge one of the two gauge nodes to obtain the framed quiver
\bes{
\label{mirr123}
    \begin{tikzpicture}
        \node[gauge,label=below:{$1$}] (1) at (0,0) {};
        \node[flavour,label=left:{$1$}] (f1) at (-2,0) {};
        \node[flavour,label=above:{$1$}] (f2) at (0,2) {};
        \node[flavour,label=right:{$1$}] (f3) at (2,0) {};
        \node (11) at (-0.66,0) {$1$};
        \node (21) at (0,0.66) {$2$};
        \node (31) at (0.66,0) {$3$};
        \draw[double] (1)--(11)--(f1);
        \draw[double] (1)--(21)--(f2);
        \draw[double] (1)--(31)--(f3);
    \end{tikzpicture}
}
which represents a U$(1)$ gauge theory with one hypermultiplet of charge one, one hypermultiplet of charge two and one hypermultiplet of charge three. In $\mathcal{N}=2$ language, let us denote by $Q_i$ and $\tilde{Q}_i$ the chiral multiplets with gauge charges $\pm i$ in each hypermultiplet of gauge charge $i$, and let us call the chiral field in the vector multiplet $\Phi$. We can draw the $\mathcal{N}=2$ quiver as follows:
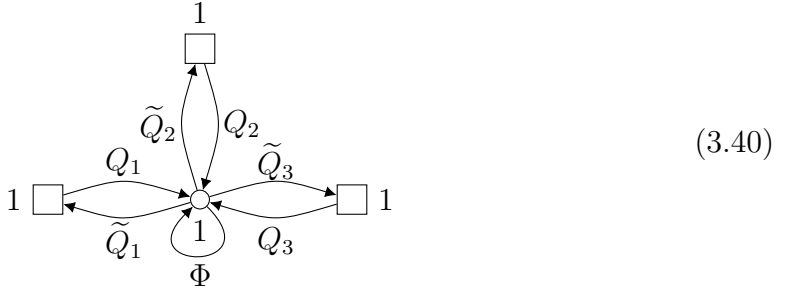
\bes{\label{N2quivermirr}
    \begin{tikzpicture}[baseline]
        \node[gauge,label=below:{$1$}] (1) at (0,0) {};
        \node[flavour,label=left:{$1$}] (f1) at (-2,0) {};
        \node[flavour,label=above:{$1$}] (f2) at (0,2) {};
        \node[flavour,label=right:{$1$}] (f3) at (2,0) {};
        \draw[->] (1) .. controls (-1,-0.3) .. (f1);
        \draw[->] (f1) .. controls (-1,0.3) .. (1);
        \node at (-1,0.5) {$Q_1$};
        \node at (-1,-0.55) {$\tilde{Q}_1$};
        \draw[<-] (1) .. controls (0.3,1) .. (f2);
        \draw[<-] (f2) .. controls (-0.3,1) .. (1);
        \node at (0.55,1) {$Q_2$};
        \node at (-0.55,1) {$\tilde{Q}_2$};
        \draw[<-] (1) .. controls (1,-0.3) .. (f3);
        \draw[<-] (f3) .. controls (1,0.3) .. (1);
        \node at (1,-0.55) {$Q_3$};
        \node at (1,0.5) {$\tilde{Q}_3$};
        \draw[<-] (1)to[out=225,in=315,loop,looseness=17](1);
        \node at (0,-1) {$\Phi$};
    \end{tikzpicture}
}
The superpotential reads
\begin{equation}
    W=\tilde{Q}_1\Phi Q_1+\tilde{Q}_2\Phi Q_2+\tilde{Q}_3\Phi Q_3\;,
\end{equation}
from which we obtain the $F$-term
\begin{equation}
    F=\frac{\partial W}{\partial \Phi}=Q_1\tilde{Q}_1+Q_2\tilde{Q}_2+Q_3\tilde{Q}_3\;.
\end{equation}

\paragraph{The Higgs branch of \eref{quivmirr1}.}
Let us list the generators of the Higgs branch below. At degree 2, we have
\begin{equation} \label{gen2}
    \begin{split}
A_1=\tilde{Q}_1Q_1~, \qquad 
A_2=\tilde{Q}_2Q_2~, \qquad
A_3=\tilde{Q}_3Q_3~,
    \end{split}
\end{equation}
subject to the following relation coming from the $F$-term 
\bes{
A_1+ A_2 +A_3 =0~.
}
At degree 3, we have
\bes{
\begin{array}{ll}
B_1 =\tilde{Q}_1\tilde{Q}_1Q_2~, & \qquad B_2 =Q_1Q_1\tilde{Q}_2~,\\
B_3 =\tilde{Q}_3Q_2Q_1~, & \qquad
B_4=Q_3\tilde{Q}_2\tilde{Q}_1~.
\end{array}
}
At degree 4, we have
\bes{
\begin{array}{ll}
C_1 =\tilde{Q}_1\tilde{Q}_1\tilde{Q}_1Q_3~, &\qquad 
C_2=Q_1Q_1Q_1\tilde{Q}_3~,\\
C_3=\tilde{Q}_1\tilde{Q}_3Q_2Q_2~, &\qquad
C_4=Q_1Q_3\tilde{Q}_2\tilde{Q}_2~.
\end{array}
}
At degree 5, we have
\begin{equation} \label{gen5}
    \begin{split}
D_1=\tilde{Q}_2\tilde{Q}_2\tilde{Q}_2Q_3Q_3~, \qquad
        D_2=Q_2Q_2Q_2\tilde{Q}_3\tilde{Q}_3~.
    \end{split}
\end{equation}
The relations are as follows. At degree 6, we have
\bes{
\begin{array}{ll}
B_2B_4 - A_1C_4=0~, & \qquad
B_1B_4 - A_2C_1=0~,\\
B_2B_3 - A_2C_2=0~, & \qquad
B_1B_3 - A_1C_3=0~,\\
A_1A_2A_3 - B_3B_4=0~, & \qquad
A_1^2A_2 - B_1B_2=0~.
\end{array}        
}
At degree 7, we have
\bes{
\begin{array}{ll}
        B_4C_4 - A_1D_1=0~, & \qquad
        B_3C_3 - A_1D_2=0~,\\
        A_1A_2B_4 - B_1C_4=0~, & \qquad
        A_1^2B_4 - B_2C_1=0~,\\
        A_1A_2B_3 - B_2C_3=0~, & \qquad
        A_1^2B_3 - B_1C_2=0~,\\
        A_2A_3B_2 - B_3C_4=0~, & \qquad
        A_1A_3B_2 - B_4C_2=0~,\\
        A_2A_3B_1 - B_4C_3=0~, & \qquad
        A_1A_3B_1 - B_3C_1=0~.\\
\end{array}        
}
At degree 8, we have
\bes{
\begin{array}{ll}
        C_4^2 - B_2D_1=0~, & \qquad
        A_2A_3C_4 - B_3D_1=0~,\\
        C_3^2 - B_1D_2=0~, & \qquad
        A_2A_3C_3 - B_4D_2=0~,\\
        A_2B_4^2 - B_1D_1=0~, & \qquad
        A_1B_4^2 - C_1C_4=0~,\\
        A_2B_3B_4 - C_3C_4=0~, & \qquad
        A_2B_3^2 - B_2D_2=0~,\\
        A_1B_3^2 - C_2C_3=0~, & \qquad
        A_3B_2^2 - C_2C_4=0~,\\
        A_3B_1^2 - C_1C_3=0~, & \qquad
        A_1^3A_3 - C_1C_2=0~.
\end{array}        
}
At degree 9, we have
\bes{
\begin{array}{ll}
        A_3B_2C_4 - C_2D_1=0~, & \qquad
        A_3B_1C_3 - C_1D_2=0~,\\
        B_4^3 - C_1D_1=0~, & \qquad
        A_2^2A_3B_4 - C_3D_1=0~,\\
        B_3^3 - C_2D_2=0~, & \qquad
        A_2^2A_3B_3 - C_4D_2=0~.\\
\end{array}
}
Finally, at degree 10, we have
\bes{
A_2^3A_3^2 - D_1D_2&=0~.
}
\paragraph{Index of \eref{quivmirr1}.}
The index of this theory is given by
\bes{
\CI_{\text{\eref{mirr123}}} = &\sum_{m \in \BZ} w^m
\oint \frac{d z}{ 2\pi i z}  \CI_\chi^1(a^{-2}; 0; x) \times \Bigg[ \prod_{s= \pm1} \CI_{\chi}^{\frac{1}{2}} \left((g z)^s ;  s m ; x\right) \\
& \qquad \qquad \quad \, \, \, \times \CI_{\chi}^{\frac{1}{2}} \left(( z^{2} u_1)^s;  2s m ; x\right) \times \CI_{\chi}^{\frac{1}{2}} \left( (z^{3} u_1 u_2)^s ;  3 s m ; x\right)\Bigg]~,
}
where $w$ is the topological fugacity and $u_{1,2}$ are the fugacities for the continuous $\U(1)^2$ flavour symmetry. In the notation of \eref{N2quivermirr}, we assign $u_1^{\pm 1}$ to $Q_2$ and $\tQ_2$ and $(u_1 u_2)^{\pm 1}$ to $Q_3$ and $\tQ_3$.  As before we turn off the background magnetic fluxes for the global symmetries. In the above, we turn on the fugacity $g$ (with $g^r=1$) associated with $(\BZ_r^{[0]})_g$ symmetry. This turns out to act on $Q_1$ and $\tQ_1$ as $g$ and $g^{-1}$ respectively.  It is interesting to point out that, in theory \eref{Trianglequivggg}, the $(\BZ_r^{[0]})_g$ zero-form symmetry acts on the monopole operators via the gauge magnetic fluxes, whereas in the mirror theory \eref{mirr123} it acts on the chiral multiplets $Q_1$ and $\tQ_1$. This is consistent with mirror symmetry, in which the Higgs and Coulomb branches of the dual theories get exchanged.

To the first few orders in $x$, the index reads
\bes{ \label{indmirr123}
\scalebox{0.91}{$
\begin{split}
\CI_{\text{\eref{mirr123}}} &=1+ (a^{-2} + 2a^2) x + \left(g^{-2} u_1 +g^{-1} u_2 + g u_2^{-1} + g^2 u_1^{-1} \right) a^3 x^{\frac{3}{2}} \\
& \quad \, \, \, \, \, +\left[a^{-4} -4 + \left(g^{-3} u_1 u_2 + g^{-1} u_1 u_2^{-1} +3+ g u_1^{-1} u_2 + g^3 u_1^{-1} u_2^{-1} \right) a^4\right] x^2 +\ldots\\
& = \text{\eref{explicitindextriangle}} ~\text{with $a \leftrightarrow a^{-1}$}~.
\end{split}
$}
}
The fact that the indices of \eref{mirr123} and \eref{Trianglequivggg} can be 
equated upon exchanging $a$ and $a^{-1}$ supports the claim that these theories are indeed mirror dual. Gauging $(\BZ_r^{[0]})_g$ can be performed as before, namely by summing $g$ over the $r$-th roots of unity and dividing by $r$. Since $(\BZ_r^{[0]})_g$ acts on $Q_1$ and $\tQ_1$ with the action $g$ and $g^{-1}$ (with $g^r=1$) respectively, such discrete gauging removes the Higgs branch generators \eref{gen2}--\eref{gen5} that carry non-trivial powers of $g$, \ie those that are not neutral under $(\BZ_r^{[0]})_g$. 

\paragraph{The Coulomb branch of \eref{quivmirr1}.} Let us discuss the Coulomb branch of theory \eref{quivmirr1}. Indeed, its Hilbert series is given by $\sum_{p\geq 0} c_p t^{2p}$, where $c_p$ is given by the coefficient of the term $a^{-p} x^{\frac{p}{2}}$ in index \eref{indmirr123} (see \cite{Razamat:2014pta}). The result is $\PE[t^2+2t^6 - t^{12}]$, which is the Hilbert series of $\BC^2/\BZ_6$, in agreement with \eref{gentriangle}. The term $t^2$ in the $\PE$ corresponds to the generator $\Phi$, where $\Phi$ is the adjoint field residing in the vector multiplet of the $\U(1)$ gauge node in \eref{mirr123}. The term $2t^6$ in the $\PE$ corresponds to the elementary monopole operators $V_\pm$ with fluxes $\pm 1$ in the $\U(1)$ gauge node in \eref{mirr123}. They satisfy the relation $V_+ V_- = \Phi^6$, which is the defining relation of $\BC^2/\BZ_6$. Furthermore, we see that the Coulomb branch does not get affected by gauging $(\BZ_r^{[0]})_g$, since $g$ appears only in the terms with positive power of $a$. This is in perfect agreement with the statement below \eref{gentriangle} upon applying mirror symmetry.

\subsection{Non-Abelian Example} \label{sec:nonAbelianexample}
Let us analyse the following non-simply laced quiver
\bes{ \label{NSLSU2pq22}
    \vcenter{\hbox{\begin{tikzpicture}
        \node[gauge,label=below:{$1$},label=above:{\red{\scriptsize $m_{4}$}}] (1) at (-2,0) {};
        \node [gauge,label=below:{$\SU(2)$},label={[label distance=-1.25cm]90:{\red{\scriptsize $m_{1,1} + m_{1,2} = 0$}}}] (2) at (0,0) {};
        \node[gauge,label=below:{$1$},label=above:{\red{\scriptsize $m_{2}$}}] (3) at (4,0) {};
        \node[gauge,label=right:{$1$},label=above:{\red{\scriptsize $m_{3}$}}] (4) at (0,2) {};
        \edge{2}{3}{2}{2}{1}
        \draw (1)--(2);
        \draw (4)--(2);
    \end{tikzpicture}}}
}
where we denote by $m_{1,1}$ and $m_{1,2}$, with $m_{1,1} + m_{1,2} = 0$, the magnetic fluxes associated with the $\SU(2)$ gauge node and by $m_{i} \in \BZ$, with $i = 2, 3, 4$, the magnetic fluxes associated with the three $\U(1)$ gauge nodes. Given
\bes{
 G=\tilde{G}/\U(1)~, \quad \text{with} \quad \tilde{G} = \U(2) \times \U(1)^3~,
}
the Coulomb branch Hilbert series can be obtained via the monopole formula \eref{CBHSref}, which in this case reads
\bes{ \label{CBHSrefSU2pq22}
\mathrm{HS}\left[\text{CB of \eref{NSLSU2pq22}}\right](t, \vec{w})=& \left(1-t^2\right) \sum_{\vec{m}\in\frac{\Gamma_{G}^{mw}}{W_G}} P_{\mathrm{U}(2)}(\vec{m}_1,t^2) \\ & \times \left[\prod_{i=2}^4 P_{\mathrm{U}(1)}(m_i,t^2) w_i^{m_{i}}\right] t^{2\Delta(\vec{m})} \delta_{m_{1,1} + m_{1,2},0} ~,
}
where $w_i$, with $i = 2, 3, 4$, are the topological fugacities associated with the three $\U(1)$ nodes.
The adjacency matrix, rank vector and shortness vector associated with \eref{NSLSU2pq22} are given respectively by
\begin{equation}
    A=\begin{pmatrix}
        -2 & 2 & 1 & 1\\
        2 & -2 & 0 & 0\\
        1 & 0 & -2 & 0\\
        1 & 0 & 0 & -2
    \end{pmatrix}\,,\;v=\begin{pmatrix}
        2\\
        1\\
        1\\
        1
    \end{pmatrix}\,,\;s=\begin{pmatrix}
        1\\
        1\\
        1\\
        1
    \end{pmatrix}~,
\end{equation}
from which, using \eref{Deltaofm}, the conformal dimension appearing in \eref{CBHSrefSU2pq22} reads
\bes{ \label{DeltaofmSU2pq22}
    2\Delta(\vec{m})=& - 2 \left|m_{1,1}-m_{1,2}\right| + 2 \left( \left|m_{1,1}-m_{2}\right| + \left|m_{1,2}-m_{2}\right| \right) \\ & \, \,+ \left( \left|m_{1,1}-m_{3}\right| + \left|m_{1,2}-m_{3}\right| \right) + \left( \left|m_{1,1}-m_{4}\right| + \left|m_{1,2}-m_{4}\right| \right)~,
}
with the constraint $m_{1,1} + m_{1,2} = 0$ coming from the delta function in \eref{CBHSrefSU2pq22}. Explicitly, if we set the topological fugacities to unity, we obtain the following expression for the Coulomb branch Hilbert series:
\bes{ \label{CBHSSU2pq22s}
\begin{split}
\mathrm{HS}\left[\text{CB of \eref{NSLSU2pq22}}\right](t, w_i=1) &= 1 + 7 t^2 + 56 t^4 + 235 t^6 + 843 t^8 + \ldots \\ &= \PE\left[7 t^2 + 28 t^4 - 45 t^6 - 242  t^8 + \ldots\right]~,
\end{split}
}
which tells us that the Coulomb branch symmetry of the theory is $\su(2)^2 \oplus \u(1)$.

Let us point out that quiver \eref{NSLSU2pq22} possesses a $(\BZ_2^{[1]})_{\hat{x}} \times (\BZ_2^{[1]})_{\hat{y}}$ one-form symmetry, where the $(\BZ_2^{[1]})_{\hat{x}}$ factor is associated with the centre of the $\SU(2)$ gauge group and $(\BZ_2^{[1]})_{\hat{y}}$ arises from the charge two $\U(1)$ matter field.\footnote{As explained in \cite[Section 2]{Mekareeya:2022spm} (see also \cite[Appendices B and C]{Nawata:2023rdx}), 3d $\CN = 4$ SQED with two hypermultiplets of charge $q$ possesses a $\BZ_q$ one-form symmetry, whose gauging leads to SQED with two hypermultiplets of charge one. As a direct consequence of this statement, gauging $(\BZ_2^{[1]})_{\hat{y}}$ in theory \eref{NSLSU2pq22} leads to \eref{NSLSU2pq21}.} The two $\BZ_2$ factors of the one-form symmetry can be gauged sequentially, giving rise to an interesting pattern of mixed 't Hooft anomalies and global symmetry extensions.
\subsubsection*{Gauging $(\BZ_2^{[1]})_{\hat{y}}$}
Let us start by gauging the $(\BZ_2^{[1]})_{\hat{y}}$ one-form symmetry of quiver \eref{NSLSU2pq22}, namely the one associated with the charge two $\U(1)$ matter field. The resulting quiver is
\bes{ \label{NSLSU2pq21}
    \vcenter{\hbox{\begin{tikzpicture}
        \node[gauge,label=below:{$1$},label=above:{\red{\scriptsize $m_{4}$}}] (1) at (-2,0) {};
        \node [gauge,label=below:{$\SU(2)$},label={[label distance=-1.25cm]90:{\red{\scriptsize $m_{1,1} + m_{1,2} = 0$}}}] (2) at (0,0) {};
        \node[gauge,label=below:{$1$},label=above:{\red{\scriptsize $m_{2}$}}] (3) at (4,0) {};
        \node[gauge,label=right:{$1$},label=above:{\red{\scriptsize $m_{3}$}}] (4) at (0,2) {};
        \edge{2}{3}{2}{1}{1}
        \draw (1)--(2);
        \draw (4)--(2);
    \end{tikzpicture}}}
}
whose Coulomb branch Hilbert series is given by \eref{CBHSrefSU2pq22}, where now the adjacency matrix, rank vector and shortness vector associated with \eref{NSLSU2pq21} are given respectively by
\begin{equation}
    A=\begin{pmatrix}
        -2 & 2 & 1 & 1\\
        1 & -2 & 0 & 0\\
        1 & 0 & -2 & 0\\
        1 & 0 & 0 & -2
    \end{pmatrix}\,,\;v=\begin{pmatrix}
        2\\
        1\\
        1\\
        1
    \end{pmatrix}\,,\;s=\begin{pmatrix}
        1\\
        2\\
        1\\
        1
    \end{pmatrix}~.
\end{equation}
It follows that the conformal dimension \eref{Deltaofm} entering in the monopole formula \eref{CBHSrefSU2pq22} of quiver \eref{NSLSU2pq21} reads
\bes{ \label{Deltaofmpq21}
    2\Delta(\vec{m}) =& - 2 \left|m_{1,1}-m_{1,2}\right| + \left( \left|2 m_{1,1}-m_{2}\right| + \left|2 m_{1,2}-m_{2}\right| \right) \\ & \, \,+ \left( \left|m_{1,1}-m_{3}\right| + \left|m_{1,2}-m_{3}\right| \right) + \left( \left|m_{1,1}-m_{4}\right| + \left|m_{1,2}-m_{4}\right| \right)~,
}
with the constraint $m_{1,1} + m_{1,2} = 0$. Upon setting $w_i = 1$, the explicit expression of the Coulomb branch Hilbert series reads
\bes{
\begin{split}
\mathrm{HS}\left[\text{CB of \eref{NSLSU2pq21}}\right](t, w_i=1) &= 1 + 9 t^2 + 88 t^4 + 405 t^6 + 1515  t^8 + \ldots \\ &= \PE\left[9 t^2 + 43 t^4 - 147 t^6 - 538  t^8 + \ldots\right]~,
\end{split}
}
from which we observe that the Coulomb branch symmetry is $\su(2)^3$. Alternatively, the same theory can be realised in an equivalent way by starting with quiver \eref{NSLSU2pq22}, with associated monopole formula \eref{CBHSrefSU2pq22} and conformal dimension \eref{DeltaofmSU2pq22}, and explicitly gauge its $(\BZ_2^{[1]})_{\hat{y}}$ one-form symmetry by summing the magnetic flux $m_{2,1}$ over half-integer values, \ie we modify the summation over $m_{2}$ in  \eref{CBHSrefSU2pq22} as follows:
\bes{ \label{gaugeZ21y}
\sum_{m_{2}\in \BZ} \longrightarrow \sum_{p = 0}^1 y^p \sum_{m_{2}\in \BZ + \frac{p}{2}}~,
}
where $y$ is the fugacity associated with the dual $(\BZ_2^{[0]})_{y}$ zero-form symmetry arising from gauging the one-form symmetry, such that $y^2 = 1$. This procedure allows us to refine the Coulomb branch Hilbert series of  \eref{NSLSU2pq21} with the fugacity $y$ as follows:
\bes{ \label{CBHSSU2pq21s}
\scalebox{0.98}{$
\begin{split}
\mathrm{HS}\left[\text{CB of \eref{NSLSU2pq21}}\right](t, y, w_i=1) = 1 &+ \left(7 + 2 y\right) t^2 + \left(56 + 32 y\right) t^4 \\&+ \left(235 + 170 y\right) t^6 + \left(843 + 672 y\right) t^8 + \ldots~.
\end{split}
$}
}
Observe that, if we sum \eref{CBHSSU2pq21s} over $y = \pm 1$ and divide by $2$, we recover \eref{CBHSSU2pq22s}:
\bes{
\mathrm{HS}\left[\text{CB of \eref{NSLSU2pq22}}\right](t, \vec w) = \frac{1}{2} \sum_{y = \pm 1} \mathrm{HS}\left[\text{CB of \eref{NSLSU2pq21}}\right](t, y, \vec w)~.
}
This is consistent with the fact that theory \eref{NSLSU2pq22} can be obtained upon gauging the $(\BZ_2^{[0]})_{y}$ symmetry of quiver \eref{NSLSU2pq21}.
\subsubsection*{Gauging $(\BZ_2^{[1]})_{\hat{x}}$}
Next, we can proceed by gauging the $(\BZ_2^{[1]})_{\hat{x}}$ one-form symmetry of quiver \eref{NSLSU2pq22}, namely the one associated with the centre of the $\SU(2)$ gauge group. Taking into account \eref{SUvmodZveqUcmodU1}, the resulting quiver is
\bes{ \label{NSLU2pq22}
\scalebox{0.8}{$
    \vcenter{\hbox{\begin{tikzpicture}
        \node[gauge,label=below:{$1$},label=above:{\red{\scriptsize $m_{4}$}}] (1) at (-2,0) {};
        \node [gauge,label=below:{$2$},label={[label distance=-1.25cm]90:{\red{\scriptsize $m_{1,1} + m_{1,2} = \{0, 1\}$}}}] (2) at (0,0) {};
        \node[gauge,label=below:{$1$},label=above:{\red{\scriptsize $m_{2}$}}] (3) at (4,0) {};
        \node[gauge,label=right:{$1$},label=above:{\red{\scriptsize $m_{3}$}}] (4) at (0,2) {};
        \edge{2}{3}{2}{2}{1}
        \draw (1)--(2);
        \draw (4)--(2);
    \end{tikzpicture}}}
\quad \text{\footnotesize$/\U(1)$} \quad = \quad
\vcenter{\hbox{\begin{tikzpicture}
        \node[gauge,label=below:{$1$},label=above:{\red{\scriptsize $m_{4}$}}] (1) at (-2,0) {};
        \node [gauge,label=below:{$2$},label={[label distance=-1.25cm]90:{\red{\scriptsize $m_{1,1}, m_{1,2}$}}}] (2) at (0,0) {};
        \node[flavour,label=below:{$1$},label=above:{\red{\scriptsize $m_{2} = 0$}}] (3) at (4,0) {};
        \node[gauge,label=right:{$1$},label=above:{\red{\scriptsize $m_{3}$}}] (4) at (0,2) {};
        \edge{2}{3}{2}{1}{1}
        \draw (1)--(2);
        \draw (4)--(2);
    \end{tikzpicture}}}
$}
}
where the equality between the two quivers is a consequence of the fact that the overall $\U(1)$ factor can be ungauged equivalently either by setting $m_{1,1} + m_{1,2} = \{0, 1\}$ or by setting one of the three $\U(1)$ magnetic fluxes to zero, for instance $m_{2} = 0$.\footnote{Note that, upon setting $m_{2} = 0$, we can trade the $(2,2)$-edge between the $\U(2)$ and $\U(1)$ nodes for a $(2,1)$-edge while leaving the conformal dimension invariant.} For definiteness, let us pick the former choice of ungauging. The Coulomb branch Hilbert series reads
\bes{ \label{CBHSrefU2pq22}
\mathrm{HS}\left[\text{CB of \eref{NSLU2pq22}}\right](t, \vec{w})=& \left(1-t^2\right) \sum_{j = 0, 1} \sum_{\vec{m}\in\frac{\Gamma_{G}^{mw}}{W_G}} P_{\mathrm{U}(2)}(\vec{m}_1,t^2) w_1^{m_{1,1}+m_{1,2}} \\ & \times \left[\prod_{i=2}^4 P_{\mathrm{U}(1)}(m_i,t^2) w_i^{m_{i}}\right] t^{2\Delta(\vec{m})} \delta_{m_{1,1} + m_{1,2},j} ~,
}
where $w_1$ is the topological fugacity associated with the $\U(2)$ gauge group and the conformal dimension is given by \eref{DeltaofmSU2pq22}, with $m_{1,1} + m_{1,2} = \{0, 1\}$. Upon setting the topological fugacities to unity, \eref{CBHSrefU2pq22} yields the Hilbert series of the closure of the next-to-minimal orbit of $\su(4)$, denoted by $\bar{\mathrm{n.min}\,A_3} \cong \bar{\mathrm{n.min}\,D_3}$, see \cite[Section 3]{Hanany:2016gbz}:
\bes{ \label{CBHSU2pq22s}
\mathrm{HS}\left[\text{CB of \eref{NSLU2pq22}}\right](t, w_i=1) &= \frac{\left(1 + t^2\right)^2 \left(1 + 5 t^2 + t^4\right)}{\left(1 - t^2\right)^8}\\&= 1 + 15 t^2 + 104 t^4 + 475 t^6 + 1659  t^8 + \ldots \\&= \PE\left[15 t^2 - 16 t^4 + 35 t^6 - 126  t^8 + \ldots\right]~.
}
Indeed, it was formerly pointed out in \cite{Hanany:2020jzl} that the Coulomb branch Hilbert series of the quiver on the right hand side of \eref{NSLU2pq22} coincides with the Hilbert series of $\bar{\mathrm{n.min}\,A_3}$. Equivalently, we can also obtain the same result by gauging explicitly the $(\BZ_2^{[1]})_{\hat{x}}$ symmetry at the level of the monopole formula. This gauging can be performed explicitly by modifying the summation over the magnetic fluxes in  \eref{CBHSrefSU2pq22} as follows:
\bes{ \label{gaugeZ21x}
\sum_{m_{1,1}\in \BZ_{\ge 0}} \, \sum_{\left(m_{2}, m_{3}, m_{4}\right)\in \BZ^3} \longrightarrow \sum_{q = 0}^1 x^q \sum_{m_{1,1}\in \BZ_{\ge 0} + \frac{q}{2}} \, \sum_{\left(m_{2}, m_{3}, m_{4}\right)\in \left(\BZ + \frac{q}{2}\right)^3}~,
}
where we denote by $x$ the fugacity associated with the $(\BZ_2^{[0]})_{x}$ zero-form symmetry, which arises from gauging the $(\BZ_2^{[1]})_{\hat{x}}$ one-form symmetry, with $x^2 = 1$. We can then refine the Coulomb branch Hilbert series \eref{CBHSU2pq22s} with respect to the fugacity $x$:
\bes{ \label{CBHSU2pq22sx}
\scalebox{0.97}{$
\begin{split}
\mathrm{HS}\left[\text{CB of \eref{NSLU2pq22}}\right](t, x, w_i=1) = 1 &+ \left(7 + 8 x\right) t^2 + \left(56 + 48 x\right) t^4 \\&+ \left(235 + 240 x\right) t^6 + \left(843 + 816 x\right) t^8 + \ldots~.
\end{split}
$}
}
Note that we recover \eref{CBHSSU2pq22s} by summing over $x = \pm 1$ and dividing by $2$:
\bes{
\mathrm{HS}\left[\text{CB of \eref{NSLSU2pq22}}\right](t, \vec w) = \frac{1}{2} \sum_{x = \pm 1} \mathrm{HS}\left[\text{CB of \eref{NSLU2pq22}}\right](t, x, \vec w)~,
}
which coincides with gauging $(\BZ_2^{[0]})_{x}$.
\subsubsection*{Gauging the whole one-form symmetry}
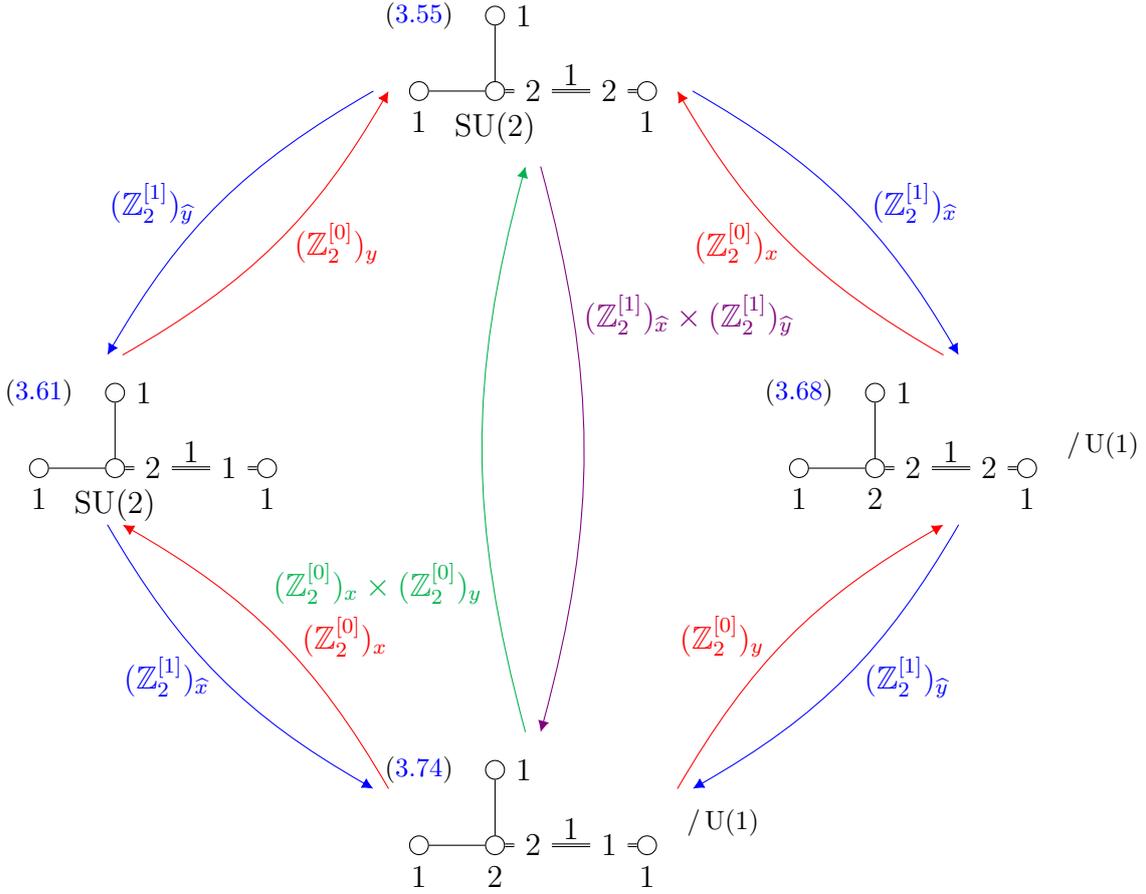
\begin{figure}[!t]
    \centering
\begin{tikzpicture}
\node[gauge,label=below:{$1$}] (1t) at (0,5) {};
        \node[gauge,label=below:{$\SU(2)$}] (2t) at (1,5) {};
        \node[gauge,label=below:{$1$}] (3t) at (3,5) {};
        \node[gauge,label=right:{$1$}] (4t) at (1,6) {};
        \edge{2t}{3t}{2}{2}{1}
        \draw (1t)--(2t);
        \draw (4t)--(2t);
        \node (x1) at (0,6) {\footnotesize \eref{NSLSU2pq22}};
\node[gauge,label=below:{$1$}] (1l) at (-5,0) {};
        \node[gauge,label=below:{$\SU(2)$}] (2l) at (-4,0) {};
        \node[gauge,label=below:{$1$}] (3l) at (-2,0) {};
        \node[gauge,label=right:{$1$}] (4l) at (-4,1) {};
        \edge{2l}{3l}{2}{1}{1}
        \draw (1l)--(2l);
        \draw (4l)--(2l);
        \node (x2) at (-5,1) {\footnotesize \eref{NSLSU2pq21}};
\node[gauge,label=below:{$1$}] (1r) at (5,0) {};
        \node[gauge,label=below:{$2$}] (2r) at (6,0) {};
        \node[gauge,label=below:{$1$}] (3r) at (8,0) {};
        \node[gauge,label=right:{$1$}] (4r) at (6,1) {};
        \node[circle,draw=white] (mod) at (9,0.3) {\text{\footnotesize$/\U(1)$}};
        \edge{2r}{3r}{2}{2}{1}
        \draw (1r)--(2r);
        \draw (4r)--(2r);
        \node (x3) at (5,1) {\footnotesize \eref{NSLU2pq22}};
\node[gauge,label=below:{$1$}] (1d) at (0,-5) {};
        \node[gauge,label=below:{$2$}] (2d) at (1,-5) {};
        \node[gauge,label=below:{$1$}] (3d) at (3,-5) {};
        \node[gauge,label=right:{$1$}] (4d) at (1,-4) {};
        \node[circle,draw=white] (modd) at (4,-4.7) {\text{\footnotesize$/\U(1)$}};
        \edge{2d}{3d}{2}{1}{1}
        \draw (1d)--(2d);
        \draw (4d)--(2d);
        \node (x4) at (0,-4) {\footnotesize \eref{NSLU2pq21}};
\draw[->,blue] (3.6,5) to [bend left=15] node[midway, right=0.2] {\blue $(\BZ_2^{[1]})_{\hat{x}}$} (7.1,1.5);
\draw[->,red] (6.9,1.5) to [bend left=15] node[midway, left] {\red$(\BZ_2^{[0]})_{x}$} (3.4,5);
\draw[->,blue] (-0.6,5) to [bend right=15] node[midway, left=0.2] {\blue $(\BZ_2^{[1]})_{\hat{y}}$} (-4.1,1.5);
\draw[->,red] (-3.9,1.5) to [bend right=15] node[midway, right=0.1] {\red $(\BZ_2^{[0]})_{y}$} (-0.4,5);
\draw[->,blue] (7.1,-0.75) to [bend left=15] node[midway, right=0.1] {\blue $(\BZ_2^{[1]})_{\hat{y}}$} (3.6,-4.25);
\draw[->,red] (3.4,-4.25) to [bend left=15] node[midway, left=0.2] {\red $(\BZ_2^{[0]})_{y}$} (6.9,-0.75);
\draw[->,blue] (-4.1,-0.75) to [bend right=15] node[midway, left] {\blue $(\BZ_2^{[1]})_{\hat{x}}$} (-0.6,-4.25);
\draw[->,red] (-0.4,-4.25) to [bend right=15] node[midway, right=0.2] {\red $(\BZ_2^{[0]})_{x}$} (-3.9,-0.75);
\draw[->,violet] (1.6,4) to [bend left=15] node[near start, right] {\violet $(\BZ_2^{[1]})_{\hat{x}} \times (\BZ_2^{[1]})_{\hat{y}}$} (1.6,-3.5);
\draw[->,new-green] (1.4,-3.5) to [bend left=15] node[near start, left] {\green $(\BZ_2^{[0]})_{x} \times (\BZ_2^{[0]})_{y}$} (1.4,4);
\end{tikzpicture}
    \caption[NSL web]{Sequential gauging of the $(\BZ_2^{[1]})_{\hat{x}} \times (\BZ_2^{[1]})_{\hat{y}}$ one-form symmetry of quiver \eref{NSLSU2pq22} and subgroups thereof. Each downwards arrow with label $\hat{F}$ denotes the gauging of the one-form symmetry $\hat{F}$, whereas each upwards arrow with label $F$ denotes the gauging of the zero-form symmetry $F$.  In each case, the manifest UV topological symmetry $\prod_{i=2}^4 \U(1)_{w_i}$ gets enhanced to the following symmetry: \eref{NSLSU2pq22}: $\u(1)_{w_2} \times \su(2)_{w_3} \times \su(2)_{w_4}$, \eref{NSLSU2pq21}: $\prod_{i = 2}^4 \su(2)_{w_i}$, \eref{NSLU2pq22}: $\su(4)$, and \eref{NSLU2pq21}: $\so(7)$.}
    \label{fig:NSLweb}
\end{figure}

Finally, let us gauge the whole $(\BZ_2^{[1]})_{\hat{x}} \times (\BZ_2^{[1]})_{\hat{y}}$ one-form symmetry of quiver \eref{NSLSU2pq22}. Equivalently, this corresponds to either gauging $(\BZ_2^{[1]})_{\hat{x}}$ in theory \eref{NSLSU2pq21} or to gauging $(\BZ_2^{[1]})_{\hat{y}}$ in theory \eref{NSLU2pq22}. This yields the affine $B_3$ quiver
\bes{ \label{NSLU2pq21}
    \vcenter{\hbox{\begin{tikzpicture}
        \node[gauge,label=below:{$1$},label=above:{\red{\scriptsize $m_{4}$}}] (1) at (-2,0) {};
        \node [gauge,label=below:{$2$},label={[label distance=-1.25cm]90:{\red{\scriptsize $m_{1,1} + m_{1,2} = \{0,1\}$}}}] (2) at (0,0) {};
        \node[gauge,label=below:{$1$},label=above:{\red{\scriptsize $m_{2}$}}] (3) at (4,0) {};
        \node[gauge,label=right:{$1$},label=above:{\red{\scriptsize $m_{3}$}}] (4) at (0,2) {};
        \edge{2}{3}{2}{1}{1}
        \draw (1)--(2);
        \draw (4)--(2);
    \end{tikzpicture}}}
    \quad \text{\footnotesize$/\U(1)$}
}
whose Coulomb branch Hilbert series,\footnote{This is given by \eref{CBHSrefU2pq22} with conformal dimension \eref{Deltaofmpq21}, where $m_{1,1} + m_{1,2} = \{0,1\}$.} once unrefined with respect to the topological fugacities, yields the Hilbert series of the closure of the minimal nilpotent orbit of $\so(7)$, which is denoted by $\bar{\mathrm{min}\,B_3}$, namely \cite[Section 4]{Hanany:2016gbz}
\bes{ \label{CBHSU2pq21s}
\mathrm{HS}\left[\text{CB of \eref{NSLU2pq21}}\right](t, w_i=1) &= \frac{\left(1 + 13 t^2 + 28 t^4 + 13 t^6 + t^8\right)}{\left(1 - t^2\right)^8}\\&= 1 + 21 t^2 + 168 t^4 + 825  t^6 + 3003  t^8 + \ldots \\&= \PE\left[21 t^2 - 63 t^4 + 377 t^6 - 2940  t^8 + \ldots\right]~.
}
This can also be achieved directly from theory \eref{NSLSU2pq22} by gauging the whole $(\BZ_2^{[1]})_{\hat{x}} \times (\BZ_2^{[1]})_{\hat{y}}$ one-form symmetry at the level of the monopole formula by combining \eref{gaugeZ21y} and \eref{gaugeZ21x}. Explicitly, we modify the summation in \eref{CBHSrefSU2pq22} as follows:
\bes{ \label{gaugeZ21xZ21y}
&\sum_{m_{1,1}\in \BZ_{\ge 0}} \, \sum_{\left(m_{2}, m_{3}, m_{4}\right)\in \BZ^3} \\&\longrightarrow \sum_{q = 0}^1 \sum_{p = 0}^1 x^q y^p \sum_{m_{1,1}\in \BZ_{\ge 0} + \frac{q}{2}} \, \sum_{m_{2}\in \BZ + \frac{q}{2} + \frac{p}{2}} \, \, \sum_{\left(m_{3}, m_{4}\right)\in \left(\BZ + \frac{q}{2}\right)^2}~.
}
In this way, we can refine the Coulomb branch Hilbert series \eref{CBHSU2pq21s} with both $\BZ_2$ fugacities $x$ and $y$:
\bes{ \label{CBHSU2pq21xys}
\scalebox{0.96}{$
\begin{split}
\mathrm{HS}\left[\text{CB of \eref{NSLU2pq21}}\right](t, x, y, w_i=1) = 1 &+ \left(7 + 8 x + 2 y + 4 x y\right) t^2 \\&+ \left(56 + 48 x + 32 y + 32 x y\right) t^4 \\&+ \left(235 + 240 x + 170 y + 180 x y\right) t^6 \\&+ \left(843 + 816 x + 672 y + 672 x y\right) t^8 + \ldots~,
\end{split}
$}
}
where we emphasise that the $\so(7)$ symmetry of \eref{NSLU2pq21} is manifest only when $x$ and $y$ are set to unity. Note that we can reproduce \eref{CBHSSU2pq21s}, \eref{CBHSU2pq22sx} and \eref{CBHSSU2pq22s} by gauging $(\BZ_2^{[0]})_{x}$, $(\BZ_2^{[0]})_{y}$ and $(\BZ_2^{[0]})_{x} \times (\BZ_2^{[0]})_{y}$ respectively, namely
\begin{subequations}  
\begin{align}
\begin{split} 
\mathrm{HS}\left[\text{CB of \eref{NSLSU2pq21}}\right](t, y, \vec w) &= \frac{1}{2} \sum_{x = \pm 1} \mathrm{HS}\left[\text{CB of \eref{NSLU2pq21}}\right](t, x, y, \vec w)~,
\end{split} \\
\begin{split} \label{gaugeZ20y}
\mathrm{HS}\left[\text{CB of \eref{NSLU2pq22}}\right](t, x, \vec w) &= \frac{1}{2} \sum_{y = \pm 1} \mathrm{HS}\left[\text{CB of \eref{NSLU2pq21}}\right](t, x, y, \vec w)~,
\end{split} \\
\begin{split} 
\mathrm{HS}\left[\text{CB of \eref{NSLSU2pq22}}\right](t, \vec w) &= \frac{1}{4} \sum_{x = \pm 1} \sum_{y = \pm 1} \mathrm{HS}\left[\text{CB of \eref{NSLU2pq21}}\right](t, x, y, \vec w)~.
\end{split} 
\end{align}
\end{subequations}
Observe that \eref{gaugeZ20y} is consistent with the well-known result of \cite{BK} (see also \cite{Hanany:2020jzl}):
\bes{
\bar{\mathrm{n.min}\,A_3} = \bar{\mathrm{min}\,B_3}/\BZ_2~.
}
The various one-form symmetry gaugings of quiver \eref{NSLSU2pq22} are summarised in Figure \ref{fig:NSLweb}, together with the dual zero-form symmetry gaugings.
\subsubsection*{Mixed 't Hooft anomalies}
The Coulomb branch Hilbert series \eref{CBHSU2pq21xys} turns out to be very interesting if, in addition to $x$ and $y$, we also turn on the topological fugacities $w_i$. Let us focus on the coefficient appearing at order $t^2$ of such Hilbert series:\footnote{For our purpose, it suffices to consider the coefficient of $t^2$ in the Hilbert series. This is due to the fact that the Coulomb branch of the theory in question is $\bar{\mathrm{min}\,B_3}$, which has precisely one set of generators appearing at order $t^2$ and transforming in the adjoint representation of $\so(7)$; see also the last line of \eref{CBHSU2pq21s}. As we will discuss below, upon turning on the fugacities $x$ and $y$, the characters that are manifest in the Hilbert series are those of $\su(2)_{w_i}$, with $i=2,3,4$.}
\bes{ \label{CBHSU2pq21xyzs} 
\scalebox{0.97}{$
\begin{split}
&\mathrm{HS}\left[\text{CB of \eref{NSLU2pq21}}\right](t, x, y, \vec{w})\\& = 1 + \left[3 + w_3 + w_3^{-1} + w_4 + w_4^{-1} + {\blue x \left(  w_2^{\frac{1}{2}} +w_2^{-\frac{1}{2}} \right)\left(  w_3^{\frac{1}{2}} +w_3^{-\frac{1}{2}} \right)\left(  w_4^{\frac{1}{2}} +w_4^{-\frac{1}{2}} \right)} \right. \\& \left. \qquad \quad \quad + {\red y \left(w_2^{\frac{1}{2}} + w_2^{-\frac{1}{2}}\right)} + {\green x y \left(  w_3^{\frac{1}{2}} +w_3^{-\frac{1}{2}} \right)\left(  w_4^{\frac{1}{2}} +w_4^{-\frac{1}{2}} \right)}\right] t^2 + \ldots~,
\end{split}
$}
}
where we highlight the topological fugacities appearing with half-integer powers in {\blue blue}, {\red red} and {\green green}. Note that the $\SO(7)$ flavour symmetry is manifest in the Hilbert series as the $\U(1)_{w_2} \times \SU(2)_{w_3} \times \SU(2)_{w_4}$ subgroup.\footnote{Naively, the character that appears in \eref{CBHSU2pq21xyzs} might seem to be that of $\SU(2)_{w_2} \times \SU(2)_{w_3} \times \SU(2)_{w_4}$, but this does not fit into the branching rule $\mathbf{21} \rightarrow (\mathbf{3},\mathbf{2},\mathbf{2})\oplus (\mathbf{3},\mathbf{1},\mathbf{1})\oplus (\mathbf{1},\mathbf{3},\mathbf{1})\oplus (\mathbf{1},\mathbf{1},\mathbf{3})$ of $\SO(7) \rightarrow \SU(2)^3$. In fact, we see that the first $\SU(2)$ factor in the aforementioned branching rule is broken to $\U(1)_{w_2}$ and we have to further use the branching rule $\mathbf{3} \rightarrow (2)\oplus (0) \oplus (-2)$ of $\SU(2) \rightarrow \U(1)_{w_2}$. We see that, in this convention, the normalisation for $w_2$ is such that $w_2^{\pm \frac{1}{2}}$ corresponds to the $\U(1)_{w_2}$ charge $\pm 2$. \label{footcharge}}  The normalisations of the fugacities are as follows: $w_2^{\pm \frac{1}{2}}$ corresponds to the $\U(1)_{w_2}$ charge $\pm 2$,\footnote{See Footnote \ref{footcharge} for the convention.} and, for $\SU(2)_{w_i}$  (with $i=3,4$), the character of the fundamental representation is $w_i^{\frac{1}{2}}+w_i^{-\frac{1}{2}}$ and that of the adjoint representation is $1+w_i+w_i^{-1}$.

The product of the brackets in {\green green}, namely $\left(  w_3^{\frac{1}{2}} +w_3^{-\frac{1}{2}} \right)\left(  w_4^{\frac{1}{2}} +w_4^{-\frac{1}{2}} \right)$, corresponds to $(\mathbf{2};\mathbf{2})$ representation of $\left[\SU(2)_{w_3} \times \SU(2)_{w_4}\right]/\BZ_2$, where the quotient by $\BZ_2$ indicates the action that there is one independent combination of the centres of $\SU(2)_{w_3}$ and $\SU(2)_{w_4}$ that acts trivially. Specifically, the aforementioned $\BZ_2$ acts by  $\left(w_3^{\frac{1}{2}} \rightarrow -w_3^{\frac{1}{2}},\, w_4^{\frac{1}{2}} \rightarrow -w_4^{\frac{1}{2}} \right)$ simultaneously. On the other hand, the product of the brackets in {\blue blue}, namely $\left(  w_2^{\frac{1}{2}} +w_2^{-\frac{1}{2}} \right)\left(  w_3^{\frac{1}{2}} +w_3^{-\frac{1}{2}} \right)\left(  w_4^{\frac{1}{2}} +w_4^{-\frac{1}{2}} \right)$, corresponds to the $(\mathbf{2}; \mathbf{2})(2) \oplus (\mathbf{2}; \mathbf{2})(-2) $ representation of $\left[\U(1)_{w_2} \times \SU(2)_{w_3} \times \SU(2)_{w_4}\right]/(\BZ_2 \times \BZ_2)$, where the quotient by $\BZ_2 \times \BZ_2$ indicates that there are two independent combinations of the centres of $\SU(2)_{w_{3,4}}$ and $\U(1)_{w_2}$ that act trivially. Specifically, the action of this $\BZ_2\times \BZ_2$ can be taken as $\left(w_2^{\frac{1}{2}} \rightarrow -w_2^{\frac{1}{2}}, \, w_3^{\frac{1}{2}} \rightarrow -w_3^{\frac{1}{2}} \right)$ and $\left(w_3^{\frac{1}{2}} \rightarrow -w_3^{\frac{1}{2}}, \, w_4^{\frac{1}{2}} \rightarrow -w_4^{\frac{1}{2}}\right)$.

The terms in colour in \eref{CBHSU2pq21xyzs} indicate the following mixed 't Hooft anomalies (see for instance \cite{Bhardwaj:2022dyt,Mekareeya:2022spm,Bhardwaj:2023zix,Comi:2023lfm,Sacchi:2023omn}).
\bi
\item The {\red red} terms in \eref{CBHSU2pq21xyzs} indicate that gauging $(\BZ_2^{[1]})_{\hat{y}}$, which leads to the dual symmetry $(\BZ_2^{[0]})_{y}$, gives rise to the half-integral power in $w_2^{\pm \frac{1}{2}}$. Indeed, the smallest magnitude of the $\U(1)_{w_2}$ charge that appears in the Hilbert series is $2$. This means that there is a mixed 't Hooft anomaly characterised by the following anomaly theory:
\bes{ \label{mixedanomZ2yU1Z2}
\exp\left(i \pi \int_{M_4} B^{(2)}_y \cup \left[c_1(\U(1)_{w_2})\,\, \mod \,\, 2\right]  \right)~,
}
where $B^{(2)}_y \in H^{2}(M_3, \BZ_2^{[1]})$ is the two-form background field for the $(\BZ_2^{[1]})_{\hat{y}}$ one-form symmetry, and $c_1(\U(1)_{w_2})$ is the first Chern-class associated with the $\U(1)_{w_2}$ symmetry. Here, $M_4$ is the 4d bulk such that $\partial M_4 = M_3$, with $M_3$ being the 3d spacetime in which the theory in question lives.
\item Similarly, the terms highlighted in {\blue blue} in \eref{CBHSU2pq21xyzs} indicate that gauging $(\BZ_2^{[1]})_{\hat{x}}$, which leads to the dual $(\BZ_2^{[0]})_{x}$ symmetry, is in conflict with the global form of the symmetry $(\U(1)_{w_2}/\BZ_2) \times \SO(3)_{w_3} \times \SO(3)_{w_4}$, due to the presence of the representation $(\mathbf{2}; \mathbf{2})(2) \oplus (\mathbf{2}; \mathbf{2})(-2) $. Thus, there is a mixed 't Hooft anomaly characterised by the following anomaly theory:
\bes{ \label{mixedanomZ2xU1Zi}
\exp\left(i \pi \int_{M_4} B^{(2)}_x \cup \mathfrak{v}_2\right)~,
}
where $B^{(2)}_x \in H^{2}(M_3, \BZ_2^{[1]})$ is the two-form background field associated with the $(\BZ_2^{[1]})_{\hat{x}}$ one-form symmetry, and $\mathfrak{v}_2$ is the second Stiefel-Whitney class that obstructs the lift of the $(\U(1)_{w_2}/\BZ_2) \times \SO(3)_{w_3} \times \SO(3)_{w_4}$ bundle to the $(\U(1)_{w_2} \times \SU(2)_{w_3} \times \SU(2)_{w_4})/(\BZ_2 \times \BZ_2)$ bundle.
\item Finally, for the {\green green} terms in \eref{CBHSU2pq21xyzs}, the fugacity $xy$ corresponds to the diagonal subgroup of $(\BZ_2^{[0]})_x \times (\BZ_2^{[0]})_y$, which we will denote by $(\BZ_2^{[0]})_{xy}$. The dual one-form symmetry is the diagonal subgroup of $(\BZ_2^{[1]})_{\hat x} \times (\BZ_2^{[1]})_{\hat y}$ which we will denote by $(\BZ_2^{[1]})_{\hat{x}\hat{y}}$. Gauging $(\BZ_2^{[1]})_{\hat{x}\hat{y}}$ is in conflict with the global form of the symmetry $\SO(3)_{w_3} \times \SO(3)_{w_4}$, due to the presence of the representation $(\mathbf{2}; \mathbf{2})$. This signals that there is a mixed 't Hooft anomaly characterised by
\bes{ \label{mixedanomZ2xZ2yU1Z2}
\exp\left(i \pi \int_{M_4} B^{(2)}_{xy} \cup  \mathfrak{w}_2 \right)~,
}
where $B^{(2)}_{xy}$ is the two-form background field associated with the $(\BZ_2^{[1]})_{\hat{x}\hat{y}}$ one-form symmetry, and $\mathfrak{w}_2$ is the second Stiefel-Whitney class that obstructs the lift of the $\SO(3)_{w_3} \times \SO(3)_{w_4}$ bundle to the $(\SU(2)_{w_3} \times \SU(2)_{w_4})/\BZ_2$ bundle.  Note that this mixed anomaly does not involve the topological symmetry associated with $w_2$.
\ei

\section{\texorpdfstring{Moduli Space of $B_n$ Instantons via Discrete Gauging}{Moduli space of Bn instantons via discrete gauging}}
\label{sec:Bn}

\subsection{\texorpdfstring{$\bar{\min B_n}$ via Discrete Gauging}{minBn via discrete gauging}}
In this section, we present a method for obtaining the Hilbert series of the minimal nilpotent orbit closure of $B_n$, denoted by $\bar{\min B_n}$, by means of discrete gauging. So far in the literature, this has been obtained in two ways. The first way is from the $\USp(2)$ gauge theory with $\frac{1}{2}(2n+1)$ flavours. Since the number of flavours is not integral, in order for the theory to be consistent, the $\USp(2)$ gauge group must possess a half-integral Chern-Simons level. The latter reduces the amount of supersymmetry and lifts the Coulomb branch. In this approach, $\bar{\min B_n}$ should be regarded as a subvariety of the total moduli space of vacua of the supersymmetric Chern-Simons theory with vanishing expectation values for monopole operators.  The second way is to apply the Coulomb branch formula \cite{Cremonesi:2013lqa} to the following affine $B_n$ quiver with the prescription described in \cite{Cremonesi:2014xha}:
\bes{ \label{quivaffineBn}
\vcenter{\hbox{\begin{tikzpicture}
        \node[gauge,label=below:{$1$}] (1) at (0,0) {};
        \node[gauge,label=below:{$2$}] (2) at (1,0) {};
        \node[gauge,label=right:{$2$}] (1a) at (1,1) {};
        \node[gauge,label=below:{$2$}] (2a) at (2,0) {};
        \node[label=below:{}] (2b) at (3,0) {$\ldots$};
        \node[gauge,label=below:{$2$}] (2c) at (4,0) {};
        \node[gauge,label=below:{$1$}] (1b) at (5,0) {};
         \draw[transform canvas={yshift=1.5pt},thick] (2c)--(1b);
         \draw[transform canvas={yshift=-1.5pt},thick] (2c)--(1b);
        \draw (1)--(2)--(1a);
        \draw (2)--(2a)--(2b)--(2c);
        \node (>) at (4.5,0) {\Large $>$};
    \end{tikzpicture}}}
}
Although this method gives the required result, quiver \eref{quivaffineBn} does not possess a conventional Lagrangian description, due to the presence of the non-simply laced edge. In the following, we provide an alternative approach to realise the closure of the minimal nilpotent orbit of $B_n$ involving only 3d $\CN=4$ theories with a Lagrangian description.

Our starting point is to consider the 3d $\CN=4$ $\USp(2)$ gauge theory with $\frac{1}{2}\left[\left(2n+1\right)+m\right]$ flavours, where $m$ is odd. 
\bes{ \label{USp2w2n+1+m}
\vcenter{\hbox{\begin{tikzpicture}
        \node[gauge,label=below:{\footnotesize $\USp(2)$}] (1) at (0,0) {};
        \node[flavour,label=below:{\footnotesize $\SO(2n+1+m)$}] (2) at (4,0) {};
        \draw (1)--(2);
\end{tikzpicture}}}}
Observe that the number of flavours is integral. The flavour symmetry algebra is $\so(2n+1+m)$, where $2n+1+m$ is even.  The Higgs branch is indeed the closure of the minimal nilpotent orbit of $D_{(2n+1+m)/2}$.
Next we rewrite this quiver in an equivalent way as 
\bes{
\vcenter{\hbox{\begin{tikzpicture}
        \node[gauge,label={below,xshift=0.22cm}:{~~\footnotesize $\USp(2)$}] (1) at (0,0) {};
        \node[flavour,label=below:{\footnotesize $\SO(2n+1)$}] (2) at (4,0) {};
        \node[flavour,label=left:{\footnotesize $\SO(1)$}] (1a) at (-2,2) {};
        \node[flavour,label=left:{\footnotesize $\SO(1)$}] (1b) at (-2,1) {};
        \node[label=left:{}] (1c) at (-2,0) {$\vdots$};
        \node[flavour,label=left:{\footnotesize $\SO(1)$}] (1d) at (-2,-1) {};
        \node[flavour,label=left:{\footnotesize $\SO(1)$}] (1e) at (-2,-2) {};
        \draw (1)--(2);
        \draw (1)--(1a);
        \draw (1)--(1b);
        \draw (1)--(1d);
        \draw (1)--(1e);
        \draw[decorate,decoration={brace, amplitude=6pt, mirror}] (-3.5,2.4) -- (-3.5,-2.4) node[midway, left=0.2cm] {$m$};
\end{tikzpicture}}}}
We then implement discrete gauging by replacing each $\SO(1)$ by a $\BZ_m$ gauge group:
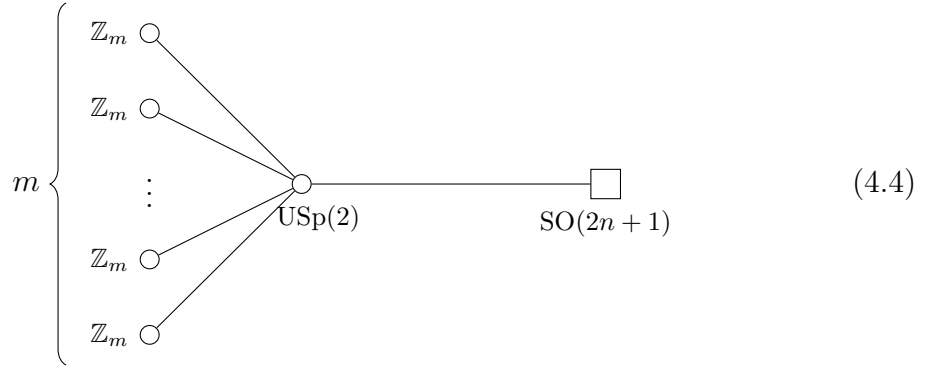
\bes{ \label{BnZm}
\vcenter{\hbox{\begin{tikzpicture}
        \node[gauge,label={below,xshift=0.22cm}:{\footnotesize $\USp(2)$}] (1) at (0,0) {};
        \node[flavour,label=below:{\footnotesize $\SO(2n+1)$}] (2) at (4,0) {};
        \node[gauge,label=left:{\footnotesize $\BZ_m$}] (1a) at (-2,2) {};
        \node[gauge,label=left:{\footnotesize $\BZ_m$}] (1b) at (-2,1) {};
        \node[label=left:{}] (1c) at (-2,0) {$\vdots$};
        \node[gauge,label=left:{\footnotesize $\BZ_m$}] (1d) at (-2,-1) {};
        \node[gauge,label=left:{\footnotesize $\BZ_m$}] (1e) at (-2,-2) {};
        \draw (1)--(2);
        \draw (1)--(1a);
        \draw (1)--(1b);
        \draw (1)--(1d);
        \draw (1)--(1e);
        \draw[decorate,decoration={brace, amplitude=6pt, mirror}] (-3.1,2.4) -- (-3.1,-2.4) node[midway, left=0.2cm] {$m$};
\end{tikzpicture}}}}
where each line connecting $\BZ_m$ and $\USp(2)$ denotes a half-hypermultiplet in the representation $(e^{2\pi i/m}; \mathbf{2})$ of $\BZ_m \times \USp(2)$. 

Our main claim is that, upon taking the limit $m \rightarrow \infty$, the Higgs branch of theory \eref{BnZm} is the closure of the minimal nilpotent orbit of $B_n$.  In other words, we propose a realisation of the transition between the following nilpotent orbits:
\bes{
\bar{\min\, D_{(2n+1+m)/2}} \,\, \longrightarrow \,\, \bar{\min\, B_n}~.
}
Let us also emphasise that, for each $m\in\BZ_{m\geq 1}$ (with $m$ odd), the Higgs branch moduli space is a $(2m+n-2)$ quaternionic dimensional hyperK\"ahler cone, whose Hilbert series has a palindromic numerator. The Coulomb branch of theory \eref{BnZm} is isomorphic to $\BC^2/\hat{D}_{(2n+1+m)/2}$, which is actually the Coulomb branch of the orginal theory \eref{USp2w2n+1+m}.  We subsequently demonstrate these claims using the superconformal index and Hilbert series.

The index of theory \eref{BnZm} is given by
\bes{ \label{indexBnZm}
\scalebox{0.98}{$
\begin{split}
&\CI_{\eref{BnZm}}(\vec f, \vec m|a,n_a; x) \\
&= \frac{1}{2} \sum_{l \in \BZ} \oint \frac{d z}{2 \pi i z} \CZ^{\USp(2)}_{\text{vec}}\left(z; l; x\right) \prod_{s = {0, \pm 1}} \CZ^{1}_{\chi} \left(z^{2 s} a^{-2}; 2 s l -2 n_a; x\right)  \\ 
& \qquad \times \frac{1}{m} \sum_{j=0}^{m-1} \prod_{s'= \pm 1} \left[ \CZ^{1/2}_{\chi} \left(e^{\frac{2\pi i}{m} j}z^{s'} a; s' l + n_a; x\right) \right]^m  \\
& \qquad \times \prod_{s_1 = \pm 1} \CZ^{1/2}_{\chi} \left(z^{s_1} a; s_1 l + n_a; x\right) \prod_{s_2 = \pm 1} \prod_{i = 1}^{n} \CZ^{1/2}_{\chi} \left(z^{s_1} f_i^{s_2} a; s_1 l + s_2 m_i + n_a; x\right) ~,
\end{split}
$}
}
where the $\USp(2)$ vector multiplet contribution is
\bes{
&\CZ^{\USp(2)}_{\text{vec}}(z; l; x)
={x^{-{\left|2 l\right|}}} \prod_{{s}={\pm{1}}}{\left({1}-{\left(-{1}\right)^{2 {s}{l}}}{z^{2 s}}{x^{2 \left|{s}{l}\right|}}\right)}~.
} 
For $m\geq 3$ and a general value of $n$, upon setting the background fluxes for the flavour symmetry and for the axial symmetry, \ie $\vec m$ and $n_a$ respectively, to zero, the series expansion of the index up to order $x^2$ reads
\bes{
\scalebox{0.92}{$
\begin{split}
&\CI_{\eref{BnZm}}(\vec f|a; x) = 1+ \left[ \chi^{B_n}_{[0,1,0,\ldots,0]}(\vec f) a^2 \right] x \\
& \quad \, \, \, \,\, +\left[\left(\chi^{B_n}_{[0,2,0,\ldots,0]}(\vec f) +R\right)a^4 +\left(1+r\right)a^{-4} - \left(1+ \chi^{B_n}_{[0,1,0,\ldots,0]}(\vec f) + m^2 \right) \right] x^2 + \ldots~,
\end{split}
$}
}
where
\bes{
R= \begin{cases} 
8 \chi^{B_n}_{[1,0,\ldots,0]} (\vec f) & \quad m=3 \\
0 & \quad m \geq 5
\end{cases}~ \quad \text{and} \quad
r= \begin{cases} 
1 & \quad n=2, m=3 \\
0 & \quad \text{otherwise}
\end{cases}~.
}
This has many interesting features which we now explain. At order $x$, we see that the continuous flavour symmetry of theory \eref{BnZm} is indeed $\so(2n+1)$, whose fugacities are $\vec f = (f_1, \ldots, f_n)$. We therefore do not refine the fugacities with respect to the $m$ bifundamental half-hypermultiplets in the gauge group $\BZ_m \times \USp(2)$, whose contribution is denoted in the second line of \eref{indexBnZm}. Note, however, that the contributions to the negative terms at order $x^2$ are as follows:
\bi
\item $1$: the $\U(1)_a$ axial symmetry current,
\item $\chi^{B_n}_{[0,1,0,\ldots,0]}$: the $\so(2n+1)$ flavour curents, and
\item $m^2$: the conserved currents from the $\BZ_m$ discrete gauging of $m$ hypermultiplets.\footnote{The index of the $\BZ_m$ discrete gauging of $m$ hypermultiplets can be written in terms of $\su(m) \times \u(1)$ representations as 
\bes{ \nonumber
&\frac{1}{m}\sum_{j=0}^{m-1} \prod_{i=1}^m \BZ^{1/2}_\chi(e^{\frac{2\pi i}{m}j} y^{-1}_i,0;x) \, \BZ^{1/2}_\chi(e^{-\frac{2\pi i}{m}j} y_i,0;x) = 1+ \left(1+[1,0,\ldots,0,1]_0 \right) x \\
& \qquad + \left\{[2,0,\ldots,0,2]_0 - \left(1+[1,0,\ldots,0,1]_0 \right)\right\} x^2 + \left( [n,0,\ldots,0]_{+n} + [0,\ldots,0,n]_{-n}\right) x^{n/2} +\ldots~,
}
where $\mathbf{R}_{r}$ denotes the $\su(m)$ character representation $\mathbf{R}$ with $\u(1)$ charge $r$.  Observe that the negative terms at order $x^2$ are of dimension $1+(m^2-1)=m^2$, as required.  This expression can also be obtained by starting from the index of $m$ hypermultiplets, namely $\prod_{i=1}^m \BZ^{1/2}_\chi(y^{-1}_i,0;x) \, \BZ^{1/2}_\chi(y_i,0;x)$, and then projecting out all representations of $\su(m)$ except those with $N$-ality $m$, \ie~ those with the Young diagram with the number of boxes equal to zero modulo $m$.
}
\ei
The representations $[0,p,0,\ldots,0]$ that appear as coefficients of $a^{2p} x^p$ in the index precisely agree with those appearing in the Hilbert series of $\bar{\min B_n}$ \cite{Benvenuti:2010pq}: 
\bes{ \label{HSminBn}
\mathrm{HS}\left[\bar{\min B_n}\right] (t, \vec f)= \sum_{p=0}^\infty \chi^{B_n}_{[0,p,0,\ldots,0]}(\vec f) t^{2p}~. 
}
For the special case of $m=3$, the additional gauge invariant operators that contribute $8 \chi^{B_n}_{[1,0,\ldots,0]} (\vec f)$ correspond to $\epsilon^{ab} \epsilon^{cd} Q^i_a q^{(\alpha_1)}_b q^{(\alpha_2)}_c q^{(\alpha_3)}_d$, where $i=1, \ldots, 2n+1$, $a,b,c,d=1,2$, and $\alpha_{1}, \alpha_{2}, \alpha_{3} =1,2,3$.\footnote{The multiplicity $8$ can be seen as follows. The tensor $\epsilon^{cd}$ imposes the antisymmetrisation of the indices $\alpha_2, \alpha_3$, and so $(\alpha_2, \alpha_3)$ can be taken as $(1,2)$, $(1,3)$, $(2,3)$. On the other hand, $\alpha_1$ runs over $1,2,3$, and so we have $3 \times 3 =9$ components that are subject to the identity $0=\epsilon_{\alpha_1 \alpha_2 \alpha_3} q^{(\alpha_1)}_b q^{(\alpha_2)}_c q^{(\alpha_3)}_d$ (since $b,c,d=1,2$). The latter implies that $\epsilon^{bc} \left( q^{(1)}_a q^{(2)}_b q^{(3)}_c - q^{(2)}_a q^{(1)}_b q^{(3)}_c + q^{(3)}_a q^{(1)}_b q^{(2)}_c   \right) =0$. We thus have 8 components, namely $(\alpha_1, \alpha_2, \alpha_3) =$ $\left\{(1,1,2)\right.$, $(2,1,2)$, $(3,1,2)$, $(1,1,3)$, $(2,1,3)$, $(3,1,3)$, $(2,2,3)$, $  \left.(3,2,3)\right\}$.} Here, $Q^i_a$ denotes the half-hypermultiplets in the bifundamental representation of $\USp(2)\times \SO(2n+1)$, and $q^{(\alpha_j)}_a$ denotes the half-hypermultiplets in the bifundamental representation of $\BZ^{(j)}_m\times \USp(2)$. Finally, the term $a^{-4}$ at order $x^2$ corresponds to the Coulomb branch operator $\tr(\varphi^2)$, where $\varphi$ is the complex scalar in the $\USp(2)$ vector multiplet.  We remark that, for the special case of $n=2$ and $m=3$, as will be seen in \eref{indexn=2}, we have a monopole operator with dimension two that contributes an extra term $a^{-4} x^2$ to the index.

From \eref{CBHBlimits}, we see that, upon setting $h=t$, the Higgs branch limit of the index \eref{indexBnZm} gives rise to the following Hilbert series:
\bes{ \label{HBHSBnZm}
\begin{split}
&\mathrm{HS}[\text{HB of \eref{BnZm}}](t,\vec{f}) \\
&=  \oint_{|z|=1} \frac{d z}{2\pi i z} (1-z^2) \times {\blue  \frac{1}{m}\sum_{j=0}^{m-1} \left(\PE \left[ (z+z^{-1}) e^{\frac{2 \pi i}{m} j}t \right] \right)^m} \\
& \qquad \times \PE \left[(z+z^{-1})\left\{ 1+\sum_{i=1}^{n} \left(f_i+f_i^{-1}\right) \right\} t - \left(z^2+1+z^{-2}\right) t^2  \right]~,
\end{split}
}
Note that the blue terms are effectively the contribution of $m$ hypermultiplets, each carrying charge $+1$ under $\BZ_m$. Since the total charge is $m \equiv 0 \, (\mod \,\, m)$, we expect that gauging the $\BZ_m$ symmetry preserves the hyperK\"ahler structure, and so the Hilbert series is expected to have a palindromic numerator for every $m \in \BZ_{\geq 1}$ and $m$ odd.  We also point out that the blue term approaches 1 as $m \rightarrow \infty$ and so, in this limit, the integral reproduces the Hilbert series \eref{HSminBn} as expected. 

On the other hand, the Coulomb branch limit gives the monopole formula of the original theory \eref{USp2w2n+1+m}, namely that of the $\USp(2)$ gauge theory with $\frac{1}{2}(2n+1+m)$ flavours. The is explicitly given in \cite[Section 5.2]{Cremonesi:2013lqa}.  The Coulomb branch of \eref{BnZm} is therefore $\BC^2/\hat{D}_{(2n+1+m)/2}$. As $m \rightarrow \infty$, the Coulomb branch Hilbert series simply reduces to $\PE[t^4] = (1-t^4)^{-1}$, with $t=c$.

We demonstrate these statements explicitly in the case of $n=2$ below. 

\subsubsection{\texorpdfstring{The Case of $n=2$}{The case of n=2}}
Let us focus on the example of $n=2$. We provide explicit expressions of the indices of \eref{BnZm} for $m=3,5,7$ below.\footnote{Below we set $f_i=1$ and $m_i=n_a=0$ for convenience.}
\bes{ \label{indexn=2}
\begin{tabular}{c|l}
\hline
$n=2$ and $m =$ & \qquad \qquad \qquad \quad index of \eref{BnZm} \\
\hline
$3$ & $1 + 10 a^2 x+\left(75 a^4+{2}{a^{-4}}-20\right) x^2$ \\ 
& $ \, \, \, \, + \left(674 a^6+{a^{-6}}-235 a^2-{10}{a^{-2}}\right) x^3$ \\ 
& $ \, \, \, \, +\left(3023 a^8+169+{3}{a^{-8}}-1791 a^4-{10}{a^{-4}}\right) x^4+\ldots$ \\
\hline
$5$ & $1 + 10 a^2 x+\left(35 a^4+{a^{-4}}-36\right) x^2$ \\ 
& $ \, \, \, \,+\left(959 a^6+{a^{-6}}-669 a^2-{25}{a^{-2}}\right) x^3$ \\ 
& $\, \, \, \, +\left(13010 a^8+593+{2}{a^{-8}}-1924 a^4-{26}{a^{-4}}\right) x^4+\ldots$ \\
\hline
$7$ & $1 + 10 a^2 x+\left(35 a^4+{a^{-4}}-60\right) x^2$ \\ 
& $ \, \, \, \,+\left(84 a^6-1245 a^2-{49}{a^{-2}}\right) x^3$ \\ 
& $\, \, \, \, +\left(23685 a^8+1709+{2}{a^{-8}}-5828 a^4-{49}{a^{-4}}\right) x^4+\ldots$ \\
\hline
\end{tabular}
}
The coefficient $10$ in $10a^2x$ corresponds to the adjoint representation $[0,2]$ of $B_2$. For $m=3$, the coefficient $75$ in $75a^4x^2$ corresponds to $[0,4]+ 8 [1,0]$. The negative terms at order $x^2$ for a generic value of $m$ is equal to $-1-10-m^2$.  These are in accordance with the general statement presented above.

The Higgs branch Hilbert series of \eref{BnZm} can be computed as in \eref{HBHSBnZm} or simply by reading off the coefficients of the terms $a^{2p} x^p$ (with $p=0,1,2,\ldots$) in the index.
\bes{
\scalebox{0.9}{$
\begin{tabular}{c|l}
\hline
$n=2$ and $m = $ & \qquad \qquad \quad Higgs branch Hilbert series of \eref{BnZm} \\
\hline
3 & $\frac{1}{(1-t)^{10} (1+t)^{10} \left(1-t+t^2\right)^6 \left(1+t+t^2\right)^6} $ \\ 
& $\times \left(1+t^2\right)(1 + 5 t^2 + 36 t^4 + 388 t^6 + 314 t^8 + 1773 t^{10} + 3055 t^{12}$ \\
& $\, \, \, \, \, \, \, \, \, \,+ 
 1421 t^{14} + 6426 t^{16} + 1421  t^{18} + \text{palindrome} + t^{32})$ \\
 & $= 1 + 10 t^2 + 75 t^4 + 674 t^6 + 3023 t^8 + 10671  t^{10}+\ldots$ \\
 \hline
5 &  $\frac{1}{(1 - t)^{14} (1 + t)^{14} (1 - t + t^2 - t^3 + t^4)^8 (1 + t + t^2 + t^3 + t^4)^8}$ \\
& $\times (1 + t^2) (1 + 3 t^2 - 13 t^4 + 892 t^6 + 6704 t^8 - 5144 t^{10} + 
   21884 t^{12}$ \\
&    $\, \, \, \, \, \, \, \, \, \,+ 197616 t^{14} - 29371 t^{16} + 285351 t^{18} + 
   877263 t^{20} + 769373 t^{22} $ \\
&   $\, \, \, \, \, \, \, \, \, \,+ 2300312 t^{24} + 619674 t^{26} + 
   3913260 t^{28} + 5787010 t^{30}  $ \\
&   $\, \, \, \, \, \, \, \, \, \,+ 3453595 t^{32} + 5569730 t^{34}+ 
   5062175 t^{36} + 9370620 t^{38} $ \\
&  $\, \, \, \, \, \, \, \, \, \,+ 5062175  t^{40}+  \text{palindrome} + t^{76})$ \\
& $= 1 + 10 t^2 + 35 t^4 + 959 t^6 + 13010 t^8 + 65799  t^{10} + \ldots$ \\
\hline
7 & $1 + 10 t^2 + 35 t^4 + 84 t^6 + 23685 t^8 + 358966  t^{10}+ \ldots$
\end{tabular} $}
}
In the case of $m=7$, the closed form is complicated and so we report only the series expansion in $t$ up to order $10$. It is instructive to compare these results with the Hilbert series of $\bar{\min\, B_2}$:
\bes{ \label{HSminB2}
\mathrm{HS}\left[\bar{\min\, B_2}\right] &= \frac{1 + 6 t^2 + t^4}{(1-t)^4 (1+t)^4} \\ &= 1 + 10 t^2 + 35 t^4 + 84 t^6 + 165 t^8 + 286  t^{10} + \ldots~.
}
From the above, we see that the Higgs branch Hilbert series of \eref{BnZm} with for a given $m$ reproduces \eref{HSminB2} correctly up to order $t^{m-1}$.

The Coulomb branch Hilbert series can be obtained by reading off the coefficients of the terms $a^{-2p} x^p$ (with $p=0,1,2,\ldots$) in the index.
\bes{
\scalebox{0.97}{$
\begin{tabular}{c|c|c}
\hline
$n=2$ and $m = $ & Coulomb branch Hilbert series of \eref{BnZm} & Coulomb branch\\
\hline
3 & $1 + 2 t^4 + t^6 + 3 t^8 + 2  t^{10} +\ldots$ &$\BC^2/\hat{D}_4$ \\&$= \PE[2 t^4 + t^6 - t^{12}]$ \\
\hline
5 & $1 + t^4 + t^6 + 2 t^8 + t^{10} + \ldots $&$\BC^2/\hat{D}_5$\\&$= \PE[t^4 + t^6 + t^8 - t^{16}]$\\
\hline
7 & $1 + t^4 + 2 t^8 + t^{10} +2  t^{12}+\ldots$ & $\BC^2/\hat{D}_6$\\&$=\PE[t^4 + t^8 + t^{10} - t^{20}]$ \\
\end{tabular} 
$}
}

\subsection{\texorpdfstring{Moduli Space of $k$ $\so(2n+1)$ Instantons}{Moduli Space of k so(2n+1) Instantons}}
The above method can be immediately generalised to study the moduli space of $k$ $\so(2n+1)$ instantons on $\BC^2$. The latter can be realised as the Higgs branch in the $m \rightarrow \infty$ limit of the following theory:
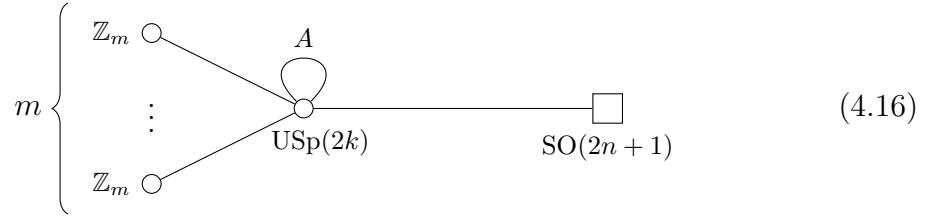
\bes{ \label{instBnZm}
\vcenter{\hbox{\begin{tikzpicture}
        \node[gauge,label={below,xshift=0.22cm}:{\footnotesize $\USp(2k)$}] (1) at (0,0) {};
        \node[flavour,label=below:{\footnotesize $\SO(2n+1)$}] (2) at (4,0) {};
        \node[gauge,label=left:{\footnotesize $\BZ_m$}] (1b) at (-2,1) {};
        \node[label=left:{}] (1c) at (-2,0) {$\vdots$};
        \node[gauge,label=left:{\footnotesize $\BZ_m$}] (1d) at (-2,-1) {};
        \draw (1)--(2);
        \draw (1)--(1b);
        \draw (1)--(1d);
        \path (1) edge [out=45,in=135,looseness=15]  node[midway,above]{\footnotesize $A$} (1);
        \draw[decorate,decoration={brace, amplitude=6pt, mirror}] (-3.1,1.4) -- (-3.1,-1.4) node[midway, left=0.2cm] {$m$};
\end{tikzpicture}}}}
where the loop around the $\USp(2k)$ node with label $A$ denotes a hypermultiplet in the antisymmetric representation.  For simplicity, we will focus on the case of $k=2$, for which the index is
\bes{
\scalebox{0.91}{$
\begin{split}
&\CI_{\eref{instBnZm}_{k=2}}
(\vec f, \vec m| y, n_y|a,n_a; x) \\
&= \frac{1}{8} \sum_{l_1, l_2 \in \BZ} \oint \frac{d z_1}{2 \pi i z_1} \oint \frac{d z_2}{2 \pi i z_2} ~ \CZ^{\USp(4)}_{\text{vec}}\left(\vec z; \vec l; x\right) \CZ^{\USp(4)}_{\chi, \text{adj}}\left(\vec z, \vec l; a, n_a; x\right) \\ 
& \quad \times \prod_{s=\pm 1} \Bigg[
\CZ^{1/2}_{\chi} \left(y^s a; sn_y + n_a; x\right) \prod_{s_1, s_2 = { \pm 1}} \CZ^{1/2}_{\chi} \left(y^s z_1^{s_1}z_2^{s_2} a; s n_y +s_1 l_1 +s_2 l_2 + n_a; x\right) \Bigg ]  \\
& \quad \times \prod_{p=1}^2 \Bigg[ \frac{1}{m} \sum_{j=0}^{m-1} \prod_{s'= \pm 1} \left[ \CZ^{1/2}_{\chi} \left(e^{\frac{2\pi i}{m} j}z_p^{s'} a; s' l_p + n_a; x\right) \right]^m  \\
& \quad \times \prod_{s_1 = \pm 1} \CZ^{1/2}_{\chi} \left(z_p^{s_1} a; s_1 l_p + n_a; x\right) \prod_{s_2 = \pm 1} \prod_{j = 1}^{n} \CZ^{1/2}_{\chi} \left(z_p^{s_1} f_j^{s_2} a; s_1 l_p + s_2 m_j + n_a; x\right) \Bigg]~,
\end{split}$}
}
where $(\vec f, \vec m)$ are fugacities/fluxes for the $\so(2n+1)$ flavour symmetry, $(y, n_y)$ are the fugacity/flux for the $\su(2)$ symmetry associated with the antisymmetric hypermultiplet, and the contribution of the $\USp(4)$ vector multiplet and of the scalar in the vector multiplet of $\USp(4)$ are given respectively by
\bes{
\CZ^{\USp(4)}_{\text{vec}}(\vec z; \vec l; x)
&=\prod_{i=1}^2{x^{-{\left|2 l_i\right|}}} \prod_{{s}={\pm{1}}}{\left({1}-{\left(-{1}\right)^{2 {s}{l_i}}}{z_i^{2 s}}{x^{2 \left|{s}{l_i}\right|}}\right)} \\&\times {x^{-{\left|l_1+l_2\right|}}} {x^{-{\left|l_1-l_2\right|}}} \prod_{s_1, s_2 = { \pm 1}}{\left({1}-{\left(-{1}\right)^{{s_1}{l_1}+{s_2}{l_2}}}{z_1^{s_1}}{z_2^{s_2}}{x^{\left|{s_1}{l_1}+{s_2}{l_2}\right|}}\right)}~,
} 
\bes{
\CZ^{\USp(4)}_{\chi, \text{adj}}\left(\vec z, \vec l; a, n_a; x\right) &=\CZ^{1}_{\chi} \left(a^{-2}; -2 n_a; x\right)^2 \prod_{s=\pm1} \prod_{i=1}^2 \CZ^{1}_{\chi} \left(z_i^{2s} a^{-2}; 2 s l_i -2 n_a; x\right) \\
& \qquad \times \prod_{s_1, s_2 = { \pm 1}} \CZ^{1}_{\chi} \left(z_1^{s_1}z_2^{s_2} a^{-2}; s_1 l_1 +s_2 l_2 -2 n_a; x\right) ~.
}
In the following, we will remove the free hypermultiplet and focus on the interacting part.  The corresponding index is
\bes{
\CI_{\text{int} \, \eref{instBnZm}_{k=2}}
(\vec f, \vec m| y, n_y|a,n_a; x)  = \frac{\CI_{\eref{instBnZm}_{k=2}}
(\vec f, \vec m| y, n_y|a,n_a; x)}{\prod_{s=\pm 1}
\CZ^{1/2}_{\chi} \left(y^s a; sn_y + n_a; x\right)}~.
}
The expressions for $\CI_{\text{int} \, \eref{instBnZm}_{k=2}}$ with $n=2$ and $m=3, 5, 7$ (with $f_i=y=1$ and $m_i = n_y = n_a =0$) are as follows.
\bes{ \label{indexinstBnZm}
\begin{tabular}{c|l}
\hline
$n=2$ and $m =$ & \qquad \qquad \qquad \qquad $\CI_{\text{int} \, \eref{instBnZm}_{k=2}}$ \\
\hline
$3$ &\quad   $1 + 13 a^2 x + 
 20 a^3 x^{
  3/2} +\left(135 a^4+{2}{a^{-4}}-23\right) x^2$  \\
  & $\, \, \, \, \quad +\left(370 a^5+{4}{a^{-3}}-40 a\right) x^{5/2}$  \\
  & $\, \, \, \, \quad  +\left(1922 a^6+{16}{a^{-2}}+{a^{-6}}-419 a^2\right) x^3+ \ldots$ \\
\hline  
$5$ & \quad $1 + 13 a^2 x + 
 20 a^3 x^{
  3/2} +\left(90 a^4+{a^{-4}}-39\right) x^2$ \\
  & $\, \, \, \, \quad + \left(200 a^5+{2}{a^{-3}}-72 a\right) x^{5/2}$  \\
  & $\, \, \, \, \quad + \left(1907 a^6+{a^{-6}}-1117 a^2-{12}{a^{-2}}\right) x^3+\ldots$\\
\hline  
$7$ & \quad $1 + 13 a^2 x + 
 20 a^3 x^{
  3/2} + \left(90 a^4+{a^{-4}}-63\right) x^2$ \\
  & $\, \, \, \, \quad +\left(200 a^5+{2}{a^{-3}}-120 a\right) x^{5/2}$  \\
  & $\, \, \, \, \quad  +\left(537 a^6-2029 a^2-{36}{a^{-2}}\right) x^3+ \ldots$
\end{tabular}
}
As before, the Higgs (resp. Coulomb) branch limit can be obtained as stated in \eref{CBHBlimits} or simply by reading off the coefficients of the terms $a^{p} x^{\frac{p}{2}}$ (resp. $a^{-p} x^{\frac{p}{2}}$) in the index. We emphasise that the closed form of the unrefined Higgs branch Hilbert series always has a palindromic numerator for each $m$.\footnote{For example, in the case of $m=3$, the Hilbert series reads
\bes{ \nonumber
& \frac{1}{(1 - t)^{22} (1 + t)^{12} (1 - t + t^2)^9 (1 + t + t^2)^11 (1 + t^3 + 
   t^6)^7} \Big( 1- 8 t+ 38 t^{2}- 111 t^{3}+ 249 t^{4}- 348 t^{5}\\
   & \quad + 881 t^{6}- 1819 t^{7}+ 
 4081 t^{8}- 1276 t^{9}- 3928 t^{10}+ 22063 t^{11}+ 7989 t^{12}- 
 46569 t^{13}+ 153376 t^{14}\\
 & \quad + 26067 t^{15}- 157449 t^{16}+ 629021 t^{17}+ 
 195232 t^{18}- 600782 t^{19}+ 2381257 t^{20}+ 599635 t^{21} \\
 & \quad - 
 1477418 t^{22}+ 7084214 t^{23}+ 1779297 t^{24}- 3244566 t^{25}+ 
 18358846 t^{26}+ 4050880 t^{27}- 5558417 t^{28} \\
 & \quad + 40440192 t^{29}+ 
 8439311 t^{30}- 8424385 t^{31}+ 78685858 t^{32}+ 14035225 t^{33}- 
 9524846 t^{34}+ 133646194 t^{35} \\
 & \quad + 20541158 t^{36}- 8041117 t^{37}+ 
 201393051 t^{38}+ 25256773 t^{39}- 2183828 t^{40}+ 268308193 t^{41}\\ 
 & \quad + 
 27444177 t^{42}+ 5877249 t^{43}+ 319396661 t^{44}+ 23667652 t^{45}+ 
 16047391 t^{46}+ 337880374 t^{47} \\
 & \quad + 16047391  t^{48} + \text{palindrome} + t^{94}  \Big)~.
}
}  It is instructive to compare this with the Hilbert series of the centred moduli space of $2$ $\so(5)$ instantons \cite[(4.23)]{Hanany:2012dm} (see also \cite{Cremonesi:2014xha}):
\bes{ \label{HS2so5inst}
&\frac{1}{(1- t)^{10} (1 + t)^6  (1 + t + t^2)^5} \Big( 1 + t + 8  t^2 + 23  t^3 + 50  t^4 + 95  t^5 + 177  t^6 \\
& \qquad + 222  t^7 + 
  236 t^8 + 222 t^9 + \text{palindrome} + t^{16} \Big) \\
&=1 + 13 t^2 + 20 t^3 + 90 t^4 + 200 t^5 + 537  t^6+\ldots~.
}
We see that the Higgs branch limit of the index for a given $m$ reproduces \eref{HS2so5inst} correctly up to order $t^{m}$. 

 The Coulomb branch limit of \eref{indexinstBnZm} is indeed equal the Coulomb branch Hilbert series of the $\USp(2k)$ gauge theory with one rank-two antisymmetric and $(2n+1+m)/2$ fundamental hypermultiplets, which is isomorphic to $\Sym^k\left(\BC^2/\hat{D}_{(2n+1+m)/2}\right)$ \cite[Section 4.2]{Cremonesi:2013lqa}. For example, the case of $k=2$, $n=2$ and $m=3$, the Hilbert series of $\Sym^2(\BC^2/\hat{D}_{4})$ reads
\bes{
&\frac{1}{2} \left( \PE\left[2 t^4 + t^6 - t^{12}\right]^2 + \PE\left[2 t^8 + t^{12} - t^{24}\right]\right) \\
&= \frac{1 - 2 t^2 + 2 t^4 - t^6 + 2 t^8 - 2 t^{10} + t^{12}}{(1 - t^2)^4 (1 + t^2)^2 (1 + t^4)} \\
&= 1 + 2 t^4 + t^6 + 6 t^8 + 4  t^{10}+ \ldots~.
}
In the limit $m\rightarrow \infty$, the Coulomb branch Hilbert series of $\eref{instBnZm}_{k=2}$, which is equal to that of $\Sym^2(\BC^2/\hat{D}_r)$ with $r\rightarrow \infty$, reads
\bes{
\frac{1}{2} \left( \PE[t^4]^2 +\PE[t^8] \right) =  \PE[t^4 + t^8]~.
}

\pagebreak

\section*{Acknowledgements}
We would like to thank Antoine Bourget, Federico Carta, Riccardo Comi, Sara Pasquetti, Matteo Sacchi and Gabi Zafrir for several useful conversations. JFG is supported by the EPSRC Open Fellowship (Schafer-Nameki) EP/X01276X/1 and the ``Simons Collaboration on Special Holonomy in Geometry, Analysis and Physics''. JFG and NM gratefully acknowledge support from the Simons Center for Geometry and Physics, Stony Brook University, at which part of this research project was conducted during the Simons Physics Summer Workshop 2024.
NM and WH are partially supported by the MUR-PRIN grant No. 2022NY2MXY.

\appendix

\section{Examples for Parametrising Magnetic Lattices}
\label{app:LatticeConstraints}

In this appendix, we illustrate various possible constraints on the $m_{i,a_i}\in\Gamma_G^{mw}$ which can be used to parametrise the $\Gamma_{\tilde{G}}^{mw}$ lattice, for several examples of quivers.

\subsection{\texorpdfstring{Example 1: Two Node Abelian Quiver with $N$ $(p,p)$-Edges}{Example1: Two node abelain quiver with N (p,p)-edges}} 
Let us consider the following quiver:
\begin{equation} \label{quivex1}
    \raisebox{-.5\height}{\begin{tikzpicture}
        \node[gauge,label=below:{$1$},label=above:{$\color{red}m_{1}$}] (1) at (0,0) {};
        \node[gauge,label=below:{$1$},label=above:{$\color{blue}m_{2}$}] (2) at (3,0) {};
        \edge{1}{2}{p}{p}{N}
    \end{tikzpicture}}
\end{equation}
with magnetic weights $({\color{red}m_{1}},{\color{blue}m_{2}})\in\Gamma_{\mathrm{U}(1)\times\mathrm{U}(1)}^{mw}$ of the naive gauge group $\tilde{G}=\mathrm{U}(1)\times\mathrm{U}(1)$.
The adjacency matrix $A$, rank vector $v$ and shortness vector $s$ are given by
\begin{equation}
    A=\begin{pmatrix}
        -2 & Np \\
        Np & -2
    \end{pmatrix}\;, \quad v =\begin{pmatrix}
        1\\
        1
    \end{pmatrix}\;,\quad s=\begin{pmatrix}
        1\\
        1
    \end{pmatrix}\;,
\end{equation}\\
from which the conformal dimension \eref{Deltaofm} is
\begin{equation}
    2\Delta({\color{red}m_{1}},{\color{blue}m_{2}})=N|p\,{\color{red}m_{1}}-p\,{\color{blue}m_{2}}|=Np|{\color{red}m_{1}}-{\color{blue}m_{2}}|\;.
\end{equation}\\
The shift vector is then
\begin{equation}
    {\color{goodgreen}\delta}=({\color{red}1},{\color{blue}1})\;,
\end{equation}\\
where ${\color{goodgreen}\delta}$ acts on $({\color{red}m_{1}},{\color{blue}m_{2}})\in\Gamma_{\mathrm{U}(1)\times\mathrm{U}(1)}^{mw}$ by translation:
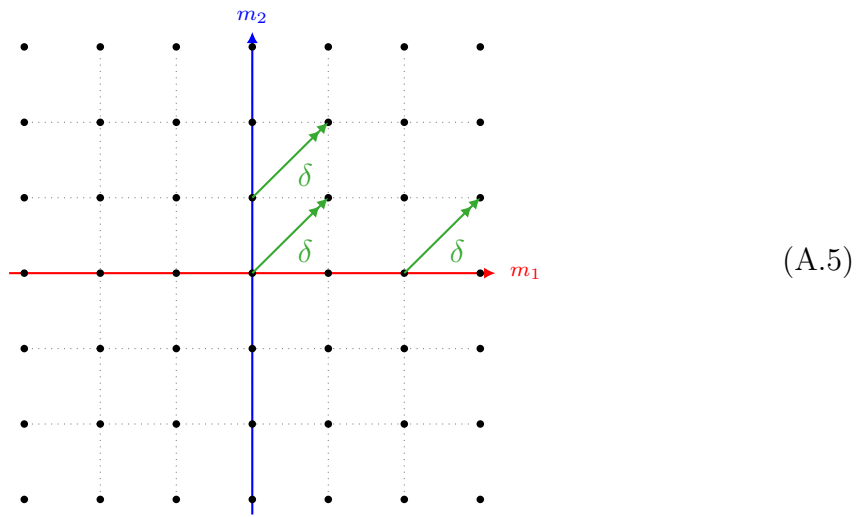
\bes{
    \latshif{a1}
}
We can indicate the $\Lambda_{\color{goodgreen}\delta}$-orbits in $\Gamma_{\mathrm{U}(1)\times\mathrm{U}(1)}^{mw}$, where $2\Delta({\color{red}m_{1}},{\color{blue}m_{2}})=\textnormal{constant}$, via dashed green lines:
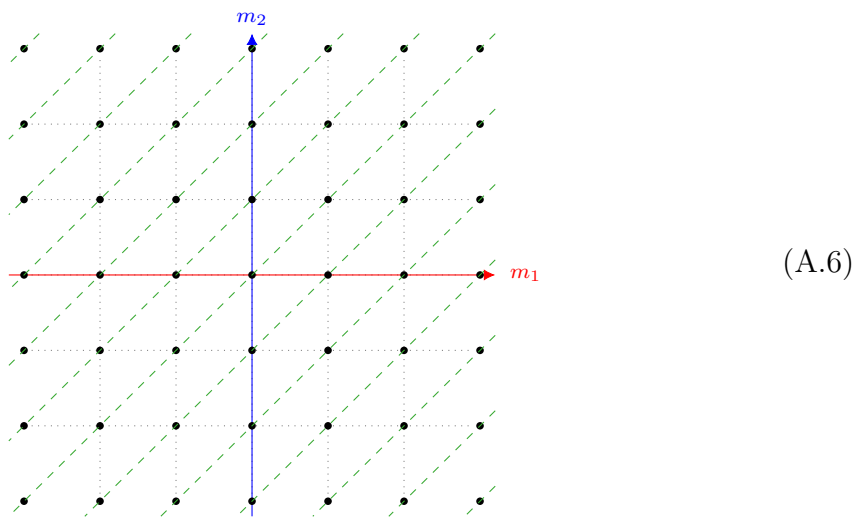
\bes{
    \latshif{a2}
}
These orbits are the magnetic lattice $\Gamma_{G}^{mw}$ of the true gauge group $G=\tilde{G}/\mathrm{U}(1)$.

We can parametrise $\Gamma_{G}^{mw}$ in many different ways. We will show a few of these here
\begin{enumerate}
    \item Fix ${\color{red}m_{1}}={\color{red}0}$. Let us depict the points in $\Gamma_{\tilde{G}}^{mw}$ obeying the constraint as red dots, and the corresponding $\Lambda_{\color{goodgreen}\delta}$-orbits as red lines:
    \bes{
        \latshif{a3}
    }
    We can see that every $\Lambda_{\color{goodgreen}\delta}$-orbit in $\Gamma_{\tilde{G}}^{mw}$ is represented exactly once, and hence we have a good parametrisation of $\Gamma_{G}^{mw}$. Note that fixing ${\color{red}m_{1}}=a$ for any $a\in\mathbb{Z}$ works just as well.
    \item Fix ${\color{blue}m_{2}}={\color{blue}0}$. Let us depict the points in $\Gamma_{\tilde{G}}^{mw}$ obeying the constraint as blue dots, and the corresponding $\Lambda_{\color{goodgreen}\delta}$-orbits as blue lines:
    \bes{
        \latshif{a4}
    }
    We can see that every $\Lambda_{\color{goodgreen}\delta}$-orbit in $\Gamma_{\tilde{G}}^{mw}$ is represented exactly once, and hence we have a good parametrisation of $\Gamma_{G}^{mw}$. Note that, just like before, fixing ${\color{blue}m_{2}}=a$ for any $a\in\mathbb{Z}$ works just as well.
    \item Fix ${\color{red}m_{1}}+{\color{blue}m_{2}}\in\{{\color{orange}0},{\color{purple}1}\}$. This is a more interesting case.
    
    Let us first depict the points in $\Gamma_{\tilde{G}}^{mw}$ obeying ${\color{red}m_{1}}+{\color{blue}m_{2}}={\color{orange}0}$ as orange dots, and the corresponding $\Lambda_{\color{goodgreen}\delta}$-orbits as orange lines:
    \bes{
        \latshif{a5}
    }
    We can see that not all $\Lambda_{\color{goodgreen}\delta}$-orbits in $\Gamma_{\tilde{G}}^{mw}$ are represented.
    
    Let us now depict the points in $\Gamma_{\tilde{G}}^{mw}$ obeying ${\color{red}m_{1}}+{\color{blue}m_{2}}={\color{purple}1}$ as purple dots, and the corresponding $\Lambda_{\color{goodgreen}\delta}$-orbits as purple lines:
    \bes{
        \latshif{a6}
    }
    We can see that all the remaining $\Lambda_{\color{goodgreen}\delta}$-orbits in $\Gamma_{\tilde{G}}^{mw}$ are represented.

    Combining the two we get
    \bes{
        \latshif{a7}
    }
    where all $\Lambda_{\color{goodgreen}\delta}$-orbits in $\Gamma_{\tilde{G}}^{mw}$ are represented exactly once, and hence we have a good parametrisation of $\Gamma_{G}^{mw}$.

    Note that fixing ${\color{red}m_{1}}+{\color{blue}m_{2}}\in\{a,b\}$ for any $a,b\in\mathbb{Z}$ with $a$ even and $b$ odd works as well.
    \item Fix ${\color{red}2m_{1}}+{\color{blue}m_{2}}\in\{{\color{orange}0},{\color{purple}1},{\color{cyan}2}\}$.
    
    Let us first depict the points in $\Gamma_{\tilde{G}}^{mw}$ obeying ${\color{red}2m_{1}}+{\color{blue}m_{2}}={\color{orange}0}$ as orange dots, and the corresponding $\Lambda_{\color{goodgreen}\delta}$-orbits as orange lines:
    \bes{
        \latshif{a8}
    }
    We can see that not all $\Lambda_{\color{goodgreen}\delta}$-orbits in $\Gamma_{\tilde{G}}^{mw}$ are represented.
    
    Let us now depict the points in $\Gamma_{\tilde{G}}^{mw}$ obeying ${\color{red}2m_{1}}+{\color{blue}m_{2}}={\color{purple}1}$ as purple dots, and the corresponding $\Lambda_{\color{goodgreen}\delta}$-orbits as purple lines:
    \bes{
        \latshif{a9}
    }
    We can see that still not all remaining $\Lambda_{\color{goodgreen}\delta}$-orbits in $\Gamma_{\tilde{G}}^{mw}$ are represented.
    
    Let us now depict the points in $\Gamma_{\tilde{G}}^{mw}$ obeying ${\color{red}2m_{1}}+{\color{blue}m_{2}}={\color{cyan}2}$ as cyan dots, and the corresponding $\Lambda_{\color{goodgreen}\delta}$-orbits as cyan lines:
    \bes{
        \latshif{a10}
    }
    We can see that now all remaining $\Lambda_{\color{goodgreen}\delta}$-orbits in $\Gamma_{\tilde{G}}^{mw}$ are represented.

    Combining the three we get
    \bes{
        \latshif{a11}
    }
    where all $\Lambda_{\color{goodgreen}\delta}$-orbits in $\Gamma_{\tilde{G}}^{mw}$ are represented exactly once, and hence we have a good parametrisation of $\Gamma_{G}^{mw}$.

    Note that fixing ${\color{red}2m_{1}}+{\color{blue}m_{2}}\in\{a,b,c\}$ for any $a,b,c\in\mathbb{Z}$ with $(a\;\mod\;3)=0$, $(b\;\mod\;3)=1$, and $(c\;\mod\;3)=2$ works just as well.
    \item Now, consider a constraint on ${\color{red}2m_{1}}$. Since ${\color{red}2m_{1}}=0$ is equivalent to ${\color{red}m_{1}}=0$, ${\color{red}2m_{1}}\in\{0\}$ is a fine constraint. However one could also use the constraint ${\color{red}2m_{1}}\in\{0,1\}$. Observe that ${\color{red}2m_{1}}=1$ implies ${\color{red}m_{1}}=\frac{1}{2}$, and, since $({\color{red}\frac{1}{2}},{\color{blue}m_{2}})\notin\Gamma_G^{mw}$, there is no additional contribution coming from ${\color{red}2m_{1}}=1$.

    It may seem silly to use the constraint ${\color{red}2m_{1}}\in\{0,1\}$ over ${\color{red}2m_{1}}\in\{0\}$. We discuss it here to explain why the general formula \eqref{eq:MagLatGenCon} applies even in such boundary cases.
    \item Similarly, consider a constraint on ${\color{red}2m_{1}}+{\color{blue}2m_{2}}$. Note that ${\color{red}2m_{1}}+{\color{blue}2m_{2}}\in\{0,1,2,3\}$ is equivalent to ${\color{red}2m_{1}}+{\color{blue}2m_{2}}\in\{0,2\}$, which is equivalent to ${\color{red}m_{1}}+{\color{blue}m_{2}}\in\{0,1\}$, which is a valid constraint as discussed before.

    We discuss the constraint ${\color{red}2m_{1}}+{\color{blue}2m_{2}}\in\{0,1,2,3\}$, since it agrees with the general formula \eqref{eq:MagLatGenCon}.
\end{enumerate}

\pagebreak
\subsection{Example 2: Single U(2) Node} 
Let us consider a single $\U(2)$ gauge node
\begin{equation}
    \raisebox{-.5\height}{\begin{tikzpicture}
        \node[gauge,label=below:{$2$},label=above:{${\color{red}m_{1,1}},{\color{red}m_{1,2}}$}] (1) at (0,0) {};
    \end{tikzpicture}}
\end{equation}
with magnetic weights $({\color{red}m_{1,1}},{\color{red}m_{1,2}})\in\Gamma_{\mathrm{U}(2)}^{mw}$ of the naive gauge group $\tilde{G}=\mathrm{U}(2)$.
The Weyl group of $\tilde{G}=\mathrm{U}(2)$ is
\begin{equation}
    W_{\mathrm{U}(2)}=S_2=\mathbb{Z}_2
\end{equation}
and we may parametrise $\Gamma_{\mathrm{U}(2)}^{mw}/\mathbb{Z}_2$ by setting
\begin{equation}
    -\infty<{\color{red}m_{1,1}}\leq{\color{red}m_{1,2}}<\infty\;.
\end{equation}
We depict the region of ${\color{red}m_{1,1}}={\color{red}m_{1,2}}$ with an orange shading, and the region of ${\color{red}m_{1,1}}<{\color{red}m_{1,2}}$ with a lighter orange shading:
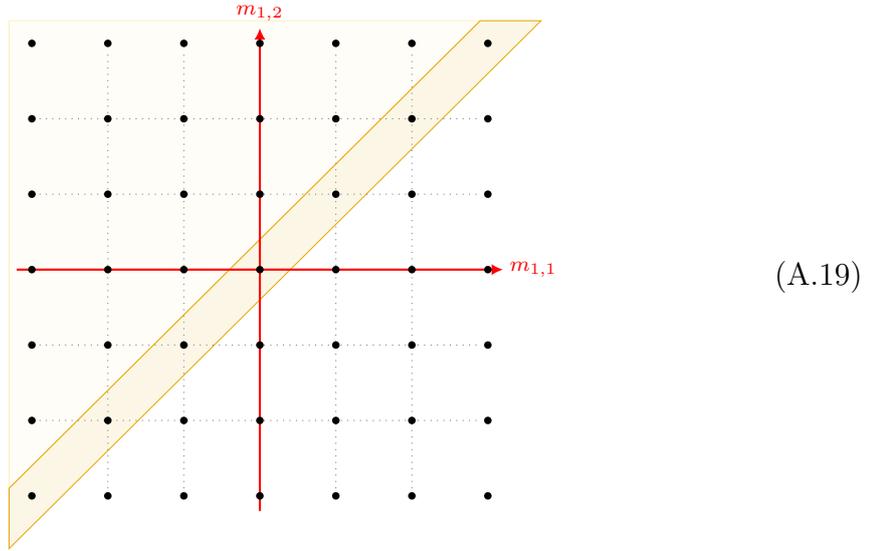
\bes{
    \latshif{b0}
}
The adjacency matrix $A$, rank vector $v$ and shortness vector $s$ are given by
\begin{equation}
    A=\begin{pmatrix}
        -2
    \end{pmatrix}\;, \quad v = \begin{pmatrix}
        2
    \end{pmatrix}\;,\quad s=\begin{pmatrix}
        1
    \end{pmatrix}\;.
\end{equation}\\
The conformal dimension \eref{Deltaofm} is then
\begin{equation}
    2\Delta({\color{red}m_{1,1}},{\color{red}m_{1,2}})=-2|{\color{red}m_{1,1}}-{\color{red}m_{1,2}}|\;.
\end{equation}\\
The shift vector is given by
\begin{equation}
    {\color{goodgreen}\delta}=({\color{red}1},{\color{red}1})\;,
\end{equation}
where ${\color{goodgreen}\delta}$ acts on $({\color{red}m_{1,1}},{\color{red}m_{1,2}})\in\Gamma_{\mathrm{U}(2)}^{mw}$ by translation:
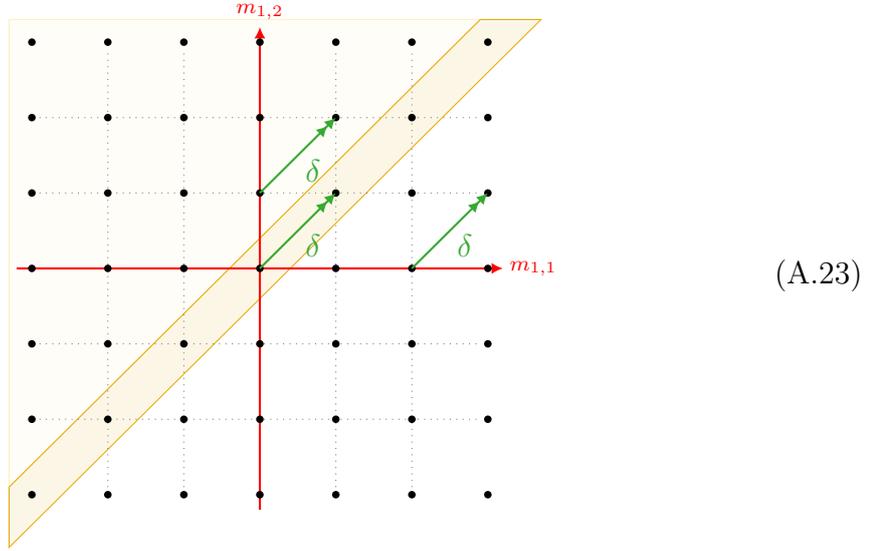
\bes{
    \latshif{b1}
}
We can indicate the $\Lambda_{\color{goodgreen}\delta}$-orbits in $\Gamma_{\mathrm{U}(2)}^{mw}$, where $2\Delta({\color{red}m_{1,1}},{\color{red}m_{1,2}})=\textnormal{constant}$, via dashed green lines:
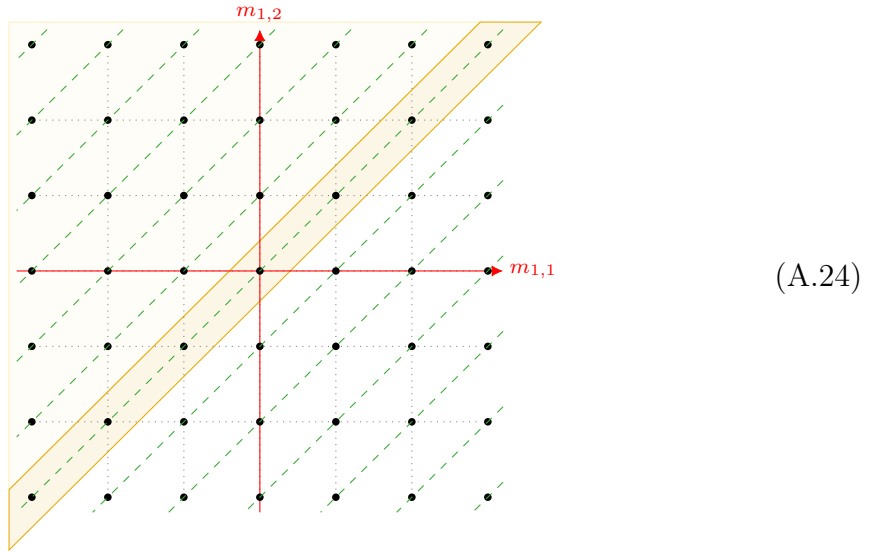
\bes{
    \latshif{b2}
}
These orbits are the magnetic lattice $\Gamma_{G}^{mw}$ of the true gauge group $G=\tilde{G}/\mathrm{U}(1)$.

We can again parametrise $\Gamma_{G}^{mw}$ in many different ways. We will show a few of these here.
\begin{enumerate}
    \item Fix ${\color{red}m_{1,1}}={\color{red}0}$. Let us depict the points in $\Gamma_{\tilde{G}}^{mw}$ obeying the constraint as red dots, and the corresponding $\Lambda_{\color{goodgreen}\delta}$-orbits as red lines:
    \bes{
        \latshif{b3}
    }
    We can see that every $\Lambda_{\color{goodgreen}\delta}$-orbit in $\Gamma_{\tilde{G}}^{mw}$ is represented exactly once, and hence we have a good parametrisation of $\Gamma_{G}^{mw}$. Furthermore, restricting to the shaded part via the constraint ${\color{red}m_{1,1}}\leq{\color{red}m_{1,2}}$, we get a good parametrisation of $\Gamma_{G}^{mw}/W_G=\Gamma_{\mathrm{U}(2)/\mathrm{U}(1)}^{mw}/\mathbb{Z}_2=\Gamma_{\mathrm{SU}(2)/\mathbb{Z}_2}^{mw}/\mathbb{Z}_2$. Note that fixing ${\color{red}m_{1,1}}=a$ for any $a\in\mathbb{Z}$ works just as well. Note also that we could instead have imposed this constraint on ${\color{red}m_{1,2}}$ instead of ${\color{red}m_{1,1}}$.
    \item Fix ${\color{red}m_{1,1}}+{\color{red}m_{1,2}}\in\{{\color{orange}0},{\color{purple}1}\}$. This is again a more interesting case.
    
    Let us first depict the points in $\Gamma_{\tilde{G}}^{mw}$ obeying ${\color{red}m_{1,1}}+{\color{red}m_{1,2}}={\color{orange}0}$ as orange dots, and the corresponding $\Lambda_{\color{goodgreen}\delta}$-orbits as orange lines:
    \bes{
        \latshif{b5}
    }
    We can see that not all $\Lambda_{\color{goodgreen}\delta}$-orbits in $\Gamma_{\tilde{G}}^{mw}$ are represented, just as in the previous example. However, this case is quite different. The $\Lambda_{\color{goodgreen}\delta}$-orbits indicated in orange make up $\Gamma_{\mathrm{SU}(2)}^{mw}$, and those located in the shaded area represent $\Gamma_{\mathrm{SU}(2)}^{mw}/\mathbb{Z}_2$. Since $\mathrm{SU}(2)$ is the double cover of $G=\mathrm{SU}(2)/\mathbb{Z}_2$, we can see that choosing ${\color{red}m_{1,1}}+{\color{red}m_{1,2}}={\color{orange}0}$ corresponds to picking a different global structure of the true gauge group.
    
    Let us now depict the points in $\Gamma_{\tilde{G}}^{mw}$ obeying ${\color{red}m_{1,1}}+{\color{red}m_{1,2}}={\color{purple}1}$ as purple dots, and the corresponding $\Lambda_{\color{goodgreen}\delta}$-orbits as purple lines:
    \bes{
        \latshif{b6}
    }
    We can see that all the remaining $\Lambda_{\color{goodgreen}\delta}$-orbits in $\Gamma_{\tilde{G}}^{mw}$ are represented.

    Combining the two we get
    \bes{
        \latshif{b7}
    }
    where all $\Lambda_{\color{goodgreen}\delta}$-orbits in $\Gamma_{\tilde{G}}^{mw}$ are represented exactly once, and hence we have a good parametrisation of $\Gamma_{G}^{mw}$.

    Note that fixing ${\color{red}m_{1,1}}+{\color{red}m_{1,2}}\in\{a,b\}$ for any $a,b\in\mathbb{Z}$ with $a$ even and $b$ odd works as well. Note further that ${\color{red}m_{1,1}}+{\color{red}m_{1,2}}=a$ for any $a\in\mathbb{Z}$ with $a$ even picks the magnetic lattice for $\mathrm{SU}(2)$ instead of $G=\mathrm{SU}(2)/\mathbb{Z}_2$.
\end{enumerate}

\subsection{Example 3: (2,3)-Edged Two Node Abelian Quiver} 
Let us consider the following quiver
\begin{equation}
\label{eq:(2,3)-edged.Quiver}
    \raisebox{-.5\height}{\begin{tikzpicture}
        \node[gauge,label=below:{$1$},label=above:{$\color{red}m_{1}$}] (1) at (0,0) {};
        \node[gauge,label=below:{$1$},label=above:{$\color{blue}m_{2}$}] (2) at (2,0) {};
        \edge{1}{2}{2}{3}{1};	
    \end{tikzpicture}}\;,
\end{equation}
with magnetic weights $({\color{red}m_{1}},{\color{blue}m_{2}})\in\Gamma_{\mathrm{U}(1)\times\mathrm{U}(1)}^{mw}$ of the naive gauge group $\tilde{G}=\mathrm{U}(1)\times\mathrm{U}(1)$.
The adjacency matrix $A$, rank vector $v$ and shortness vector $s$ are given by
\begin{equation}
    A=\begin{pmatrix}
        -2 & 2 \\
        3 & -2
    \end{pmatrix}\;,\quad v=\begin{pmatrix}
        1\\
        1
    \end{pmatrix}\;,\quad s=\begin{pmatrix}
        3\\
        2
    \end{pmatrix}\;.
\end{equation}\\
It follows that the conformal dimension \eref{Deltaofm} is
\begin{equation}
    2\Delta({\color{red}m_{1}},{\color{blue}m_{2}})=|{2\color{red}m_{1}}-{3\color{blue}m_{2}}|\;.
\end{equation}\\
The shift vector is given by
\begin{equation}
    {\color{goodgreen}\delta}=({\color{red}3},{\color{blue}2})\;,
\end{equation}\\
where ${\color{goodgreen}\delta}$ acts on $({\color{red}m_{1}},{\color{blue}m_{2}})\in\Gamma_{\mathrm{U}(1)\times\mathrm{U}(1)}^{mw}$ by translation:
\bes{
    \latshif{c1}
}
We can indicate the $\Lambda_{\color{goodgreen}\delta}$-orbits in $\Gamma_{\mathrm{U}(1)\times\mathrm{U}(1)}^{mw}$, where $2\Delta({\color{red}m_{1}},{\color{blue}m_{2}})=\textnormal{constant}$, via dashed green lines:
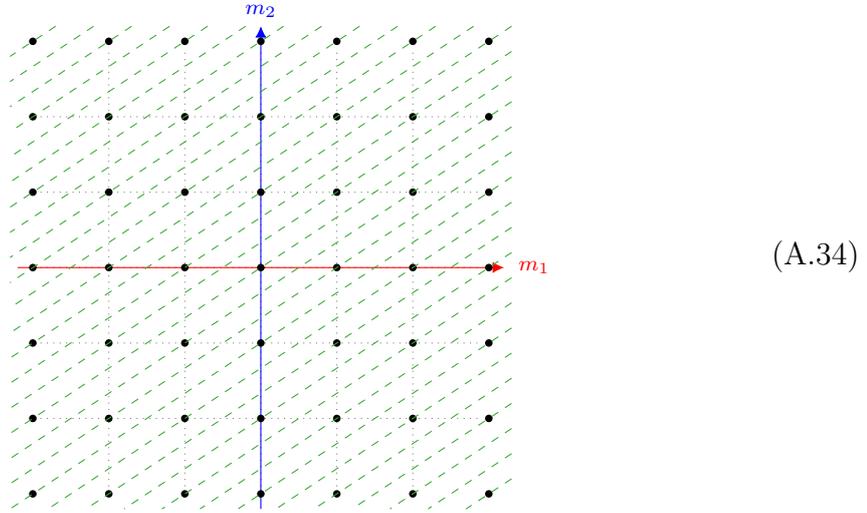
\bes{
    \latshif{c2}
}
These orbits are the magnetic lattice $\Gamma_{G}^{mw}$ of the true gauge group $G=\tilde{G}/\mathrm{U}(1)$.

We can parametrise $\Gamma_{G}^{mw}$ in many different ways. We will show a few of these here
\begin{enumerate}
    \item Fix ${\color{red}m_{1}}\in\{0,1,2\}$. We depict the points in $\Gamma_{\tilde{G}}^{mw}$ obeying ${\color{red}m_{1}}\in\{0,1,2\}$ as red dots, and the corresponding $\Lambda_{\color{goodgreen}\delta}$-orbits as red lines:
    \begin{enumerate}[label=(\alph*)]
        \item ${\color{red}m_{1}}=0$:
        \bes{
            \latshif{c3}
        }
        \item ${\color{red}m_{1}}=1$:
        \bes{
            \latshif{c4}
        }
        \item ${\color{red}m_{1}}=2$:
        \bes{
            \latshif{c5}
        }
    \end{enumerate}
    Combining the three we see that all $\Lambda_{\color{goodgreen}\delta}$-orbits in $\Gamma_{\tilde{G}}^{mw}$ are represented exactly once, and hence we have a good parametrisation of $\Gamma_{G}^{mw}$.
    \item Fix ${\color{blue}m_{2}}\in\{0,1\}$. We depict the points in $\Gamma_{\tilde{G}}^{mw}$ obeying ${\color{blue}m_{2}}\in\{0,1\}$ as blue dots, and the corresponding $\Lambda_{\color{goodgreen}\delta}$-orbits as blue lines:
    \begin{enumerate}[label=(\alph*)]
        \item ${\color{blue}m_{2}}=0$:
        \bes{
            \latshif{c6}
        }
        \item ${\color{blue}m_{2}}=1$:
        \bes{
            \latshif{c7}
        }
    \end{enumerate}
    Combining both we see that all $\Lambda_{\color{goodgreen}\delta}$-orbits in $\Gamma_{\tilde{G}}^{mw}$ are represented exactly once, and hence we have a good parametrisation of $\Gamma_{G}^{mw}$.
    \item Fix ${\color{red}m_{1}}+{\color{blue}m_{2}}\in\{0,1,2,3,4\}$. In Figure \ref{fig:(2,3)-edged.m11+m21.lattices}, we depict the points in $\Gamma_{\tilde{G}}^{mw}$ obeying ${\color{red}m_{1}}+{\color{blue}m_{2}}\in\{0,1,2,3,4\}$ as orange dots, and the corresponding $\Lambda_{\color{goodgreen}\delta}$-orbits as orange lines. Combining all five we see that all $\Lambda_{\color{goodgreen}\delta}$-orbits in $\Gamma_{\tilde{G}}^{mw}$ are represented exactly once, and hence we have a good parametrisation of $\Gamma_{G}^{mw}$.
\end{enumerate}

\begin{landscape}
    \begin{figure}
    \centering
    \begin{subfigure}{0.34\linewidth}
        \centering
        \latshif{c8}
        \caption{${\color{red}m_{1}}+{\color{blue}m_{2}}={\color{orange}0}$}
    \end{subfigure}%
    ~ 
    \begin{subfigure}{0.33\linewidth}
        \centering
        \latshif{c9}
        \caption{${\color{red}m_{1}}+{\color{blue}m_{2}}={\color{orange}1}$}
    \end{subfigure}%
    ~ 
    \begin{subfigure}{0.33\linewidth}
        \centering
        \latshif{c10}
        \caption{${\color{red}m_{1}}+{\color{blue}m_{2}}={\color{orange}2}$}
    \end{subfigure}
    \begin{subfigure}{0.5\linewidth}
        \centering
        \latshif{c11}
        \caption{${\color{red}m_{1}}+{\color{blue}m_{2}}={\color{orange}3}$}
    \end{subfigure}%
    ~ 
    \begin{subfigure}{0.5\linewidth}
        \centering
        \latshif{c12}
        \caption{${\color{red}m_{1}}+{\color{blue}m_{2}}={\color{orange}4}$}
    \end{subfigure}
    \caption{For \eqref{eq:(2,3)-edged.Quiver}: depiction of points in $\Gamma_{\tilde{G}}^{mw}$ obeying ${\color{red}m_{1}}+{\color{blue}m_{2}}\in\{{\color{orange}0},{\color{orange}1},{\color{orange}2},{\color{orange}3},{\color{orange}4}\}$ as orange dots, and the corresponding $\Lambda_{\color{goodgreen}\delta}$-orbits as orange lines.}
    \label{fig:(2,3)-edged.m11+m21.lattices}
\end{figure}
\end{landscape}

\section{\texorpdfstring{Quotients $\mathbb{H}^n/\mathbb{Z}_k$}{Quotients Hn/Zk}}
\label{app:AbelianQuotients}

In this section we produce quivers which are unframed, simply-laced, loop-free and unitary, with one special unitary node, whose Coulomb branch is the quotient $\mathbb{H}^n/\mathbb{Z}_k$.

The quotients $\mathbb{H}^n/\mathbb{Z}_k$ we study here are specified not only by $n$ and $k$, but also by an $n$-tuple of charges $q=(q_1\ldots,q_n)$, where $q_i\in\{1,\ldots,\lfloor\frac{k}{2}\rfloor\}$. After picking a complex structure $\mathbb{H}^n=\mathbb{C}^{2n}$, with $(x_1,\ldots,x_n,y_1,\ldots,y_n)$ the coordinates of $\mathbb{C}^{2n}$, a quotient is defined by the action:
\begin{equation}
    \begin{split}
        \mathbb{Z}_k\ni1:\quad\mathbb{C}^{2n}&\rightarrow \mathbb{C}^{2n}\\
        x_i&\mapsto \omega_k^{q_i}x_i \\
        y_i&\mapsto \omega_k^{-q_i}y_i~,\quad\textnormal{where }\omega_k=e^{\frac{2 \pi i}{k}}
    \end{split}
\end{equation}
We can introduce a $\lfloor\frac{k}{2}\rfloor$-tuple $\mathfrak{q}=(\mathfrak{q}_1,\ldots,\mathfrak{q}_{\lfloor\frac{k}{2}\rfloor})$ where $\mathfrak{q}_j$ is the number of times $j$ appears in $q$. This allows for the following notation of the quotient:
\begin{equation}
    h_{n,k,\mathfrak{q}}=\mathbb{H}^n/\mathbb{Z}_k\textnormal{ with charges given by }\mathfrak{q}~,
\end{equation}
with the shorthand notation:
\begin{equation}
    h_{n,k}:=h_{n,k,(n,0,\ldots,0)}~.
\end{equation}
The Hilbert series of the quotient is computed via the Molien sum:
\begin{equation}
    \mathrm{HS}(h_{n,k,\mathfrak{q}};t)=\frac{1}{k}\sum_{s=1}^{k}\prod_{j=1}^{\lfloor\frac{k}{2}\rfloor}\frac{1}{(1-\omega_k^{sj}t)^{\mathfrak{q}_j}(1-\omega_k^{-sj}t)^{\mathfrak{q}_j}}~.
\end{equation}
The global symmetry $G$ of the quotient (as a hyper-K\"ahler space) is
\begin{equation}
    G=\left\{ \begin{matrix}
    \bigoplus_{j=1}^{\lfloor\frac{k}{2}\rfloor}\mathfrak{u}(\mathfrak{q}_j) & k\textnormal{ odd} \\
    \bigoplus_{j=1}^{\frac{k}{2}-1}\mathfrak{u}(\mathfrak{q}_j)\oplus \usp(2\mathfrak{q}_{\frac{k}{2}}) & k\textnormal{ even}
    \end{matrix}\right.~.
\end{equation}

\subsection{Producing Quivers}

One method of producing such a quiver is the following.

\begin{enumerate}
    \item Take a unitary quiver whose Coulomb branch is $\mathbb{H}^n$. For example:
            \begin{equation}
                \mathrm{CB}\left(\raisebox{-.5\height}{\begin{tikzpicture}
                    \node[gauge,label=below:{$1$}] (1) at (1,0) {};
                    \node[gauge,label=below:{$2$}] (2) at (2,0) {};
                    \node[gauge,label=below:{$2$}] (3) at (3,0) {};
                    \node[gauge,label=below:{$2$}] (4) at (4,0) {};
                    \node[gauge,label=right:{$1$}] (5u) at (5,1) {};
                    \node[gauge,label=right:{$1$}] (5d) at (5,-1) {};
                    \draw (1)--(2)--(3)--(4)--(5u) (4)--(5d);
                \end{tikzpicture}}\right)=\mathbb{H}^8
            \end{equation}
    \item Replacing one $\U(k)$ node with an $\SU(k)$ node,\footnote{Replacing an $\U(k)$ node with an $\SU(k)$ is implemented by picking a specific sublattice of the magnetic lattice of the original quiver. One could pick other sublattices to produce different discrete quotients.} the Coulomb branch of the resulting quiver will be a $\mathbb{Z}_k$ quotient of the one before. For example:
            \begin{equation}
                \mathrm{CB}\left(\raisebox{-.5\height}{\begin{tikzpicture}
                    \node[gauge,label=below:{$1$}] (1) at (1,0) {};
                    \node[gauge,label=below:{$2$}] (2) at (2,0) {};
                    \node[gauge,label=below:{SU($2$)}] (3) at (3,0) {};
                    \node[gauge,label=below:{$2$}] (4) at (4,0) {};
                    \node[gauge,label=right:{$1$}] (5u) at (5,1) {};
                    \node[gauge,label=right:{$1$}] (5d) at (5,-1) {};
                    \draw (1)--(2)--(3)--(4)--(5u) (4)--(5d);
                \end{tikzpicture}}\right)=\mathbb{H}^8/\mathbb{Z}_2
            \end{equation}
    \item The resulting Coulomb branch may have a free part, i.e.\ it is of the form $\mathbb{H}^d\times\mathbb{H}^{n-d}/\mathbb{Z}_k$. Strap off the free part by replacing the {\color{orange} $\U(k)$} nodes with a {\color{orange}$-1$} balance by $\U(k-1)$ nodes iteratively, until there are no more unitary nodes of balance $-1$. For example:
            \begin{equation}
                \begin{tikzpicture}[baseline=-95]
                    \node[gauge,label=below:{$1$}] (1) at (1,0) {};
                    \node[gaugeo,label=below:{$2$}] (2) at (2,0) {};
                    \node[gauge,label=below:{SU($2$)}] (3) at (3,0) {};
                    \node[gauge,label=below:{$2$}] (4) at (4,0) {};
                    \node[gauge,label=right:{$1$}] (5u) at (5,1) {};
                    \node[gauge,label=right:{$1$}] (5d) at (5,-1) {};
                    \draw (1)--(2)--(3)--(4)--(5u) (4)--(5d);
                    
                    \begin{scope}[shift={(0,-3)}]
                        \node[gaugeo,label=below:{$1$}] (11) at (1,0) {};
                        \node[gauge,label=below:{$1$}] (12) at (2,0) {};
                        \node[gauge,label=below:{SU($2$)}] (13) at (3,0) {};
                        \node[gauge,label=below:{$2$}] (14) at (4,0) {};
                        \node[gauge,label=right:{$1$}] (15u) at (5,1) {};
                        \node[gauge,label=right:{$1$}] (15d) at (5,-1) {};
                        \draw (11)--(12)--(13)--(14)--(15u) (14)--(15d);
                    \end{scope}
                    
                    \begin{scope}[shift={(0,-6)}]
                        {\color{gray}\node[gauge,label=below:{$0$}] (21) at (1,0) {};}
                        \node[gauge,label=below:{$1$}] (22) at (2,0) {};
                        \node[gauge,label=below:{SU($2$)}] (23) at (3,0) {};
                        \node[gauge,label=below:{$2$}] (24) at (4,0) {};
                        \node[gauge,label=right:{$1$}] (25u) at (5,1) {};
                        \node[gauge,label=right:{$1$}] (25d) at (5,-1) {};
                        {\color{gray}\draw (21)--(22);}
                        \draw (22)--(23)--(24)--(25u) (24)--(25d);
                    \end{scope}
                    \draw[->] (3,-1.4)--(3,-2.4);
                    \draw[->] (3,-4.4)--(3,-5.4);
                    
                    {\color{orange}\node[] (a) at (2.2,0.3) {$-1$};
                    \node[] (b) at (1.2,-2.7) {$-1$};}
                \end{tikzpicture}
            \end{equation}
        \item The Coulomb branch of the final quiver should be a symplectic singularity of the form $\mathbb{H}^{n-d}/\mathbb{Z}_k$. For example:
            \begin{equation}
                \mathrm{CB}\left(\raisebox{-.5\height}{\begin{tikzpicture}
                    \node[gauge,label=below:{$1$}] (2) at (2,0) {};
                    \node[gauge,label=below:{SU($2$)}] (3) at (3,0) {};
                    \node[gauge,label=below:{$2$}] (4) at (4,0) {};
                    \node[gauge,label=right:{$1$}] (5u) at (5,1) {};
                    \node[gauge,label=right:{$1$}] (5d) at (5,-1) {};
                    \draw (2)--(3)--(4)--(5u) (4)--(5d);
                \end{tikzpicture}}\right)=\mathbb{H}^6/\mathbb{Z}_2=c_6
            \end{equation}
\end{enumerate}

\subsection{Examples}

\paragraph{Families.}
For instance, we have
\begin{equation}
\label{eq:QuotQuivFam1}
    \mathrm{CB}\left(\raisebox{-.5\height}{\scalebox{0.78}{\begin{tikzpicture}
        \node[gauge,label=below:{$1$}] (1) at (1,0) {};
        \node[gauge,label=below:{$2$}] (2) at (2,0) {};
        \node (3) at (3,0) {$\ldots$};
        \node[gauge,label={[xshift=-5]below:{\small$N-1$}}] (4) at (4,0) {};
        \node[gauge,label=below:{SU($N$)}] (5) at (5.3,0) {};
        \node[gauge,label=below:{$N$}] (6) [right=8mm of 5] {};
        \node (7) [right=4mm of 6] {$\ldots$};
        \node[gauge,label=below:{$N$}] (8) at (8,0) {};
        \node[gauge,label=right:{$1$}] (8u) at (8,1) {};
        \node[gauge,label={[xshift=5]below:{\small$N-1$}}] (9) at (9,0) {};
        \node (10) at (10,0) {$\ldots$};
        \node[gauge,label=below:{$2$}] (11) at (11,0) {};
        \node[gauge,label=below:{$1$}] (12) at (12,0) {};
        \draw (1)--(2)--(3)--(4)--(5)--(6)--(7)--(8)--(9)--(10)--(11)--(12) (8)--(8u);
		\draw [decorate,decoration={brace,amplitude=5pt}] (8.3,-0.7)--(4.7,-0.7);
		\node at (6.5,-1.2) {$m$};
    \end{tikzpicture}
    }}\right)=h_{(N+m-1)N,N}
\end{equation}
and
\begin{equation}
\mathrm{CB}\left(\raisebox{-.5\height}{\scalebox{0.9}{\begin{tikzpicture}
        \node[gauge,label=below:{$1$}] (1) at (1,0) {};
        \node[gauge,label=below:{$2$}] (2) at (2,0) {};
        \node (3) at (3,0) {$\ldots$};
        \node[gauge,label={[xshift=-8]below:{\small$2N-1$}}] (4) at (4,0) {};
        \node[gauge,label=below:{\small $\SU(2N)$}] (5) at (5.5,0) {};
        \node[gauge,label=below:{$2N$}] (6) at (7,0) {};
        \node (7) at (8,0) {$\ldots$};
       \node[gauge,label=below:{$2N$}] (8) at (9,0) {};
       \node[gauge,label=below:{$N$}] (9) at (10.2,0) {};
        \draw (1)--(2)--(3)--(4)--(5)--(6)--(7)--(8);
        \draw[transform canvas={yshift=-1.5pt}] (8)--(9);
       \draw[transform canvas={yshift=1.5pt}] (8)--(9);
       \node (>) at (9.6,0) {\rotatebox{0} {\Large $>$}};
		 \draw [decorate,decoration={brace,amplitude=5pt}] (9.3,-0.7)--(4.7,-0.7);
		\node at (7,-1.2) {$m$};
\end{tikzpicture}}}\right)
=h_{2(N+m)N-1,2N}
\end{equation}

\paragraph{Special quivers.}
Let us construct some special quivers. We start with an $E_n$ affine Dynkin quiver, whose Coulomb branch is $\overline{\mathrm{min}.E_n}$, and we lower the rank of a single node by $1$. This produces a quiver whose Coulomb branch is freely generated. We can turn one of the unitary nodes in the resulting quiver into a special unitary node, then strap off the free part as explained before. The resulting quiver is either part of the \eqref{eq:QuotQuivFam1} family, or it is given in Table \ref{tab:Quotients} below.

\footnotesize
\begin{longtable}{|c|c|c|}
    \hline
    $\mathsf{Q}$ & $\mathrm{CB}(\mathsf{Q})$ & Global symmetry \\
    & & of $\mathrm{CB}(\mathsf{Q})$ \\
    \hline
    \hline
    \raisebox{-.5\height}{\scalebox{0.9}{\begin{tikzpicture}
        \node[gauge,label=below:{$\SU(2)$}] (2) at (2,0) {};
        \node[gauge,label=below:{$3$}] (3) at (3,0) {};
        \node[gauge,label=below:{$4$}] (4) at (4,0) {};
        \node[gauge,label=below:{$5$}] (5) at (5,0) {};
        \node[gauge,label=below:{$6$}] (6) at (6,0) {};
        \node[gauge,label=right:{$3$}] (6u) at (6,1) {};
        \node[gauge,label=below:{$4$}] (7) at (7,0) {};
        \node[gauge,label=below:{$2$}] (8) at (8,0) {};
        \draw (2)--(3)--(4)--(5)--(6)--(7)--(8) (6)--(6u);
        \node at (0.5,1.5) {};
        \node at (8.5,-1) {};
    \end{tikzpicture}}}      & $h_{28,2}$ & $\usp(56)$ \\ \hline
    \raisebox{-.5\height}{\scalebox{0.9}{\begin{tikzpicture}
        \node[gauge,label=below:{$1$}] (2) at (2,0) {};
        \node[gauge,label=below:{$\SU(3)$}] (3) at (3,0) {};
        \node[gauge,label=below:{$4$}] (4) at (4,0) {};
        \node[gauge,label=below:{$5$}] (5) at (5,0) {};
        \node[gauge,label=below:{$6$}] (6) at (6,0) {};
        \node[gauge,label=right:{$3$}] (6u) at (6,1) {};
        \node[gauge,label=below:{$4$}] (7) at (7,0) {};
        \node[gauge,label=below:{$2$}] (8) at (8,0) {};
        \draw (2)--(3)--(4)--(5)--(6)--(7)--(8) (6)--(6u);
        \node at (0.5,1.5) {};
        \node at (8.5,-1) {};
    \end{tikzpicture}}}     & $h_{27,3}$ & $\mathfrak{u}(27)$ \\ \hline
    \raisebox{-.5\height}{\scalebox{0.9}{\begin{tikzpicture}
        \node[gauge,label=below:{$1$}] (2) at (2,0) {};
        \node[gauge,label=below:{$2$}] (3) at (3,0) {};
        \node[gauge,label=below:{$\SU(4)$}] (4) at (4,0) {};
        \node[gauge,label=below:{$5$}] (5) at (5,0) {};
        \node[gauge,label=below:{$6$}] (6) at (6,0) {};
        \node[gauge,label=right:{$3$}] (6u) at (6,1) {};
        \node[gauge,label=below:{$4$}] (7) at (7,0) {};
        \node[gauge,label=below:{$2$}] (8) at (8,0) {};
        \draw (2)--(3)--(4)--(5)--(6)--(7)--(8) (6)--(6u);
        \node at (0.5,1.5) {};
        \node at (8.5,-1) {};
    \end{tikzpicture}}}     & $h_{26,4,(16,10)}$ & $\mathfrak{u}(16)\oplus\usp(20)$ \\ \hline
    \raisebox{-.5\height}{\scalebox{0.9}{\begin{tikzpicture}
        \node[gauge,label=below:{$1$}] (1) at (1,0) {};
        \node[gauge,label=below:{$2$}] (2) at (2,0) {};
        \node[gauge,label=below:{$3$}] (3) at (3,0) {};
        \node[gauge,label=below:{$\SU(4)$}] (4) at (4,0) {};
        \node[gauge,label=below:{$4$}] (5) at (5,0) {};
        \node[gauge,label=below:{$4$}] (6) at (6,0) {};
        \node[gauge,label=right:{$2$}] (6u) at (6,1) {};
        \node[gauge,label=below:{$2$}] (7) at (7,0) {};
        \node[gauge,label=below:{$1$}] (8) at (8,0) {};
        \draw (1)--(2)--(3)--(4)--(5)--(6)--(7)--(8) (6)--(6u);
        \node at (0.5,1.5) {};
        \node at (8.5,-1) {};
    \end{tikzpicture}}}     & $h_{22,4,(16,6)}$ & $\mathfrak{u}(16)\oplus\usp(12)$ \\ \hline
    \raisebox{-.5\height}{\scalebox{0.9}{\begin{tikzpicture}
        \node[gauge,label=below:{$1$}] (2) at (2,0) {};
        \node[gauge,label=below:{$2$}] (3) at (3,0) {};
        \node[gauge,label=below:{$3$}] (4) at (4,0) {};
        \node[gauge,label=below:{$\SU(5)$}] (5) at (5,0) {};
        \node[gauge,label=below:{$6$}] (6) at (6,0) {};
        \node[gauge,label=right:{$3$}] (6u) at (6,1) {};
        \node[gauge,label=below:{$4$}] (7) at (7,0) {};
        \node[gauge,label=below:{$2$}] (8) at (8,0) {};
        \draw (2)--(3)--(4)--(5)--(6)--(7)--(8) (6)--(6u);
        \node at (0.5,1.5) {};
        \node at (8.5,-1) {};
    \end{tikzpicture}}}     & $h_{25,5,(10,15)}=h_{25,5,(15,10)}$ & $\mathfrak{u}(10)\oplus\mathfrak{u}(15)$ \\ \hline
    \raisebox{-.5\height}{\scalebox{0.9}{\begin{tikzpicture}
        \node[gauge,label=below:{$1$}] (1) at (1,0) {};
        \node[gauge,label=below:{$2$}] (2) at (2,0) {};
        \node[gauge,label=below:{$3$}] (3) at (3,0) {};
        \node[gauge,label=below:{$4$}] (4) at (4,0) {};
        \node[gauge,label=below:{$\SU(5)$}] (5) at (5,0) {};
        \node[gauge,label=below:{$5$}] (6) at (6,0) {};
        \node[gauge,label=right:{$2$}] (6u) at (6,1) {};
        \node[gauge,label=below:{$3$}] (7) at (7,0) {};
        \node[gauge,label=below:{$1$}] (8) at (8,0) {};
        \draw (1)--(2)--(3)--(4)--(5)--(6)--(7)--(8) (6)--(6u);
        \node at (0.5,1.5) {};
        \node at (8.5,-1) {};
    \end{tikzpicture}}}     & $h_{25,5,(10,15)}=h_{25,5,(15,10)}$ & $\mathfrak{u}(10)\oplus\mathfrak{u}(15)$\\
    \hline
    \raisebox{-.5\height}{\begin{tikzpicture}
        \node[gauge,label=below:{$1$}] (2) at (2,0) {};
        \node[gauge,label=below:{$2$}] (3) at (3,0) {};
        \node[gauge,label=below:{$3$}] (4) at (4,0) {};
        \node[gauge,label=below:{$4$}] (5) at (5,0) {};
        \node[gauge,label=below:{$\SU(6)$}] (6) at (6,0) {};
        \node[gauge,label=right:{$3$}] (6u) at (6,1) {};
        \node[gauge,label=below:{$4$}] (7) at (7,0) {};
        \node[gauge,label=below:{$2$}] (8) at (8,0) {};
        \draw (2)--(3)--(4)--(5)--(6)--(7)--(8) (6)--(6u);
        \node at (0.5,1.5) {};
        \node at (8.5,-1) {};
    \end{tikzpicture}}    & $h_{24,6,(6,12,6)}$ & $\mathfrak{u}(6)\oplus\mathfrak{u}(12)\oplus\usp(12)$ \\ \hline
    \raisebox{-.5\height}{\begin{tikzpicture}
        \node[gauge,label=below:{$1$}] (1) at (1,0) {};
        \node[gauge,label=below:{$2$}] (2) at (2,0) {};
        \node[gauge,label=below:{$3$}] (3) at (3,0) {};
        \node[gauge,label=below:{$4$}] (4) at (4,0) {};
        \node[gauge,label=below:{$5$}] (5) at (5,0) {};
        \node[gauge,label=below:{$\SU(6)$}] (6) at (6,0) {};
        \node[gauge,label=right:{$2$}] (6u) at (6,1) {};
        \node[gauge,label=below:{$4$}] (7) at (7,0) {};
        \node[gauge,label=below:{$2$}] (8) at (8,0) {};
        \draw (1)--(2)--(3)--(4)--(5)--(6)--(7)--(8) (6)--(6u);
        \node at (0.5,1.5) {};
        \node at (8.5,-1) {};
    \end{tikzpicture}}     & $h_{28,6,(18,0,10)}$ & $\mathfrak{u}(18)\oplus\usp(20)$ \\ \hline
    \raisebox{-.5\height}{\begin{tikzpicture}
        \node[gauge,label=below:{$1$}] (1) at (1,0) {};
        \node[gauge,label=below:{$2$}] (2) at (2,0) {};
        \node[gauge,label=below:{$3$}] (3) at (3,0) {};
        \node[gauge,label=below:{$4$}] (4) at (4,0) {};
        \node[gauge,label=below:{$5$}] (5) at (5,0) {};
        \node[gauge,label=below:{$\SU(6)$}] (6) at (6,0) {};
        \node[gauge,label=right:{$3$}] (6u) at (6,1) {};
        \node[gauge,label=below:{$3$}] (7) at (7,0) {};
        \node[gauge,label=below:{$1$}] (8) at (8,0) {};
        \draw (1)--(2)--(3)--(4)--(5)--(6)--(7)--(8) (6)--(6u);
        \node at (0.5,1.5) {};
        \node at (8.5,-1) {};
    \end{tikzpicture}}     & $h_{27,6,(12,15,0)}$ & $\mathfrak{u}(12)\oplus\mathfrak{u}(15)$\\ \hline
    \raisebox{-.5\height}{\begin{tikzpicture}
        \node[gauge,label=below:{$1$}] (2) at (2,0) {};
        \node[gauge,label=below:{$2$}] (3) at (3,0) {};
        \node[gauge,label=below:{$3$}] (4) at (4,0) {};
        \node[gauge,label=below:{$4$}] (5) at (5,0) {};
        \node[gauge,label=below:{$5$}] (6) at (6,0) {};
        \node[gauge,label=right:{$2$}] (6u) at (6,1) {};
        \node[gauge,label=below:{$\SU(4)$}] (7) at (7,0) {};
        \node[gauge,label=below:{$2$}] (8) at (8,0) {};
        \draw (2)--(3)--(4)--(5)--(6)--(7)--(8) (6)--(6u);
        \node at (0.5,1.5) {};
        \node at (8.5,-1) {};
    \end{tikzpicture}}     & $h_{22,4,(12,10)}$ & $\mathfrak{u}(12)\oplus\usp(20)$ \\ \hline
    \raisebox{-.5\height}{\begin{tikzpicture}
        \node[gauge,label=below:{$1$}] (1) at (1,0) {};
        \node[gauge,label=below:{$2$}] (2) at (2,0) {};
        \node[gauge,label=below:{$3$}] (3) at (3,0) {};
        \node[gauge,label=below:{$4$}] (4) at (4,0) {};
        \node[gauge,label=below:{$5$}] (5) at (5,0) {};
        \node[gauge,label=below:{$6$}] (6) at (6,0) {};
        \node[gauge,label=right:{$3$}] (6u) at (6,1) {};
        \node[gauge,label=below:{$\SU(4)$}] (7) at (7,0) {};
        \node[gauge,label=below:{$1$}] (8) at (8,0) {};
        \draw (1)--(2)--(3)--(4)--(5)--(6)--(7)--(8) (6)--(6u);
        \node at (0.5,1.5) {};
        \node at (8.5,-1) {};
    \end{tikzpicture}}     & $h_{28,4}$ & $\mathfrak{u}(28)$ \\ \hline
    \raisebox{-.5\height}{\begin{tikzpicture}
        \node[gauge,label=below:{$1$}] (3) at (3,0) {};
        \node[gauge,label=below:{$2$}] (4) at (4,0) {};
        \node[gauge,label=below:{$3$}] (5) at (5,0) {};
        \node[gauge,label=below:{$4$}] (6) at (6,0) {};
        \node[gauge,label=right:{$2$}] (6u) at (6,1) {};
        \node[gauge,label=below:{$3$}] (7) at (7,0) {};
        \node[gauge,label=below:{$\SU(2)$}] (8) at (8,0) {};
        \draw (3)--(4)--(5)--(6)--(7)--(8) (6)--(6u);
        \node at (0.5,1.5) {};
        \node at (8.5,-1) {};
    \end{tikzpicture}}     & $h_{16,2}$ & $\usp(32)$ \\ \hline
    \raisebox{-.5\height}{\begin{tikzpicture}
        \node[gauge,label=below:{$1$}] (2) at (2,0) {};
        \node[gauge,label=below:{$2$}] (3) at (3,0) {};
        \node[gauge,label=below:{$3$}] (4) at (4,0) {};
        \node[gauge,label=below:{$4$}] (5) at (5,0) {};
        \node[gauge,label=below:{$5$}] (6) at (6,0) {};
        \node[gauge,label=right:{$\SU(3)$}] (6u) at (6,1) {};
        \node[gauge,label=below:{$3$}] (7) at (7,0) {};
        \node[gauge,label=below:{$1$}] (8) at (8,0) {};
        \draw (2)--(3)--(4)--(5)--(6)--(7)--(8) (6)--(6u);
        \node at (0.5,1.5) {};
        \node at (8.5,-1) {};
    \end{tikzpicture}}     & $h_{21,3}$ & $\mathfrak{u}(21)$\\
    \hline
    \raisebox{-.5\height}{\begin{tikzpicture}
        \node[gauge,label=below:{$1$}] (1) at (1,0) {};
        \node[gauge,label=below:{$2$}] (2) at (2,0) {};
        \node[gauge,label=below:{$3$}] (3) at (3,0) {};
        \node[gauge,label=below:{SU($4$)}] (4) at (4,0) {};
        \node[gauge,label=right:{$2$}] (4u) at (4,1) {};
        \node[gauge,label=below:{$2$}] (5) at (5,0) {};
        \node[gauge,label=below:{$1$}] (6) at (6,0) {};
        \draw (1)--(2)--(3)--(4)--(5)--(6) (4)--(4u);
    \end{tikzpicture}} & $h_{14,4,(8,6)}$ & $\mathfrak{u}(8)\oplus\usp(12)$ \\
    \hline
    \caption{Quivers with $\mathbb{H}^n/\mathbb{Z}_k$ Coulomb branches and their global symmetry.}
    \label{tab:Quotients}
    \end{longtable}

\bibliographystyle{JHEP}
\bibliography{bibli.bib}

\end{document}